\documentclass[twocolumn]{aastex62}
\newcommand{\gt}{>}

\accepted{ApJ}

\shorttitle{STUDIES III: Properties of 450\,\lowercase{$\micron$} Sources}
\shortauthors{C.-F. Lim et al.}
\let\oldAA\AA
\renewcommand{\AA}{\text{\normalfont\oldAA}}

\graphicspath {{figures/}}
\begin{document}

\title{SCUBA-2 ULTRA DEEP IMAGING EAO SURVEY (STUDIES) III: Multi-wavelength properties, luminosity functions and preliminary source catalog of 450-$\micron$-selected galaxies}

\correspondingauthor{Chen-Fatt Lim}
\email{chenfatt.lim@gmail.com}

\author{Chen-Fatt Lim}
\affiliation{Graduate Institute of Astrophysics, National Taiwan University, Taipei 10617, Taiwan}
\affiliation{Academia Sinica Institute of Astronomy and Astrophysics (ASIAA), No. 1, Section 4, Roosevelt Road, Taipei 10617, Taiwan}

\author{Wei-Hao Wang}
\affiliation{Academia Sinica Institute of Astronomy and Astrophysics (ASIAA), No. 1, Section 4, Roosevelt Road, Taipei 10617, Taiwan}

\author{Ian Smail}
\affiliation{Centre for Extragalactic Astronomy, Department of Physics, Durham University, South Road, Durham, DH1 3LE, UK}

\author{Douglas Scott}
\affiliation{Department of Physics \& Astronomy, University of British Columbia, BC, Canada}

\author{Chian-Chou Chen}
\affiliation{European Southern Observatory, Karl Schwarzschild Strasse 2, Garching, Germany}
\affiliation{Academia Sinica Institute of Astronomy and Astrophysics (ASIAA), No. 1, Section 4, Roosevelt Road, Taipei 10617, Taiwan}

\author{Yu-Yen Chang}
\affiliation{Academia Sinica Institute of Astronomy and Astrophysics (ASIAA), No. 1, Section 4, Roosevelt Road, Taipei 10617, Taiwan}

\author{James M. Simpson}
\affiliation{Academia Sinica Institute of Astronomy and Astrophysics (ASIAA), No. 1, Section 4, Roosevelt Road, Taipei 10617, Taiwan}

\author{Yoshiki Toba}
\affiliation{Department of Astronomy, Kyoto University, Kitashirakawa-Oiwake-cho, Sakyo-ku, Kyoto 606-8502, Japan}
\affiliation{Academia Sinica Institute of Astronomy and Astrophysics (ASIAA), No. 1, Section 4, Roosevelt Road, Taipei 10617, Taiwan}
\affiliation{Research Center for Space and Cosmic Evolution, Ehime University, 2-5 Bunkyo-cho, Matsuyama, Ehime 790-8577, Japan}

\author{Xinwen Shu}
\affiliation{Department of Physics, Anhui Normal University, Wuhu, Anhui, 241000, People’s Republic of China}

\author{Dave Clements}
\affiliation{Blackett Lab, Imperial College, London, Prince Consort Road, London SW7 2AZ, UK}

\author{Josh Greenslade}
\affiliation{Kapteyn Astronomical Institute, University of Groningen, Postbus 800, 9700 AV, Groningen, The Netherlands}

\author{YiPing Ao}
\affiliation{Purple Mountain Observatory, Chinese Academy of Sciences, Nanjing 210033, People’s Republic of China}

\author{Arif Babul}
\affiliation{Department of Physics and Astronomy, University of Victoria, Elliott Building, 3800 Finnerty Road, Victoria, BC V8P 5C2, Canada}

\author{Jack Birkin}
\affiliation{Centre for Extragalactic Astronomy, Department of Physics, Durham University, South Road, Durham, DH1 3LE, UK}

\author{Scott C. Chapman}
\affiliation{Department of Physics and Astronomy, University of British Columbia, 6225 Agricultural Road, Vancouver, BC, V6T 1Z1, Canada}
\affiliation{National Research Council, Herzberg Astronomy and Astrophysics, 5071 West Saanich Road, Victoria, BC, V9E 2E7, Canada}
\affiliation{Department of Physics and Atmospheric Science, Dalhousie University, Halifax, NS B3H 4R2, Canada}

\author{Tai-An Cheng}
\affiliation{Blackett Lab, Imperial College, London, Prince Consort Road, London SW7 2AZ, UK}

\author{Brian S. Cho}
\affiliation{Astronomy Program, Department of Physics and Astronomy, Seoul National University, Gwanak-gu, Seoul 151-742, Republic of Korea}

\author{Helmut Dannerbauer}
\affiliation{Instituto de Astrofisica de Canarias (IAC), E-38205 La Laguna, Tenerife, Spain}
\affiliation{Universidad de La Laguna, Dpto. Astrofisica, E-38206 La Laguna, Tenerife, Spain}

\author{Ugn\.e Dudzevi\v{c}i\={u}t\.e}
\affiliation{Centre for Extragalactic Astronomy, Department of Physics, Durham University, South Road, Durham, DH1 3LE, UK}

\author{James Dunlop}
\affiliation{Institute for Astronomy, University of Edinburgh, Blackford Hill, Edinburgh, EH9 3HJ, UK}

\author{Yu Gao}
\affiliation{Purple Mountain Observatory, Chinese Academy of Sciences, Nanjing 210033, People’s Republic of China}
\affiliation{Purple Mountain Observatory PMO/Key Lab of Radio Astronomy, Department of Astronomy, Xiamen University, Xiamen, Fujian 361005, People's Republic of China}

\author{Tomotsugu Goto}
\affiliation{National Tsing Hua University, No. 101, Section 2, Kuang-Fu Road, Hsinchu 30013, Taiwan}

\author{Luis C. Ho}
\affiliation{Kavli Institute for Astronomy and Astrophysics, Peking University, Beijing 100871, People’s Republic of China}
\affiliation{Department of Astronomy, School of Physics, Peking University, Beijing 100871, People’s Republic of China}

\author{Li-Ting Hsu}
\affiliation{Academia Sinica Institute of Astronomy and Astrophysics (ASIAA), No. 1, Section 4, Roosevelt Road, Taipei 10617, Taiwan}

\author{Ho Seong Hwang}
\affiliation{Korea Astronomy and Space Science Institute, 776 Daedeokdae-ro, Yuseong-gu, Daejeon 34055, Republic of Korea}

\author{Woong-Seob Jeong}
\affiliation{Korea Astronomy and Space Science Institute, 776 Daedeokdae-ro, Yuseong-gu, Daejeon 34055, Republic of Korea}
\affiliation{Korea University of Science and Technology, 217 Gajeong-ro, Yuseong-gu, Daejeon 34113, Republic of Korea}

\author{Maciej Koprowski}
\affiliation{Institute of Astronomy, Faculty of Physics, Astronomy and Informatics, Nicolaus Copernicus University, Grudziadzka 5, 87-100 Torun, Poland}

\author{Chien-Hsiu Lee}
\affiliation{NSF's National Optical-Infrared Astronomy Research Laboratory, 950 North Cherry Avenue, Tucson, AZ 85719, USA}

\author{Ming-Yi Lin}
\affiliation{Academia Sinica Institute of Astronomy and Astrophysics (ASIAA), No. 1, Section 4, Roosevelt Road, Taipei 10617, Taiwan}

\author{Wei-Ching Lin}
\affiliation{Graduate Institute of Physics, National Taiwan University, Taipei 10617, Taiwan}
\affiliation{Academia Sinica Institute of Astronomy and Astrophysics (ASIAA), No. 1, Section 4, Roosevelt Road, Taipei 10617, Taiwan}

\author{Micha{\l} J.~Micha{\l}owski}
\affiliation{Astronomical Observatory Institute, Faculty of Physics, Adam Mickiewicz University, 60-286 Pozna$\acute{n}$, Poland}

\author{Harriet Parsons}
\affiliation{Joint Astronomy Centre, 660 North A’ohoku Place, University Park, Hilo, HI 96720, USA}

\author{Marcin Sawicki}
\affiliation{Department of Astronomy \& Physics and the Institute for Computational Astrophysics, Saint Mary’s University, Halifax, NS B3H 3C3, Canada}

\author{Raphael Shirley}
\affiliation{Astronomy Centre, Department of Physics and Astronomy, University of Sussex, Falmer, Brighton BN1 9QH, UK}

\author{Hyunjin Shim}
\affiliation{Department of Earth Science Education, Kyungpook National University, Deagu 41566, Republic of Korea}

\author{Sheona Urquhart}
\affiliation{School of Physical Sciences, The Open University, Walton Hall, Milton Keynes, MK7 6AA, UK}

\author{Jianfa Wang}
\affiliation{Purple Mountain Observatory, Chinese Academy of Sciences, Nanjing 210033, People’s Republic of China}

\author{Tao Wang} %[0000-0002-2504-2421]
\affiliation{Institute of Astronomy, Graduate School of Science, The University of Tokyo, 2-21-1 Osawa, Mitaka, Tokyo 181-0015, Japan}
\affiliation{National Astronomical Observatory of Japan, Mitaka, Tokyo 181-8588, Japan}

%%==============
%%   abstract
%%==============
\begin{abstract}

We construct a SCUBA-2 450-$\micron$ map in the COSMOS field that covers an area of 300\,arcmin$^{2}$ and reaches a 1$\sigma$ noise level of 0.65\,mJy in the deepest region. We extract 256 sources detected at 450\,$\micron$ with signal-to-noise ratios $>$ 4.0 and analyze the physical properties of their multi-wavelength counterparts. We find that most of the sources are at $z\lesssim3$, with a median of $z = 1.79^{+0.03}_{-0.15}$. About $35^{+32}_{-25}$\% of our sources are classified as starburst galaxies based on their total star-formation rates (SFRs) and stellar masses ($M_{\ast}$). By fitting the far-infrared spectral energy distributions, we find that our 450-$\micron$-selected sample has a wide range of dust temperatures (20\,K $ \lesssim T_{\rm d} \lesssim$ 60\,K), with a median of ${T}_{\rm d} = 38.3^{+0.4}_{-0.9}$\,K.  We do not find a redshift evolution in dust temperature for sources with $L_{\rm IR} > 10^{12}\,\rm L_\sun$ at $z<3$. However, we find a moderate correlation where the dust temperature increases with the deviation from the SFR--$M_{\ast}$ relation. The increase in dust temperature also correlates with optical morphology, which is consistent with merger-triggered starbursts in sub-millimeter galaxies. Our galaxies do not show the tight IRX--$\beta_{\rm UV}$ correlation that has been observed in the local Universe. We construct the infrared luminosity functions of our 450-$\micron$ sources and measure their comoving SFR densities (SFRDs). The contribution of the $L_{\rm IR} > 10^{12}\,\rm L_\sun$ population to the SFRD rises dramatically from $z$ = 0 to 2 ($\propto$ ($1+z$)$^{3.9\pm1.1}$) and dominates the total SFRD at $z \gtrsim 2$. 

\end{abstract}

%%  keywords

%%==============
%%   introduction 
%%==============
\section{Introduction} \label{sec:Introduction}

In the past two decades, intensive work has revealed that most sub-millimeter galaxies (SMGs; \citealt{Smail:1997aa, Barger:1998aa, Barger:1999aa, Hughes:1998aa, Eales:1999aa}) lie at $z$ $\sim$ 1.5--3.5 \citep{Barger:2000aa, Chapman:2003ab, Chapman:2005aa, Pope:2006aa, Aretxaga:2007aa, Michaowski:2012ab, Yun:2012aa, Simpson:2014aa, Simpson:2017ab, Chen:2016ab, Dunlop:2017aa, Michaowski:2017aa}, occupying the same putative peak epoch of star formation \citep{Madau:2014aa} and active galactic nucleus (AGN) activity \citep{Schmidt:1995aa, Hasinger:2005aa, Wall:2008aa, Assef:2011aa}. The SMGs also dominate the massive-end of the star formation main sequence \citep{Swinbank:2004aa, Tacconi:2006aa, Tacconi:2008aa, Hainline:2011aa, Michaowski:2012aa, Michaowski:2017aa, da-Cunha:2015aa, Dunlop:2017aa} with star-formation rates (SFRs) ranging from 100 to $>\,1000\,\rm M_{\sun}\,yr^{-1}$ \citep{Michaowski:2010aa, Hainline:2011aa, Barger:2012aa, Simpson:2015aa}. Furthermore, clustering analyses have revealed that SMGs reside in high-mass (10$^{12}$--10$^{13}$\,$h^{-1}$\,$\rm M_{\sun}$) dark matter halos \citep{Blain:2004aa, Farrah:2006aa, Magliocchetti:2007aa, Hickox:2012aa, Chen:2016aa, Wilkinson:2017aa}, suggesting that SMGs may be the progenitors of elliptical galaxies in the local Universe \citep{Miller:2018aa}. Despite this progress, our understanding of this population is still incomplete in number counts \citep{Karim:2013aa}, stellar masses \citep{Michaowski:2012aa, Michaowski:2014aa, Zhang:2018aa}, and the triggering mechanism of the star formation \citep{Targett:2011aa, Targett:2013aa, Hodge:2016aa}, especially at the faint and high-redshift ends.

The peak of the rest-frame spectral energy distribution (SED) of typical SMGs is at $\lambda_{\rm rest}\simeq100\,\micron$. Space observations such as \emph{Spitzer}/MIPS (24, 70, and 160\,$\micron$; \citealt{Rieke:2004aa}), \emph{AKARI}/FIS (65, 90, 140, and 160\,$\micron$; \citealt{Murakami:2007aa}), \emph{Herschel}/PACS (70, 100, and 160\,$\micron$; \citealt{Poglitsch:2010aa}), and \emph{Herschel}/SPIRE (250, 350, and 500\,$\micron$; \citealt{Griffin:2010aa}) can constrain the SED of SMGs near the peak of the modified blackbody emission. However, the insufficient resolution of far-infrared (FIR) or single-dish sub-millimeter surveys ($15\arcsec$--$35\arcsec$) limits our ability to detect and identify sources below the confusion limit. For instance, the confusion limits of \emph{Herschel}/SPIRE are $S_\mathrm{250\,\micron} \simeq 12$\,mJy, $S_\mathrm{350\,\micron} \simeq 14$\,mJy, and $S_\mathrm{500\,\micron} \simeq 15$\,mJy \citep{Casey:2012aa}, corresponding to the SFR range of $\simeq$ 500--1500$\,\rm M_{\sun}$\,yr$^{-1}$ for an SMG with a dust temperature ($T_{\rm d}$) of 20--50\,K at $z\simeq2$. Although \emph{Herschel}/SPIRE can be pushed significantly deeper than the above confusion limits with de-blending methods (e.g., DESPHOT, \citealt{Roseboom:2010aa, Roseboom:2012aa}; T-PHOT, \citealt{Merlin:2015aa}; XID+, \citealt{Hurley:2017aa}), the results are dependent on the depths of the positional priors and thus limit our understanding to sources that are already detected in high-resolution shorter-wavelength observations.

With ground-based observations, our understanding of SMGs primarily comes from 850-$\micron$- and 1-mm-selected samples \citep{Smail:1997aa, Barger:1998aa, Hughes:1998aa, Scott:2002aa, Scott:2008aa, Borys:2003aa, Webb:2003aa, Wang:2004aa, Coppin:2005aa, Laurent:2005aa, Mortier:2005aa, Bertoldi:2007aa, Greve:2008aa, WeiB:2009aa, Vieira:2010aa, Aretxaga:2011aa, Hatsukade:2011aa, Chen:2013aa, Chen:2013ab, Geach:2013aa, Geach:2017aa, Mocanu:2013aa, Marsden:2014aa, Staguhn:2014aa, Hsu:2016aa, Cowie:2017aa, Cowie:2018aa, Simpson:2019aa} because of the atmospheric windows. These wavebands are offset from the peaks of the SED of typical SMGs, even for redshifts of $z = 1$--3. Another available window is located at 450\,$\micron$, which is closer to the redshifted SED peak; however, the atmospheric transmission is only about half of that for the 850-$\micron$ window even at the best sites. Several efforts have been made to obtain shorter wavelength sub-millimeter measurements to sample the rest-frame peak of dust emission. Follow up 350-$\micron$ observations of 850-$\micron$ sources were conducted using the second-generation Sub-millimeter High Angular Resolution Camera (SHARC-2) at the Caltech Submillimeter Observatory \citep{Kovacs:2006aa, Coppin:2008aa}. 450-$\micron$ observations of the 850-$\micron$ population were made with the Sub-millimeter Common User Bolometric Array (SCUBA) on the James Clerk Maxwell Telescope (JCMT; \citealt{Chapman:2002aa, Smail:2002aa}) but limited to a population that is extremely bright at sub-millimeter wavelengths. Although interferometric observations with the Atacama Large Millimeter/sub-millimeter Array (ALMA) have detected SMGs with SFRs $< 100\,\rm M_{\sun}$\,yr$^{-1}$ \citep{Aravena:2016aa, Hatsukade:2016aa, Hatsukade:2018aa, Dunlop:2017aa, Franco:2018aa}, it is extremely time-consuming to obtain large samples with ALMA due to its limited field of view.

Efficient 450-$\micron$ imaging surveys were enabled by the Sub-millimeter Common User Bolometric Array-2 \citep[SCUBA-2; ][]{Holland:2013aa} on the 15-m JCMT. SCUBA-2 contains 5000 pixels (field of view $\simeq45$\,arcmin$^{2}$) in each of the 450- and 850-$\micron$ detector arrays, meaning that it can efficiently survey large areas of sky at 450\,$\micron$ and 850\,$\micron$ simultaneously. The beam size at 450\,$\micron$ (7${\farcs}$9) is nearly two times smaller than that at 850\,$\micron$ ($13\arcsec$). This provides an important advantage for multi-wavelength counterpart identification, as the maps are less confused. For example, comparing to the 36$\arcsec$ resolution at 500\,$\micron$ for \emph{Herschel}, the confusion limit of SCUBA-2 at 450\,$\micron$ is about 20 times lower. The 450-$\micron$ SMG surveys, despite being more challenging, can probe more typical dusty galaxies at $z\simeq 1$--2, the peak epoch of both star formation and AGN activity.

To date, there have only been a handful of studies of 450-$\micron$-selected SMGs. The deepest SCUBA-2 450-$\micron$ blank-field surveys, with detection limits of 3--5\,mJy, have resolved 20--50\% of the 450-$\micron$ extragalactic background light \citep{Casey:2013aa, Geach:2013aa, Wang:2017aa, Zavala:2017aa}. Lensing cluster surveys have reached intrinsic unlensed 450-$\micron$ flux densities of $\lesssim1$\,mJy and have nearly fully resolved the 450-$\micron$ extragalactic background light \citep{Chen:2013ab, Chen:2013aa, Hsu:2016aa}. The physical properties of the 450-$\micron$ population have been examined in several studies based on shallow 450-$\micron$ maps with noise levels of $\sigma_{450\,\micron}$ = 1.0--4.2\,mJy \citep{Casey:2013aa, Roseboom:2013aa, Bourne:2017aa, Cowie:2017aa, Zavala:2018aa}. Such 450-$\micron$-selected galaxies occupy similar parameter spaces to 850-$\micron$ sources in infrared luminosity ($L_{\rm IR}$), SFR, and stellar mass, with typical ranges of 10$^{11.5}$--10$^{13}$\,$\rm L_{\sun}$, 100--1000\,$\rm M_{\sun}$\,yr$^{-1}$, and 10$^{10.5}$--10$^{11.5}$\,$\rm M_{\sun}$, respectively. However, 450-$\micron$ sources have dust temperatures higher than those of 850-$\micron$ sources by roughly 10\,K \citep{Casey:2013aa, Roseboom:2013aa}. They are also at somewhat lower redshifts, with a peak of the redshift distribution at $z = 1.5$--2.0, (\citealp{Casey:2013aa, Simpson:2014aa, Bourne:2017aa, Zavala:2018aa}; see $z = 2.5$--3.0 for 850-$\micron$ sources). Current studies are limited by the sample size of 450-$\micron$ sources ($\lesssim100$ SMGs in each of the aforementioned studies), and the samples are biased toward relatively bright sources ($L_{\rm IR} = 10^{11.5}$--$10^{13}\,\rm L_{\sun}$). We push the sensitivity limit of 450-$\micron$ imaging by initiating a new 450-$\micron$ imaging survey in the Cosmic Evolution Survey (COSMOS; \citealt{Scoville:2007aa}) field, called the SCUBA-2 Ultra Deep Imaging EAO Survey (STUDIES; \citealt{Wang:2017aa}), and by combining it with all archival SCUBA-2 data in the COSMOS field. We have obtained by far the deepest single-dish image at 450\,$\micron$ ($\sigma_{450\,\micron}$ = 0.65\,mJy). In \cite{Chang:2018aa}, we analyzed the structural parameters and morphological properties of 450-$\micron$-selected SMGs from this survey. We found that the irregular/merger fractions are similar for SMGs and for normal star-forming galaxies matched in stellar mass and SFR, and the fractions depend on the SFRs. In this paper, we analyze the multi-wavelength properties of 256 450-$\micron$-selected SMGs with signal-to-noise ratios (S/Ns) $>$ 4. By combining the rich multi-wavelength data in the COSMOS field, we can probe the physical properties of a much fainter SMG population with $L_{\rm IR} \simeq 10^{11}\,\rm L_{\sun}$.

This paper is structured as follows. In \S\ref{sec:MultiWavelengthData}, we describe the observations, data reduction techniques, source extraction procedure, and the multi-wavelength data in the COSMOS field. In \S\ref{sec:SourceExtractionCounterpartID}, we describe the method we use for counterpart identification. In \S\ref{sec:PhysicalParameters}, we analyze the physical properties of our sample, including stellar mass, $L_{\rm IR}$, SFR, extinction, and $T_{\rm d}$. We present the results of our analyses in \S\ref{sec:Results}. We derive the infrared luminosity functions (LFs) in \S\ref{sec:IR_LFs} and estimate the obscured cosmic star-formation history in \S\ref{sec:SFRD}. We summarize our findings in \S\ref{sec:Summary}. Throughout this work, the standard errors of our sample medians are estimated from bootstrap analysis. We adopt the cosmological parameters $H_{0}$ = 70 km\,s$^{-1}$\,Mpc$^{-1}$, $\Omega_{\Lambda}$ = 0.70, and $\Omega_{\rm m}$ = 0.30. We adopt the \cite{Kroupa:2003aa} initial mass function (IMF). When the occasion arises, we rescale the stellar masses (SFRs) from the \cite{Chabrier:2003aa} or \cite{Salpeter:1955aa} IMF to the \cite{Kroupa:2003aa} IMF by multiplying a constant factor of 1.08 (1.06) or 0.66 (0.67), respectively (conversion factors adopted from \citealt{Madau:2014aa}).

%%==============
%%   Data
%%==============
\section{Multi-wavelength Data} \label{sec:MultiWavelengthData}

%%%%%%%%%%%%%%%%%%%%%%%%%%%%%%%%%%%%%%%%%%%%%%%%%%%%%%%%%%%%%%%%%%%%%%

\subsection{SCUBA-2 Data}  \label{subsec:TheSCUBA2Data}
The SCUBA-2 data presented in this paper come from three sources: STUDIES \citep{Wang:2017aa}, \citealt{Casey:2013aa}'s work (hereafter C13) in the COSMOS field, and the SCUBA-2 Cosmology Legacy Survey (S2CLS; \citealt{Geach:2013aa, Geach:2017aa}). We combine these observations to produce an extremely deep 450-$\micron$ map.

\subsubsection{Observations}  \label{subsubsec:Observations}
STUDIES is a multi-year JCMT Large Program that aims to reach the confusion limit at 450\,$\micron$ within the CANDELS \citep{Grogin:2011aa, Koekemoer:2011aa} footprint in the COSMOS field. The standard \texttt{CV DAISY} mapping pattern \citep{Holland:2013aa} is used for this survey. The \texttt{CV DAISY} scan mode maximizes the exposure time at the center of the image and creates a circular map with a radius of $R \simeq 6\arcmin$ and increasing depth toward the center. The final goal of STUDIES is to make a single \texttt{CV DAISY} map that reaches the confusion limit of r.m.s $\sim0.6$\,mJy at its center. The pointing center of STUDIES is ${\rm R.A.} = 10^{\rm h}00^{\rm m}30{\rm \fs}7$ and ${\rm decl.} = +02\degr26\arcmin40\arcsec$. By 2018 March, 56\% of the total allocated integration (330\,hr) of STUDIES had been taken, and the total on-sky integration time was 184\,hours. The current instrumental noise levels in the deepest region of the 450-$\micron$ and 850-$\micron$ images are 0.75 and 0.11\,mJy, respectively. The data collection for the STUDIES program is still ongoing, and the sensitivity of STUDIES will be increased in the future.

Several deep sub-millimeter imaging observations had been carried out by various teams in the COSMOS field with SCUBA-2, and we combine their data with the STUDIES data. The work of C13 was a wider and uniform blank-field survey taken between 2011 December 26 and 2012 December 21. The pointing center of C13 is ${\rm R.A.} = 10^{\rm h}00^{\rm m}28{\rm \fs}0$ and ${\rm decl.} = +02\degr24\arcmin00\arcsec$, which is located south ($\simeq 2\farcm7$) of the STUDIES pointing. The total on-sky time is 38\,hr. The survey of C13 used the \texttt{PONG}-900 scan pattern, which covers a scan area of approximately 15\arcmin $\times$ 15\arcmin. The noise levels of C13 are 3.6 and 0.63\,mJy at 450\,$\micron$ and 850\,$\micron$, respectively.

The S2CLS was a cosmological survey carried out with SCUBA-2 over 4\,yr from 2011 December to 2015 November. The S2CLS program covered several well-studied extragalactic legacy fields. In this study, we include the S2CLS data in the COSMOS field. The mapping strategy of S2CLS in the COSMOS field was a mosaic consisting of two \texttt{CV DAISY} maps offset by 2$\arcmin$ in decl. from the central pointing of ${\rm R.A.} = 10^{\rm h}00^{\rm m}30{\rm \fs}7$ and ${\rm decl.}= +02\degr22\arcmin40\arcsec$, with some overlap. The corresponding central pointing is located $\simeq 4\arcmin$ south of the STUDIES map center within the CANDELS area, and the total on-sky integration is 150\,hr. The noise levels in the deepest regions of the S2CLS maps are 0.95 and 0.14\,mJy at 450\,$\micron$ and 850\,$\micron$, respectively. 

The majority of the observations described above were conducted under the best sub-millimeter weather on Mauna Kea (``Band 1,''  $\tau_{\rm 225\,GHz}$ $<$ 0.05, where $\tau_{\rm 225\,GHz}$ is the zenith sky opacity at 225\,GHz). The sky opacity was constantly monitored during the observations, and the pointing, focus, and flux standards were also observed frequently.

\begin{figure*}
\centering
%%% 450um maps %%%
\includegraphics[width=0.9\paperwidth]{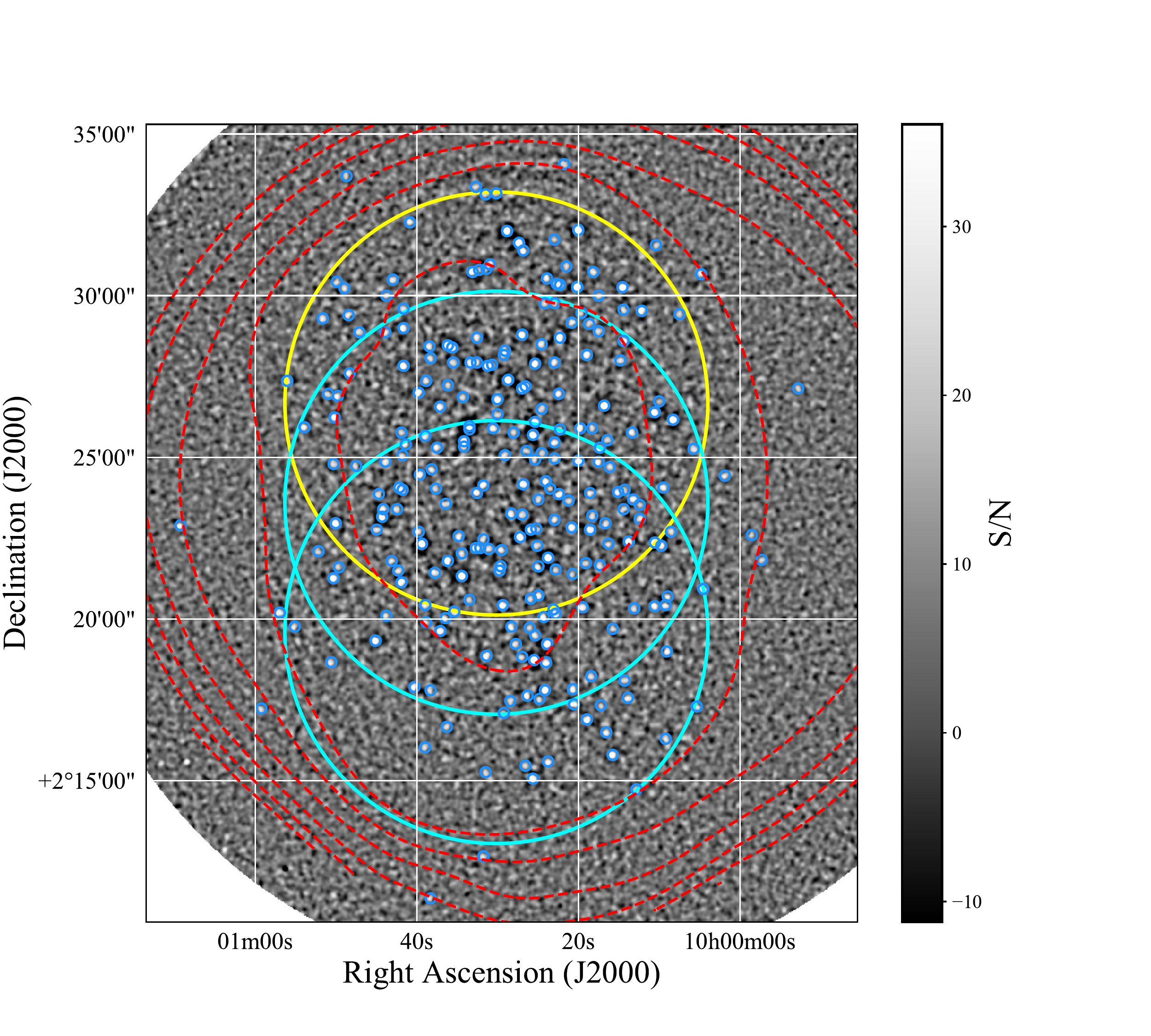}
\caption{The JCMT SCUBA-2 450-$\micron$ S/N image, with the positions of the 256 S/N $> 4$ sources (blue circles). The large yellow and cyan circles indicate the deep scan regions of STUDIES and S2CLS, respectively. The deep area of C13 covers the entire area of this image. The overlapping region is the deepest area ever observed in the 450-$\micron$ waveband. The red dashed contours show the instrumental noise, with contour levels of 1, 4, 7, 10, 13, 16, and 19\,mJy.}
\label{fig:450umImage}
\end{figure*}

\begin{figure*}
\centering
%%% 850um maps %%%
\includegraphics[width=0.9\paperwidth]{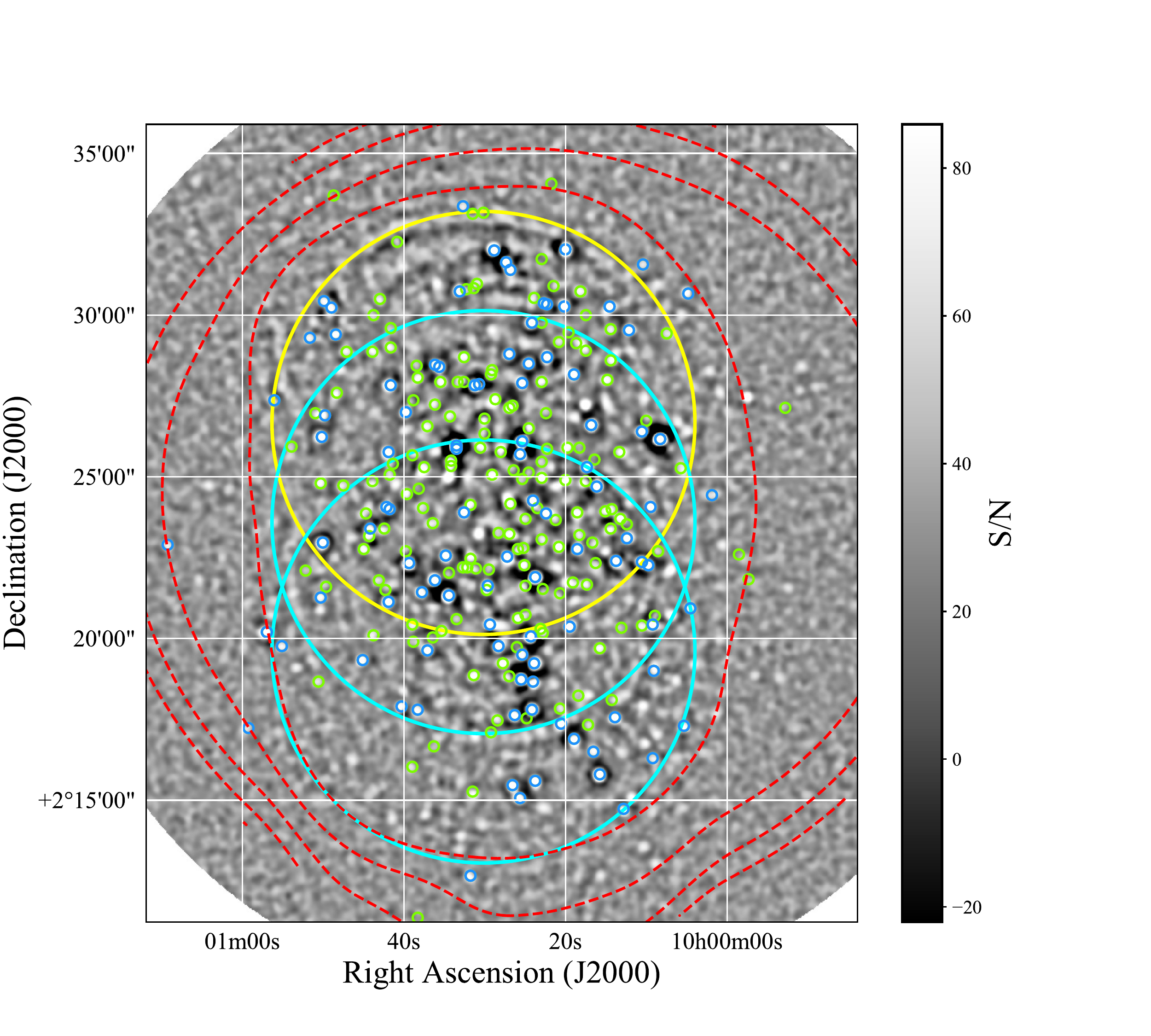}
\caption{The JCMT SCUBA-2 850-$\micron$ S/N image. All 450\,$\micron$ detected sources are circled on this map, while the 157 sources without high-significance 850-$\micron$ flux densities ($<$ 2.1\,mJy) are shown in green. The high values of S/N will decrease by a factor of roughly 3 when taking into account confusion noise ($\sigma_{\rm c} = 0.42$\,mJy; see Appendix \ref{sec:ConfusionNoise}). The red dashed contours show the instrumental noise with contour levels of 0.7, 1.4, 2.1, and 2.8\,mJy. The meaning of the large yellow and cyan circles is the same as in Figure \ref{fig:450umImage}. }
\label{fig:850umImage}
\end{figure*}

%%%%%%%%%%%%%%%%%%%%%%%%%%%%%%%%%%%%%%%%%%%%%%%%%%%%%%%%%%%%%%%%%%%%%%

\subsubsection{Data reduction} \label{subsubsec:DataReduction}
Our data reduction procedure is similar to that described in \cite{Wang:2017aa}. We reduced the data by adopting the Sub-Millimeter Common User Reduction Facility (SMURF; \citealt{Chapin:2013aa}) and the PIpeline for Combining and Analyzing Reduced Data (PICARD; \citealt{Jenness:2008aa}). Individual roughly 30-minute time streams were reduced by using the \texttt{Dynamic Iterative Map-Maker} (DIMM) routine of SMURF. We adopted the standard ``blank field'' recipe, which is a map-making configuration ideal for detecting faint point sources in deep-field surveys. 
 
To obtain flux calibration, we measured the flux conversion factors (FCFs) from a subset of sub-millimeter calibrators observed under Band-1 weather during the corresponding survey campaigns. We then calibrated the individual reduced scans into units of flux density by using the weighted mean FCFs of 476$\pm$95 and 518$\pm$33\,Jy\,beam$^{-1}$\,pW$^{-1}$ for 450\,$\micron$ and 850\,$\micron$, respectively. These FCFs are consistent with the standard values for SCUBA-2 at both 450 and 850\,$\micron$, namely 491$\pm$67 and 537$\pm$26\,Jy\,beam$^{-1}$\,pW$^{-1}$ \citep{Dempsey:2013aa}, and not-yet-published values of 535$\pm$70 and 524$\pm$26\,Jy\,beam$^{-1}$\,pW$^{-1}$ that were derived from an analysis of all of the calibrator data taken since 2011 (S. Mairs et al. 2019, in preparation). 

We adopted the \texttt{MOSAIC\_JCMT\_IMAGES} recipe from PICARD to combine all of the individual calibrated scans into a final map. To optimize the detection of point sources, we convolved the map with a broad Gaussian kernel of full width at half-maximum (FWHM) = $20\arcsec$ and $30\arcsec$ for 450 and 850\,$\micron$ and subtracted the convolved map from the original maps to remove any large-scale structure in the sky background. Then, we convolved the subtracted map with a Gaussian kernel that is matched to the instrumental point-spread function (PSF; FWHM of 7${\farcs}$9 and 13$\arcsec$ for 450 and 850\,$\micron$; \citealt{Dempsey:2013aa}). We used the PICARD recipe \texttt{SCUBA2\_MATCHED\_FILTER} for this procedure.

To verify the flux recovery capability of SMURF and PICARD, we inserted idealized point sources (FWHM = 7${\farcs}$9 and 13$\arcsec$ for 450 and 850\,$\micron$, respectively) with fluxes uniformly distributed between 0.05 and 0.5\,Jy in the 30-minute data streams during the map-making process. The noise levels of the 30-minute data streams at 450 and 850\,$\micron$ are roughly 50 and 20\,mJy, respectively. Our adopted brightness range of 0.05--0.5\,Jy will make synthetic sources with S/N $\simeq 3\sigma$--10$\sigma$ in both 450- and 850-$\micron$ images. After that, we followed the same procedure of applying a matched filter and measured the recovered flux density at the peak position of each inserted source. We repeated this procedure 100 times. The averaged results from sources with S/N $>3\sigma$ suggest that we should apply upward corrections of $5.1\% \pm0.3$\% at 450\,$\micron$ and $10.9\% \pm0.02$\% at 850\,$\micron$. We verify that the corrections do not depend on the inserted flux density. For the 450-$\micron$ image, this adjustment is slightly less than the 10\% correction reported by \cite{Geach:2013aa} and \cite{Chen:2013aa}, but the difference is within the commonly accepted 10\% calibration uncertainty.

Finally, we constructed an extremely deep 450-$\micron$ image and a confusion-limited 850-$\micron$ image with the STUDIES, C13, and S2CLS data combined. Figures~\ref{fig:450umImage} and~\ref{fig:850umImage} are the 450-$\micron$ and 850-$\micron$ S/N maps, respectively. Our images cover a region of approximately 300\,arcmin$^{2}$. The instrumental noise levels at 450\,$\micron$ and 850\,$\micron$ in the deepest regions are 0.65\,mJy and 0.10\,mJy, respectively. The apparent S/N in the 850-$\micron$ image (Figure~\ref{fig:850umImage}) is overestimated by a factor of roughly three due to not including the 850-$\micron$ confusion noise ($\sigma_{\rm c} = 0.42$\,mJy; see Appendix \ref{sec:ConfusionNoise}).

\subsection{Ancillary Data} \label{subsec:AncillaryData}

The source coordinates from radio and near-/mid-infrared observations are key ingredients for identifying counterpart galaxies to our 450-$\micron$ detected sources (\S\ref{subsec:CounterpartID}). We use the Very Large Array (VLA)-COSMOS Large Project survey conducted with the Jansky VLA at 3\,GHz \citep{Smolcic:2017aa}. The survey covers the entire 2\,deg$^{2}$ COSMOS field with a noise of 2.3\,$\mu$Jybeam$^{-1}$ that is uniform across the field with an angular resolution of 0$\farcs$7. The catalog contains approximately 10,000 sources above 5$\sigma$ (11\,$\mu$Jybeam$^{-1}$). In the near-/mid-infrared, we use the S-COSMOS infrared imaging survey carried out with the \emph{Spitzer Space Telescope} \citep{Sanders:2007aa}. The survey covers the entire 2\,deg$^{2}$ of the COSMOS field uniformly in all seven \emph{Spitzer} bands (IRAC: 3.6, 4.5, 5.6, 8.0\,$\micron$; MIPS: 24, 70, 160\,$\micron$). We employ the archival IRAC catalog published by \cite{Sanders:2007aa} that includes all sources with measured flux densities at 3.6\,$\micron$ above $1\,\mu$Jy and has an angular resolution of 1.7$\arcsec$ at 3.6\,$\micron$. On the other hand, the archival MIPS 24-$\micron$ catalog published by \cite{Sanders:2007aa} contains only sources with $S_\mathrm{24\,\micron} > 150\,\mu$Jy. This catalog does not reach the sensitivity limit of the map and is insufficient for identifying counterpart galaxies of our 450-$\micron$-selected sample. Therefore, in this work, we generated our own 24-$\micron$ catalog using \texttt{SExtractor} \citep{Bertin:1996aa} and recalibrated the fluxes to the \emph{Spitzer} General Observer (GO) Cycle 3 total fluxes released by the S-COSMOS team. We use the S-COSMOS 24-$\micron$ image \citep{Sanders:2007aa} in their GO2 + GO3 data delivery in 2008 to run \texttt{SExtractor}. The catalog has a 3.5$\sigma$ detection limit of 57\,$\mu$Jy without using positional priors from other wavelengths.

We also adopt the band-merged COSMOS2015 ($z^{++}JHK_{s}$ stack-selected) photometric catalog compiled by \cite{Laigle:2016aa}, which contains 30+ bands of photometric data points from the X-ray, near-ultraviolet, and optical to the FIR. This catalog includes redshift information and stellar-population parameters, which are used to understand the physical properties of our sample (\S\ref{sec:PhysicalParameters}). 

For the FIR photometry, we adopt the \emph{Herschel}/PACS (100 and 160\,$\micron$) flux densities from the COSMOS2015 catalog. We further extract the \emph{Herschel}/SPIRE 250-$\micron$ flux densities of the 450-$\micron$ sources by using the probabilistic de-blending software \texttt{XID+} \citep{Hurley:2017aa}. We do not extend this to wavelengths longer than 250\,$\micron$ for \emph{Herschel}/SPIRE photometry, since the \emph{Herschel}/SPIRE 350 and 500\,$\micron$ suffer from confusion effects and small-scale clustering \citep{Bethermin:2017aa}, which positively bias the measured 350-/500-$\micron$ fluxes. Moreover, our 450-$\micron$ data provide the constraints at these longer FIR wavebands (similar to that in \citealt{Bourne:2017aa}). We use our 450-$\micron$ sources with S/N $> 3.5$ (\S \ref{sec:SourceExtractionCounterpartID}) as positional priors for \texttt{XID+} extraction. We visually inspected the  \emph{Herschel} 250-$\micron$ image and found that there is a strong one-to-one correspondence between 250-$\micron$ detections and our 450-$\micron$ sources. This indicates that the majority of the 250-$\micron$ fluxes arise from 450-$\micron$-detected sources. Therefore, we conclude that our 450-$\micron$ catalog, which almost reaches the confusion limit (see Appendix \ref{sec:ConfusionNoise}), is sufficient for the de-blending procedure. To reduce the computing time, for each source, we crop our map to a $100\arcsec$ radius centered at the 450-$\micron$ position. These cropped maps typically contain around 20 450-$\micron$ sources, including the source of interest. Then \texttt{XID+} generates a mock map by probabilistically assigning a flux to each source in the cropped map using a Markov Chain Monte Carlo (MCMC) approach and attempts to minimize the residuals between the true map and the mock map. For each source, this procedure produces a posterior distribution of the flux density in the SPIRE band, including information on the correlation between nearby sources. We simply adopt the medians and standard deviations in the flux density distributions. We summarize our adopted public data in Table \ref{tbl:OriginOfBands}.

\section{Source Extraction and Counterpart Identification} \label{sec:SourceExtractionCounterpartID}

%%%%%%%%%%%%%%%%%%%%%%%%%%%%%%%%%%%%%%%%%%%%%%%%%%%%%%%%%%%%%%%%%%%%%%

\subsection{Source Extraction} \label{subsec:SourceExtraction}
For source extraction, we generated a synthetic PSF by averaging the 10 highest-S/N sources in our final map. We verify that the difference between the synthetic and expected PSF (matched Gaussian kernel) is insignificant, although we expect that the seeing and pointing error may make the observed PSF slightly broader. We used a source extraction method similar to the CLEAN algorithm that is widely used in radio interferometry for the deconvolution of radio images. This procedure was adopted to deal with blended sources. We searched for the peak pixel in the S/N map and subtracted 5\% of a peak-scaled synthetic PSF from the image at its position. In this step, we recorded the subtracted flux and coordinates. The next peak in the image was identified and the subtraction was iterated until the process met the S/N threshold, which we set to be 3.5$\sigma$. Finally, we summed up the subtracted flux density and remaining 3.5$\sigma$ flux density and considered this to be the final flux density for each source. In total, we detected 357 sources in the 450-$\micron$ image above 3.5$\sigma$. 

Our source list can suffer from several observational biases: detection incompleteness, flux boosting caused by noise and confusing faint sources, and spurious sources. Therefore, we performed Monte Carlo simulations to estimate these observational biases. In brief, we created a ``true noise'' map, which approximates the instrumental noise, by using the jackknife technique (similar to that described in \citealt{Cowie:2002aa}; see Appendix \ref{sec:MonteCarloSimulation}). We randomly inserted the scaled synthetic PSF into this true noise map with an assumed source count (Schechter function) in the flux range of 1--50\,mJy. Our assumed counts are consistent with the observed counts (see Appendix \ref{sec:MonteCarloSimulation}). We then ran the source extraction procedure on the simulated image and repeated this 200 times. By comparing the source counts and flux ratios between the input and output catalogs in the simulations, we can compute the completeness, flux boosting, and spurious source corrections. The completeness, flux boosting, and spurious source fractions are roughly 73\%, 30\%, and 9\% at $4\sigma$, respectively (see Appendix \ref{sec:MonteCarloSimulation} for details). 

In this work, we only focus on the 256 450-$\micron$-selected sources that have S/N $> 4$ due to the relatively high fraction of spurious sources ($> 14\%$) at S/N $< 4$. To obtain the 850-$\micron$ flux densities, we directly read the flux values from the 850-$\micron$ image at the 450-$\micron$ positions. To determine if a 450-$\micron$ source is detected at 850\,$\micron$, we require its 850-$\micron$ flux density to be higher than five times the confusion noise at 850\,$\micron$. The confusion noise is estimated to be $\sigma_{\rm c} = 0.42$\,mJy (see Appendix \ref{sec:ConfusionNoise}). Estimates from other fields are comparable with this value ($\sigma_{\rm c} = 0.33$\,mJy, \citealt{Cowie:2017aa}; $\sigma_{\rm c}$ = 0.40\,mJy, \citealt{Zavala:2017aa}; and $\sigma_{\rm c}$ = 0.40\,mJy, Simpson et al. 2019). We therefore consider a 450-$\micron$ source to be detected at 850\,$\micron$ if it is brighter than 2.1\,mJy at 850\,$\micron$. In total, we have 256 450-$\micron$-selected sources, of which 99 sources have 850-$\micron$ detections. 

%%%%%%%%%%%%%%%%%%%%%%%%%%%%%%%%%%%%%%%%%%%%%%%%%%%%%%%%%%%%%%%%%%%%%%

\subsection{Counterpart Identification} \label{subsec:CounterpartID}

Thanks to the abundant multi-wavelength data in the COSMOS field and the relatively high angular resolution of SCUBA-2 at 450\,$\micron$ (FWHM = 7${\farcs}$9), we are able to identify most of the optical counterparts for our 450-$\micron$ SMGs. 

We first cross-matched our 450-$\micron$ catalog with the VLA-COSMOS 3\,GHz catalog using a $4\arcsec$ search radius that is expected to produce false matches for $\simeq3$ sources. We find that 134 450-$\micron$ sources have radio counterparts, and all of them are significant in the corrected-Poissonian probability identification technique ($p$-value $< 0.05$; see \citealt{Downes:1986aa}). This moderately high fraction of radio counterpart identifications (134 out of 256; 52\%) is expected, given the empirical correlation between the FIR and radio luminosities of normal galaxies, the so-called ``FIR--Radio correlation'' \citep{Helou:1985aa, Condon:1992aa}. We then cross-matched the radio positions with \emph{Spitzer} IRAC mid-infrared coordinates \citep{Sanders:2007aa} with a $1\arcsec$ search radius ($\simeq 4$ expected false matches). With this method, the detection rate of IRAC counterparts is 94\% (124 out of 134), and there are no radio sources with multiple IRAC counterparts within such a small search radius. 

For the remaining 122 450-$\micron$ sources that do not have radio counterparts, we cross-matched them with the MIPS 24-$\micron$ catalog with a search radius of $4\arcsec$ ($\simeq 7$ expected false matches). We found that 76 450-$\micron$ sources have MIPS 24-$\micron$ detections with $p$-values $<0.05$. Using the 24-$\micron$ positions, we then searched for IRAC mid-infrared sources within $2\arcsec$ ($\simeq 2$ expected false matches). The detection rate of IRAC counterparts is 92\% (70 out of 76), and the fraction of MIPS sources with multiple IRAC counterparts is zero. By increasing the search radius to $3\arcsec$ and $4\arcsec$ from the MIPS 24-$\micron$ positions, we can increase the detection rates of IRAC counterparts to 99\% and 100\%, respectively; however, the fractions of multiple counterparts also dramatically increase, to 12\% and 34\%, respectively. This suggests that such large search radii lead to misidentifications. We therefore adopt a conservative search radius of $2\arcsec$ when searching for IRAC counterparts to the 24-$\micron$ sources.

There are 113 (out of 134) radio-identified sources having 24-$\micron$ detections within $4\arcsec$ from the 450-$\micron$ positions. Among these 113 radio-identified sources, 11 sources have 24-$\micron$ positional offset by more than $2\arcsec$ from the radio positions (i.e., radio and 24-$\micron$ sources lead to different identifications). If we assume that all radio-identified sources that have 24-$\micron$ detections are secure SMG counterparts, this result suggests that the misidentification fraction of just using the 24-$\micron$ sources is about 10\% (11/113). 

\begin{figure}
\centering
%%% Pie Chart %%%
\includegraphics[width=\columnwidth]{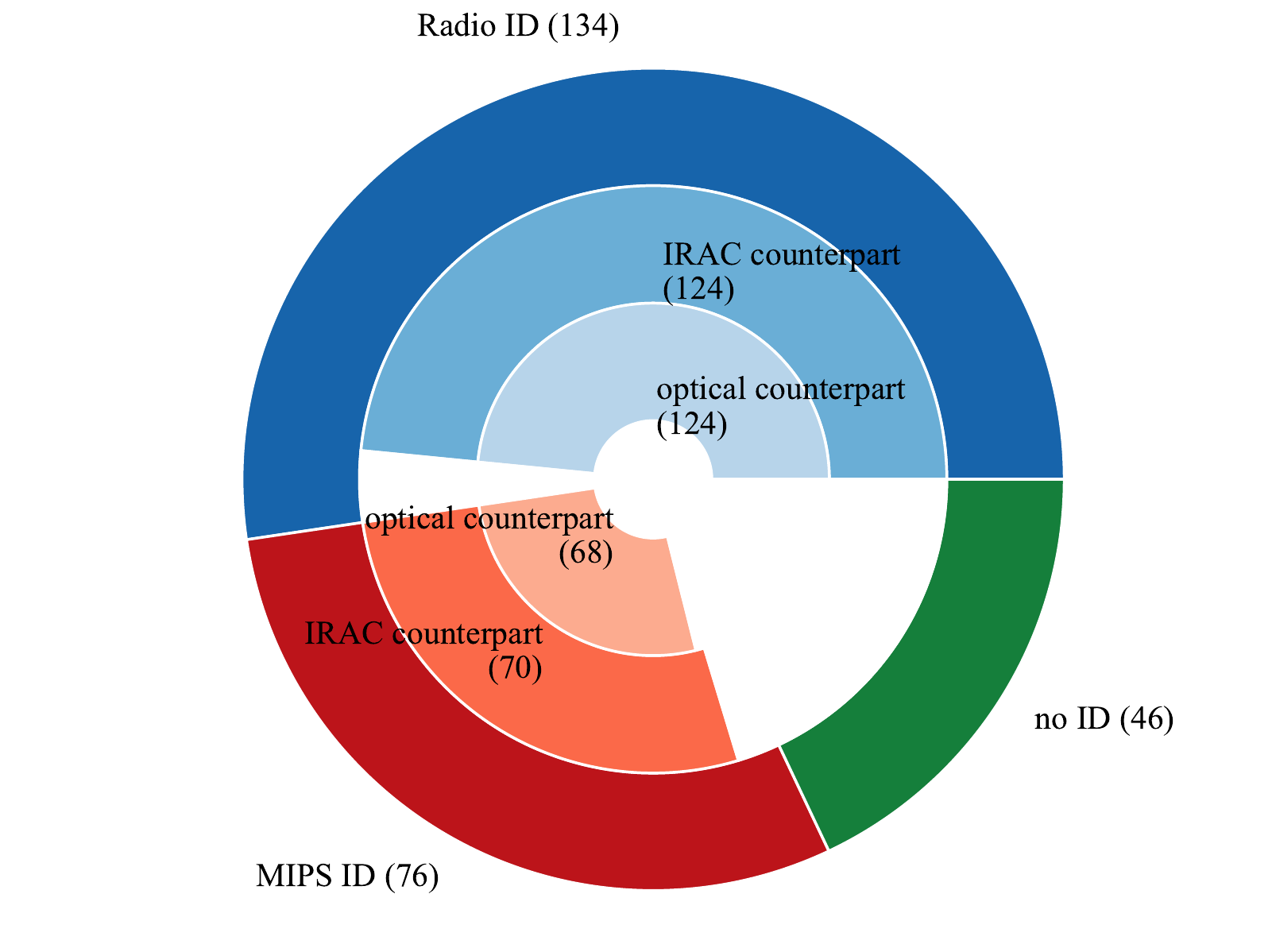}
\caption{Doughnut chart showing the breakdown of 450-$\micron$ sources into different classes of counterpart identifications. The outer ring shows the number of 450-$\micron$ sources identified at VLA 3\,GHz and MIPS 24\,$\micron$, and the unidentified sources, while the middle ring shows the number of IRAC counterparts, which are identified based on their corresponding radio or MIPS positions. Most of these IRAC counterparts have optical counterparts in the COSMOS2015 catalog (inner ring).}
\label{fig:PieChart}
\end{figure}

Figure~\ref{fig:PieChart} shows a doughnut chart summarizing the breakdown of 450-$\micron$ sources into different classes of counterpart identifications. In summary, we can identify a significant fraction (210 out of 256, 82\%) of our 450-$\micron$ sources using radio or 24-$\micron$ data. We present these sources and their derived properties (\S\ref{sec:PhysicalParameters}) in Tables \ref{tbl:Catalog1} and \ref{tbl:Catalog2}. A notable fraction of them (194 out of 256, 76\%) have IRAC detections. Among the 194 IRAC sources, 192 have optical counterparts in the COSMOS2015 catalog. We present the multi-wavelength photometries of the counterparts in Tables \ref{tbl:PhotometricCatalog1} and  \ref{tbl:PhotometricCatalog2}. 

\subsection{Unidentified sources}\label{subsec:UnidentifiedSources}

There are still 46 450-$\micron$ sources without any radio or 24-$\micron$ identifications. They are listed in Table \ref{tbl:PhotometricCatalog3}. We do not further cross-match these sources with \emph{Herschel}/PACS or \emph{Herschel}/SPIRE catalogs, since our main focus in this work is on the sources with optical counterparts (\S\ref{subsec:CounterpartID}). Nevertheless, all these 46 sources have photometry at 250-$\micron$ from \emph{Herschel}, which is extracted with \texttt{XID+}. Furthermore, some of them even have 850-$\micron$ detections by SCUBA-2. The 850-$\micron$ photometry allows us to further investigate the nature of these sources.

\begin{figure}
\centering
%%% S(850)/S(450) vs redshift %%%
\includegraphics[width=\columnwidth]{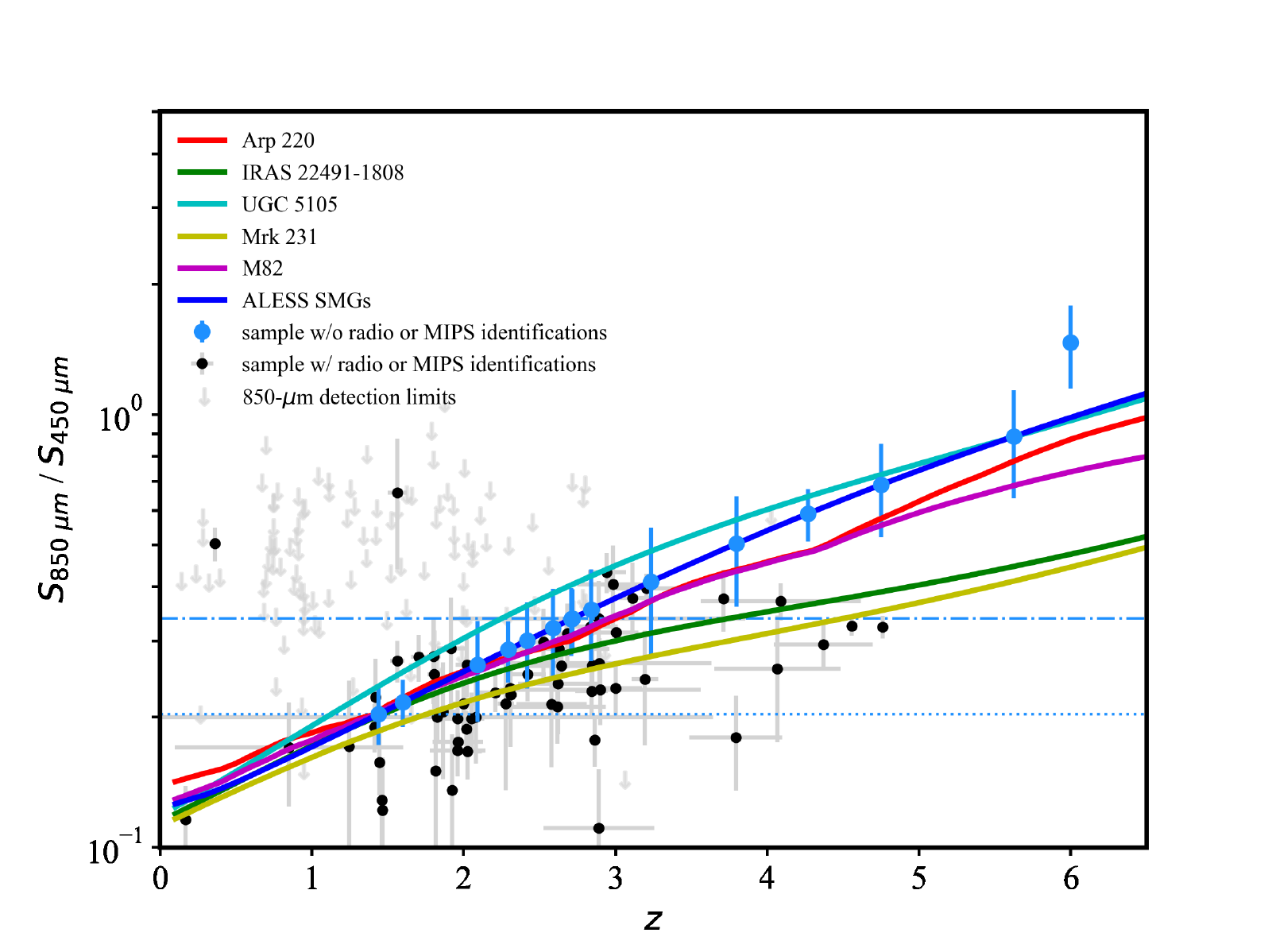}
\caption{Flux density ratio of 850--450\,$\micron$ versus redshift. The colored curves represent the flux density ratios derived from various SED templates: Arp 220$^{*}$ (ULIRG), IRAS 22491-1808$^{*}$ (ULIRG), UGC 5105$^{*}$ (ULIRG with AGN), Mrk 231$^{*}$ (ULIRG with AGN), M82$^{*}$ (luminous starburst galaxies), and averaged SED from ALESS (\citealt{da-Cunha:2015aa}). For the 15 sources that do not have any radio or 24-$\micron$ counterparts but do have 850-$\micron$ detections; we place them along the averaged ALESS SED track up to $z \simeq 6$, according to their observed flux ratios. The median of the flux density ratio is $0.34^{+0.07}_{-0.02}$ (blue dotted--dashed line), indicating that they may be SMGs at $z \gtrsim 3$. The smallest flux ratio is $0.20\pm0.03$ (blue dotted line), still placing the SMG at $z \gtrsim 1.5$. We also show the $S_\mathrm{850\,\micron}$/$S_\mathrm{450\,\micron}$ flux density ratios of our identified sources (black points) that have redshift determinates (\S\ref{subsec:Redshift}) and the upper limits considering an 850-$\micron$ flux threshold given by 2.1\,mJy (confusion limit) from the rest of the sample. Our sample is in broad agreement with most of the SED tracks. $^{*}$Spectral templates of nearby infrared-luminous galaxies are from the Spectral Atlas of Infrared Luminous Galaxies (\url{http://www.stsci.edu/hst/observatory/crds/non-stellar.html}).}
\label{fig:S84_vs_redshift}
\end{figure}

Among the 46 unidentified sources, 15 sources have 850-$\micron$ flux densities larger than the confusion limit of 2.1\,mJy at the 450-$\micron$ positions. The lack of radio or 24-$\micron$ counterparts may be due to being at high redshifts and consequently faint at these wavelengths. At the same time, the strong negative $K$-correction at 850\,$\micron$ leads to their strong 850-$\micron$ detections. To gain insight on the possible redshifts of this sub-sample, we place all 15 850-$\micron$ detected sources along the averaged ALMA LABOCA ECDFS Sub-millimeter Survey (ALESS; \citealt{da-Cunha:2015aa}) SED track (blue curve in Figure~\ref{fig:S84_vs_redshift}) up to $z \simeq 6$ according to their $S_\mathrm{850\,\micron}$/$S_\mathrm{450\,\micron}$ flux density ratios in Figure~\ref{fig:S84_vs_redshift}. Among these sources, the median flux ratio of $S_\mathrm{850\,\micron}$/$S_\mathrm{450\,\micron}$ is $0.34^{+0.07}_{-0.02}$ (dotted--dashed line in Figure~\ref{fig:S84_vs_redshift}), which corresponds to an SMG at $z \gtrsim 3$. The lowest flux ratio for these sources is $S_\mathrm{850\,\micron}$/$S_\mathrm{450\,\micron}=0.20\pm0.03$ (dotted line in Figure \ref{fig:S84_vs_redshift}), still placing an SMG at $z \gtrsim 1.5$. We also present the $S_\mathrm{850\,\micron}$/$S_\mathrm{450\,\micron}$ flux density ratios of our identified sources that have redshift determinations (see \S\ref{subsec:Redshift}) in Figure~\ref{fig:S84_vs_redshift}. Our sample is in broad agreement with most of the SED tracks. The 450-$\micron$ sources without radio or 24-$\micron$ identifications are likely at higher redshifts.

For the remaining 31 450-$\micron$ sources that do not have radio or 24-$\micron$ identifications or 850-$\micron$ detections, 27 sources have counterpart candidates in the IRAC 3.6-$\micron$ image within a search radius of $4\arcsec$ from the 450-$\micron$ positions. However, only a small fraction of this sub-sample (six out of 27) have $p$-values small enough ($< 0.05$) to be considered as reliable matches. The large beam size of SCUBA-2 compared to that of IRAC at 3.6\,$\micron$ suggests that this procedure can suffer severely from misidentifications and/or source blending. We verify that about 23\% (60/256) of our entire 450-$\micron$ sample has multiple IRAC sources within $4\arcsec$. Therefore, to be conservative, we do not include these six $p$-identified SMG candidates in our subsequent analyses. For the remaining 21 sources without robust IRAC counterparts, we employed the stacking technique in the \emph{Herschel} 250-$\micron$ image based on their 450-$\micron$ positions. These 21 sources have a stacked flux density of $4.2\pm0.8$\,mJy at 250 $\micron$, indicating that the 450-$\micron$ detections are likely to be real. In this work, we will not further discuss these sources, but we note that interferometric observations will help to reveal the origins of these FIR sources. 

This still leaves us with four sources that do not have any potential counterparts in the near-infrared, mid-infrared, 850-$\micron$, and radio images. All of these sources have S/N $<4.3$ at 450\,$\micron$ and therefore a high probability of being false detections. In our entire sample, we have 26 sources in the range of 4.0$\sigma$--4.3$\sigma$. The finding of these four sources is consistent with the spurious fraction in the range of 4.0$\sigma$--4.3$\sigma$ (about 9\%; Appendix \ref{sec:MonteCarloSimulation}). In conclusion, the unidentified sources are consistent with being at high redshifts and affected by source blending, with a small fraction of them being spurious.

\section{Deriving physical parameters}\label{sec:PhysicalParameters}

%%%%%%%%%%%%%%%%%%%%%%%%%%%%%%%%%%%%%%%%%%%%%%%%%%%%%%%%%%%%%%%%%%%%%%

\subsection{AGN contamination} \label{subsec:AGNContamination}

The main purpose of this paper is to analyze the nature of 450-$\micron$ sources that are mainly powered by star formation rather than AGNs. Therefore, we first examine the AGN contamination in our sample. Here 12 sources have X-ray detections with X-ray luminosities above 10$^{43}$\,erg\,s$^{-1}$ in the 2--10\,keV band, which can be considered AGNs. We adopt this limit instead of the widely used dividing line of $L_{\rm 2-10\,keV} = 10^{42}$\,erg\,s$^{-1}$ (e.g. \citealt{Zezas:1998aa, Ranalli:2003aa, Szokoly:2004aa}), since a star-forming galaxy with SFR of $\simeq 100\,\rm M_{\sun}\,yr^{-1}$ can also produce the X-ray luminosity $> 10^{42}$\,erg\,s$^{-1}$ in the 2--10\,keV band \citep{Aird:2017aa}. Three additional sources in our sample can be considered as AGNs if the threshold of $L_{\rm 2-10\,keV} > 10^{42}$\,erg\,s$^{-1}$ is adopted.

For the identification of mid-infrared AGNs, we simply cross-matched our sample with a public catalog of infrared AGNs from \cite{Chang:2017aa}. In brief, the authors derived AGN properties with SED fitting using MAGPHYS \citep{da-Cunha:2015aa} based on a sample of mid-infrared selected galaxies in the COSMOS field. They defined obscured AGNs as those with AGN contributions to the mid-infrared luminosity $> 50\%$ from the SED fitting. According to the authors, this definition recovers 54\% of X-ray detected AGNs. In total, we found two mid-infrared AGN candidates from our sample. 

To determine AGN contamination at radio wavelengths, we follow Equation~1 in \cite{Delvecchio:2017aa}, which describes a redshift-dependent threshold in radio-excess. In brief, they first excluded the luminous AGN populations according to X-ray, mid-infrared, or optical-to-FIR SED decomposition from the VLA 3\,GHz catalog. For the remaining 3-GHz sources, they then set a threshold of $\geqslant$ 3$\sigma$ for the radio emission compared to that expected from the star formation derived from $L_{\rm IR}$. With this method, four additional sources can be classified as radio-excess galaxies (the derivations of the parameters are described in \S\ref{subsec:InfraredLuminosities} and \S\ref{subsec:RadioPower}), but none of them exceed the threshold for radio-loud AGNs \citep{Evans:2005aa}. 

There are one, zero, zero, and zero sources having X-ray + radio-excess, mid-infrared + radio-excess, X-ray + mid-infrared, and X-ray + mid-infrared + radio-excess AGN identifications, respectively. In summary, only a small fraction of our sources (18 out of 192, $9\%\pm2$\%) are likely to be AGNs. We exclude all of these possible AGN candidates from our subsequent analyses. 

The AGN fraction in our sample is lower than the range of the potential AGN fraction of about 20--40\% from earlier studies in the literature (e.g., \citealt{Alexander:2005ab, Laird:2010aa, Georgantopoulos:2011aa, Johnson:2013aa}). We note that the sub-millimeter catalogs used in these works are biased toward brighter SMGs, since their catalogs are all from single-dish sub-millimeter surveys that have a typical angular resolution of $\simeq$ 10$\arcsec$--20$\arcsec$ and require radio counterparts. On the other hand, our estimated AGN fraction is more consistent with the ALMA-based estimate from ALESS (17$^{+16}_{-6}$\%; \citealt{Wang:2013aa}), ALMA follow-up observations in the S2CLS UDS program AS2UDS (8--28\%; \citealt{Stach:2019aa}), and ALMA follow-up observations in the SCUBA-2 850-$\micron$ survey ($\sim6$\%; \citealt{Cowie:2018aa}). A trend of a higher AGN fraction for an SMG population with brighter 870-$\micron$ flux density was previously observed \citep{Wang:2013aa}, which would imply that brighter SMGs are more likely to host AGNs. This may partially explain the discrepancy between our estimated AGN fraction and the results from previous single-dish studies, since our sub-millimeter observations are deeper and probe a fainter SMG population.

\subsection{Redshift} \label{subsec:Redshift}

We use public redshift data for our identified sources. For photometric redshifts, \cite{Laigle:2016aa} used the \texttt{LE PHARE} code \citep{Arnouts:1999aa, Ilbert:2006aa} and released their results in the COSMOS2015 catalog. For the fitting process, the authors included a set of 31 templates, including spiral and elliptical galaxies from \cite{Polletta:2007aa} and a set of 12 templates of young blue star-forming galaxies using the \cite{Bruzual:2003aa} (hereafter BC03) model. They set the extinction as a free parameter with a maximum value of {\it E(B$-$V)} = 0.5. 

Several spectroscopic redshift catalogs are also available in the COSMOS field \citep{Lilly:2007aa, Lilly:2009aa,Trump:2009aa, Coil:2011aa, Zahid:2014aa, Le-Fevre:2015aa}. In this paper, we adopt the spectroscopic redshift catalog of \citet{Hasinger:2018aa}, who compiled all of the spectroscopic data of about 10,000 objects that were observed through multi-slit spectroscopy with the Deep Imaging Multi-Object Spectrograph on the Keck II telescope. In addition, we also use the data from the zCOSMOS survey \citep{Lilly:2009aa} conducted with the VIMOS spectrograph on the Very Large Telescope and the hCOSMOS redshift survey \citep{Damjanov:2018aa} observed with the Hectospec spectrograph on the MMT (formerly Multiple Mirror Telescope). 

\subsubsection{Sources without redshifts} \label{subsubsec:MissingSources}
\begin{figure}
\centering
%%% FIR redshift %%%
\includegraphics[width=\columnwidth]{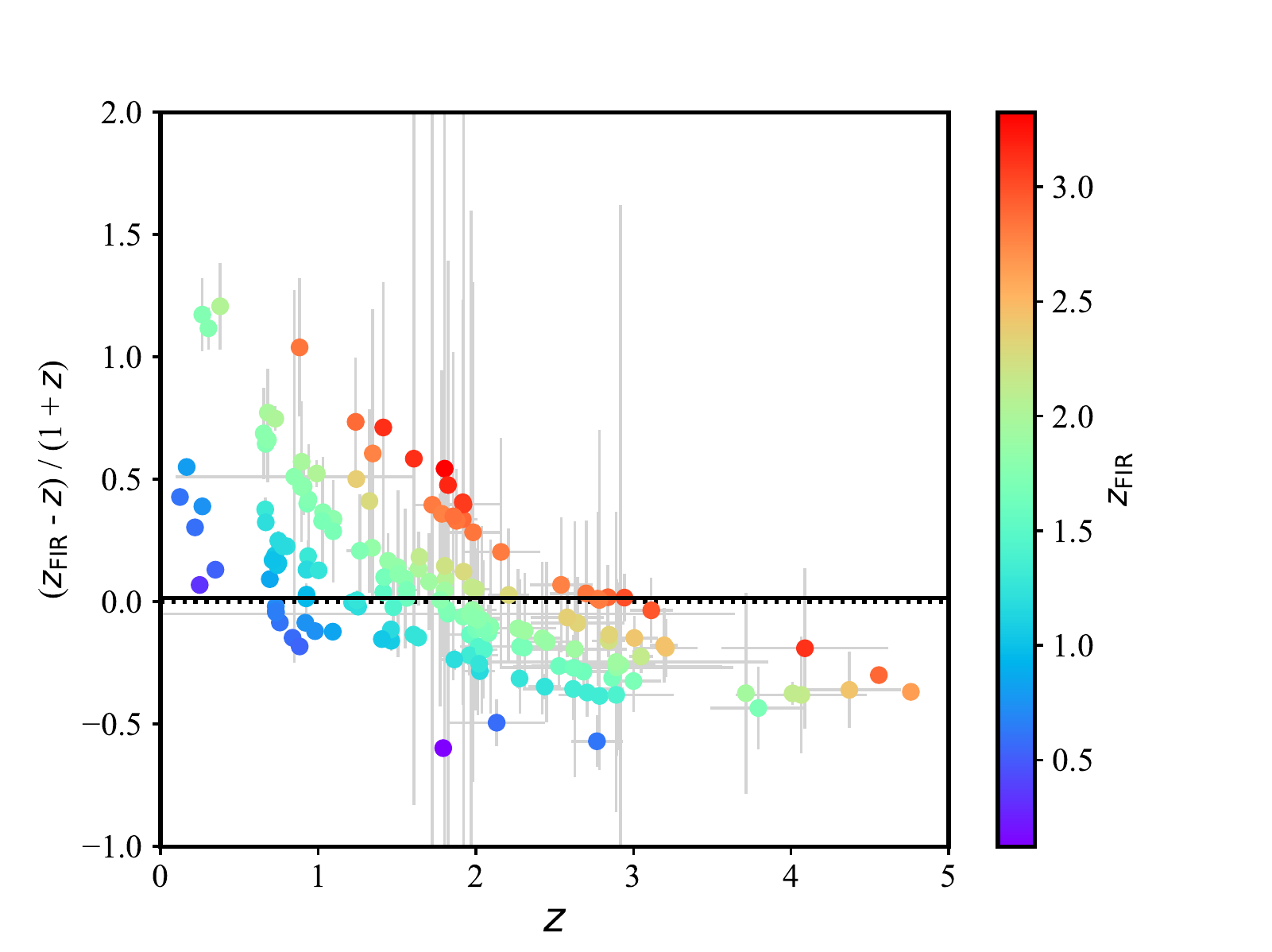}
\caption{Comparisons of the FIR-derived redshifts ($z_{\rm FIR}$) with the optical redshifts ($z$), color-coded with $z_{\rm FIR}$. The $z_{\rm FIR}$ estimates are consistent with the optical redshifts with a median offset of $0.01^{+0.02}_{-0.03}$ (solid horizontal line). The trend of $z_{\rm FIR}$ being systematically higher than the optical redshift at $z < 2$ and lower at $z > 2$ is probably caused by our single-temperature assumption of the SED template and/or the effect of the luminosity--$T_{\rm d}$ correlation (\S\ref{subsec:Td-LIR}), plus selection effects. }
\label{fig:FIRRedshift}
\end{figure}

Out of the 210 sources identified with 24-$\micron$ or radio sources, less than 10\% (20 sources) do not have either spectroscopic or photometric redshift information. As mentioned in \S\ref{subsec:CounterpartID}, 18 of them do not have optical/near-infrared counterparts in the COSMOS2015 catalog, and consequently, they do not have redshifts. The remaining two objects can be matched to the COSMOS2015 catalog, but their photometric redshifts are not reliable, because one object is only detected in $H$ and $K$, while the other object is marginally detected in $V$, $i$, and $z$ with very low S/N. Nevertheless, their detections in mid-infrared-to-radio wavelengths provide us with information for estimating their redshifts (hereafter $z_{\rm FIR}$). We derive their $z_{\rm FIR}$ by using an averaged ultraviolet-to-radio SED template from the ALESS \citep{da-Cunha:2015aa}. This template allows us to conduct a fit that only adopts the mid-infrared-to-radio photometry, i.e., 24\,$\micron$ to 20\,cm. Using sources with optical redshifts, we find that the $z_{\rm FIR}$ are consistent with the optical redshifts and have a median value of $(z_{\rm FIR} - z) / (1 + z) = 0.01^{+0.02}_{-0.03}$ (solid horizontal line in Figure \ref{fig:FIRRedshift}). 

A trend can be seen in Figure \ref{fig:FIRRedshift}, where the $z_{\rm FIR}$ estimates are systematically higher than the optical redshifts at $z < 2$ and lower than the optical redshifts at $z > 2$. A similar trend was also found in previous studies \citep{Ivison:2016aa, Michaowski:2017aa, Zavala:2018aa}. Our adopted FIR SED template, which has been calculated for galaxies at a median redshift of $ z \simeq 2$, is represented by a single temperature. We might expect that a cooler (warmer) dust SED will result in a higher (lower) value of redshift estimation, since there is a degeneracy between redshift and $T_{\rm d}$ in the observed SED. Therefore, a cooler (warmer) $T_{\rm d}$ appears to be needed to correct this effect at lower redshift (higher redshift). However, this does not necessarily imply that $T_{\rm d}$ evolves with redshift. Rather, this is perhaps simply due to the correlation between $T_{\rm d}$ and $L_{\rm IR}$ (\S\ref{subsec:Td-LIR}) and the fact that our survey is more sensitive to low-luminosity systems (\S\ref{subsec:InfraredLuminosities}). 

Among the 20 sources without optical redshift determinations, 19 sources have $z_{\rm FIR}$ estimates with a median value of $z_{\rm FIR} = 1.9^{+0.2}_{-0.1}$ with a 16th-to-84th percentile range of 1.1--2.8. The only source that has neither optical redshift nor $z_{\rm FIR}$ estimates is securely detected at 450\,$\micron$ (S/N $= 5.3$). This source is a radio-identified source with an IRAC counterpart but does not have optical detection. We visually inspected it and verified that this source has multiple detections in the IRAC images (within 2$\arcsec$); therefore, its FIR photometries may be less reliable for $z_{\rm FIR}$ estimation because of source blending. In conclusion, the lack of optical redshifts for these 20 sources is likely to be mainly caused by their high redshifts and thus faintness at optical wavebands.

In this work, we do not attempt to constrain the $z_{\rm FIR}$ for sources without radio or 24-$\micron$ counterparts (\S \ref{subsec:UnidentifiedSources}). Their FIR photometry may be less reliable, since a high fraction of this sub-sample could suffer from the effects of source blending. We will not further discuss these possibly blended systems, but we note that interferometric observations will help to reveal their nature.

\subsubsection{Redshift Distribution} \label{subsubsec:RedshiftDistribution}
\begin{figure}
\centering
%%% Redshift Distribution %%%
\includegraphics[width=\columnwidth]{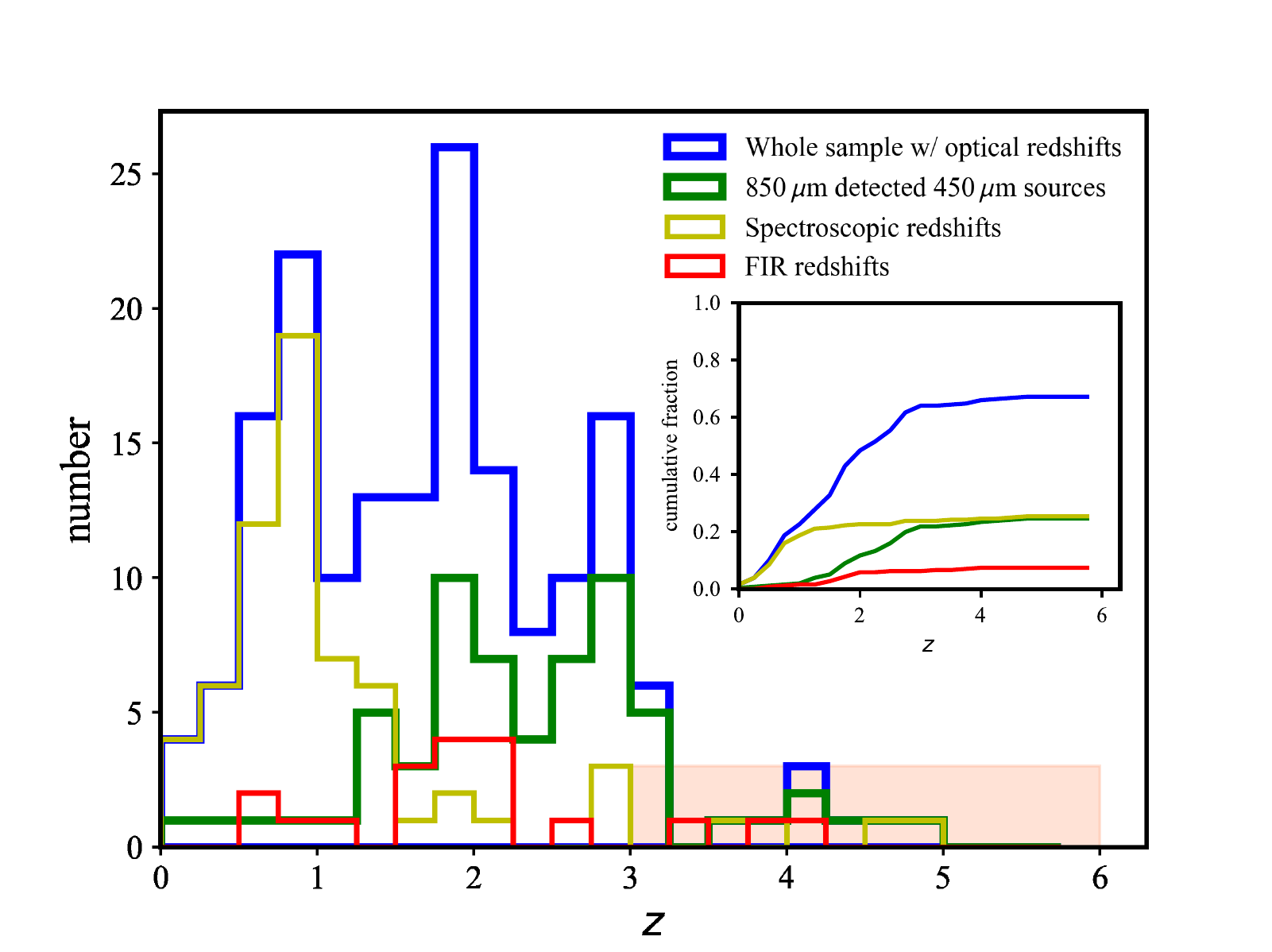}
\caption{Redshift distribution of our sample with optical redshifts (blue histogram) and 850-$\micron$ detections (green histogram). The medians of these two distributions are $z = 1.79^{+0.03}_{-0.15}$ and $z = 2.30^{+0.27}_{-0.26}$, respectively. The vast majority of our sample lies at $z \lesssim 3$. The distribution of the 65 galaxies with spectroscopic redshifts is shown with the yellow histogram, which is clearly biased toward low redshifts. The distribution of the 19 sources with $z_{\rm FIR}$ is shown with the red histogram. A significant fraction (46/256) of our sources do not have redshift determinations, since they do not have 24-$\micron$ or radio identifications and consequently lack optical counterparts. If they are at $z>3$ (see \S\ref{subsec:UnidentifiedSources}) and have a flat redshift distribution between $z=3$ and 6, they are represented by the orange shaded area in this diagram. The small panel shows the cumulative redshift completeness in each sub-sample among the 256 4$\sigma$ sources.}
\label{fig:RedshiftDistribution}
\end{figure}

Among our 174 sources that have optical counterparts (without AGN contamination), 172 sources have redshifts, and 65 of them are spectroscopic. The spectroscopic redshifts from this sub-sample are in good agreement with the photometric redshifts with a median value of $\Delta z / (1 + z_{\rm s}) = 0.01^{+0.04}_{-0.02}$, and a catastrophic outlier fraction \citep{Bernstein:2010aa} of $\simeq$~10\%. Among this, 10\%, 16\%, 16\%, and 50\% are contributed from sources at $z=0-1$, $1-2$, $2-3$, and $>3$, respectively. We replace the photometric redshifts with the spectroscopic redshifts when available. The redshift distribution of our 450-$\micron$-selected sample is shown in Figure \ref{fig:RedshiftDistribution}. The median redshift of the sample with optical redshifts is $z = 1.79^{+0.03}_{-0.15}$ with a 16th-to-84th percentile range of 1.7--1.9. Their redshifts range from $z=0.12$ to 4.76, with the majority at $z \lesssim 3$. On the other hand, the median redshift of 850-$\micron$ detected 450-$\micron$ sources is $z = 2.30^{+0.27}_{-0.26}$ with a 16th-to-84th percentile range of 1.6--3.0, which is in good agreement with previous studies of 850-$\micron$ sources \citep{Simpson:2014aa, Simpson:2017ab, Dunlop:2017aa, Cowie:2018aa}. To gain insight into the redshift distribution of our entire sample, we assume that the 46 sources without 24-$\micron$ or radio identifications are at a median redshift of $z=3$ (\S\ref{subsec:UnidentifiedSources}). After including these redshifts and the 20 sources with $z_{\rm FIR}$ (all assumed to be at $z_{\rm FIR}=1.9$; see \S\ref{subsubsec:MissingSources}), the median redshift of our complete sample of 238 SMGs (without AGNs) slightly increases to $z = 1.9\pm0.1$ with a 16th-to-84th percentile range of 0.9--3.0.

%%%%%%%%%%%%%%%%%%%%%%%%%%%%%%%%%%%%%%%%%%%%%%%%%%%%%%%%%%%%%%%%%%%%%%

\subsection{Stellar Mass} \label{subsec:StellarMass}

We directly adopt the stellar-mass measurements in the COSMOS2015 catalog, which were fitted using the \texttt{LE PHARE} code \citep{Arnouts:1999aa, Ilbert:2006aa}. The stellar masses were determined from a library of synthetic spectra generated using the stellar population synthesis (SPS) model from BC03, matching to the ultraviolet/optical/near-infrared photometry. This assumed a \cite{Chabrier:2003aa} IMF and the exponentially-decreasing star formation history ($\tau^{-2}te^{-t/\tau}$), and two metallicities (solar and half-solar) were considered. In this work, we rescaled their stellar-mass measurements to a \cite{Kroupa:2003aa} IMF in order to easily compare with other studies. The uncertainties on stellar masses computed by \texttt{LE PHARE} are derived by minimizing the $\chi^2$ function using the photometric errors in each of the wavebands. According to \cite{Laigle:2016aa}, the stellar-mass estimate is robust out to $z \simeq 4$. Above $z \simeq 4$, the rest-frame $K$ band lies below the Balmer break and consequently does not reliably constrain the stellar mass. 

The COSMOS2015 catalog provides 165 stellar masses in our sample, and the median is log($M_{\ast}$) = 10.75$^{+0.04}_{-0.05}\,\rm M_{\rm \sun}$ with a 16th-to-84th percentile range of 10.38--11.10$\,\rm M_{\rm \sun}$. Among the nine (out of 174) sources without stellar-mass estimates, two do not have redshift estimates (\S\ref{subsubsec:MissingSources}). All of the remaining sources are undetected in the near-infrared and therefore do not have reliable stellar masses. In this work, we further add an uncertainty of a factor of 3 in quadrature to the uncertainty of the adopted stellar masses, since the typical uncertainty caused by the unknown star formation history of SMGs is about a factor of 3 \citep{Hainline:2011aa}. To estimate the likely stellar mass distribution of our entire sample, we adopt the absolute $K$-band magnitudes ($M_K$) from the COSMOS2015 catalog for our sample and determine their mass-to-light ratios. In our sample, there are 27 sources with 24-$\micron$ or radio identifications but without stellar masses and 46 sources without reliable identifications. For those with redshifts and $M_K$ estimates, we assumed the median mass-to-light ratios from other sources with stellar masses at similar redshifts. For those without redshifts and reliable identifications, we assumed $z=3$ and the median stellar masses from other sources at similar redshifts. By doing so, the median stellar mass for our complete sample is log($M_{\ast}$) = $10.90\pm0.01\,\rm M_{\rm \sun}$ with a 16th-to-84th percentile range of 10.5--11.0\,$\rm M_{\rm \sun}$.

%%%%%%%%%%%%%%%%%%%%%%%%%%%%%%%%%%%%%%%%%%%%%%%%%%%%%%%%%%%%%%%%%%%%%%

\subsection{Infrared Luminosity} \label{subsec:InfraredLuminosities}

\begin{figure*}
\centering
%%% SEDs %%%
\includegraphics[width=0.7\paperwidth]{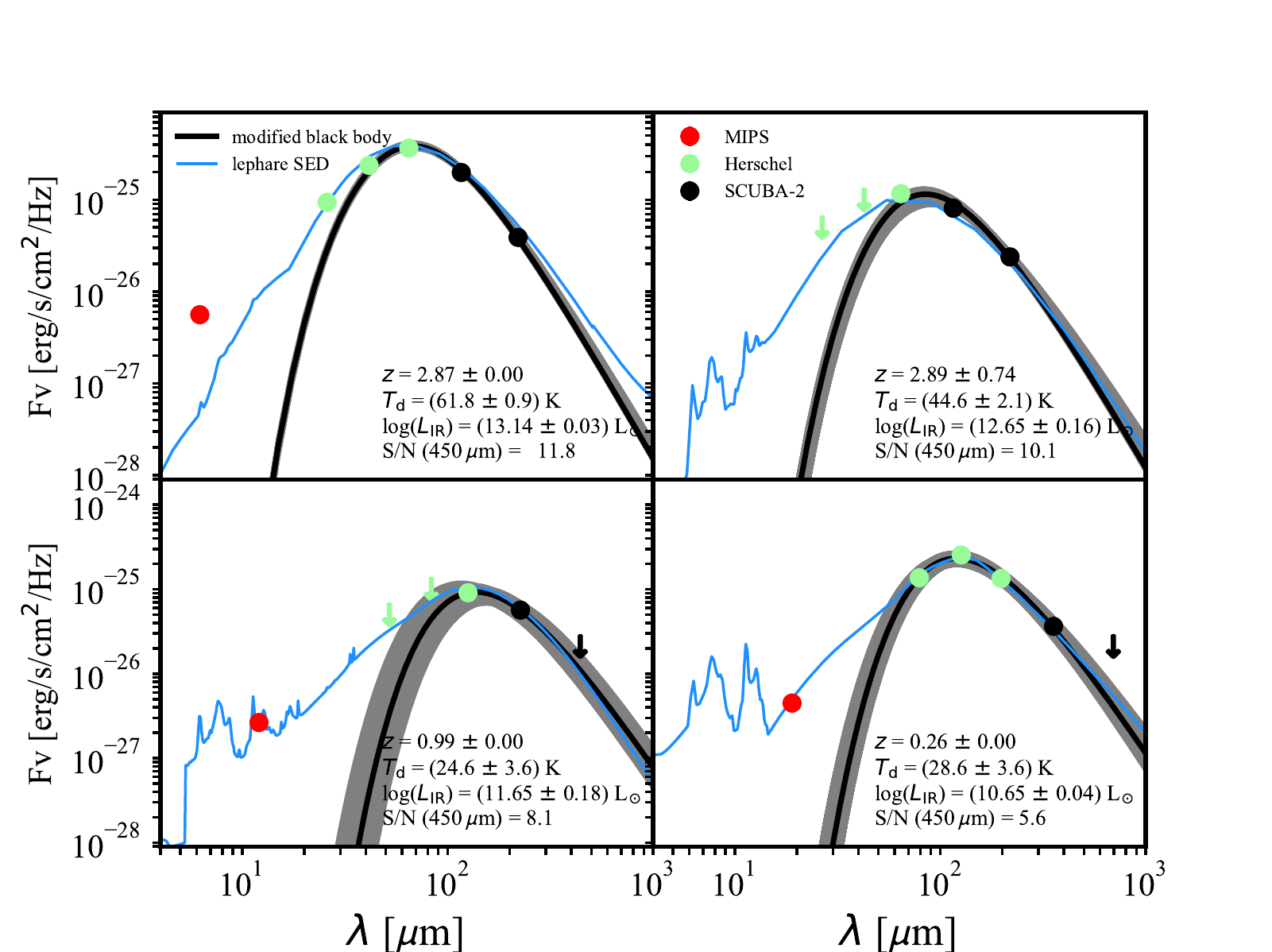}
\caption{Examples of rest-frame FIR-to-submillimeter SEDs fitted by the \texttt{LE PHARE} code (blue curves). The observed flux densities for MIPS, \emph{Herschel}, and SCUBA-2 are shown as red, green, and black points. The best-fit modified blackbodies are shown as black curves, while the errors are shown as dark shaded regions.} 
\label{fig:SEDs}
\end{figure*}

We adopt the \texttt{LE PHARE} code to derive the $L_{\rm IR}$ for our sample. For the fitting, we include the 105 FIR templates from \cite{Chary:2001aa}, the 64 FIR templates from \cite{Dale:2002aa}, the 46 FIR templates from \cite{Lagache:2003aa}, and the 25 FIR/star-forming galaxy templates from \cite{Rieke:2009aa}. We do not adopt SED templates with infrared-luminous AGNs and all of the above adopted SED models are constructed based on purely star-forming infrared galaxies in different luminosity classes. We fitted the infrared photometry from \emph{Spitzer}/MIPS (24 and 70\,$\micron$), \emph{Herschel}/PACS (100 and 160\,$\micron$), XID+ deblended \emph{Herschel}/SPIRE (250\,$\micron$), and JCMT/SCUBA-2 (450 and 850\,$\micron$). A predefined redshift for each source is required by \texttt{LE PHARE}. The uncertainties in our $L_{\rm IR}$ are derived from the maximum-likelihood function by including the photometric errors in each waveband but without including the redshift uncertainties. Examples of the rest-frame FIR-to-submillimeter SEDs fitted by the \texttt{LE PHARE} code are shown as blue curves in Figure \ref{fig:SEDs}. The $L_{\rm IR}$ values are computed by the integrating the best-fit galaxy templates between 8 and 1000\,$\micron$ in the rest-frame at their fixed redshifts.

\begin{figure}
\centering
%%% LIR vs redshift %%%
\includegraphics[width=\columnwidth]{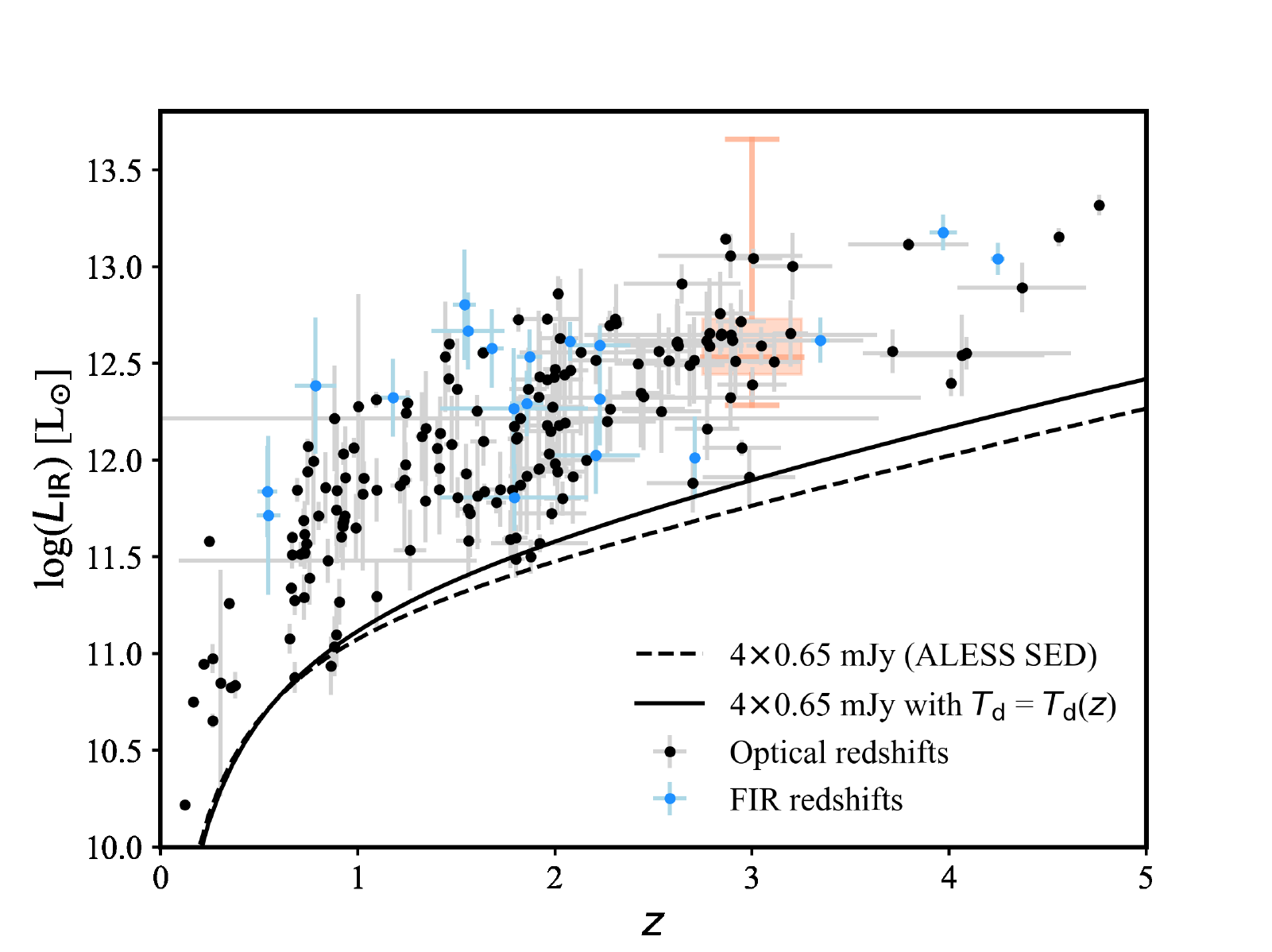}
\caption{The $L_{\rm IR}$ as a function of redshift. Our samples that have optical redshifts are shown as black points, while the 19 sources that have $z_{\rm FIR}$ estimates are shown as blue points. The orange box shows the infrared luminosity range of log($L_{\rm IR}$)=12.3--13.7\,$\rm L_{\sun}$ for the 46 sources without 24-$\micron$ or radio identifications, if we place them at their plausible median redshift of $z=3$ (see \S\ref{subsec:UnidentifiedSources}) and scale their 450-$\micron$ flux densities to $L_{\rm IR}$ using the averaged ALESS SMG SED \citep{da-Cunha:2015aa}. The detection limits corresponding to noise levels of 0.65\,mJy are shown as a dashed curve. For this, we converted the 450-$\micron$ detection limits to $L_{\rm IR}$ limits using the averaged ALESS SMG SED \citep{da-Cunha:2015aa}. We also adopt the weak evolution of $T_{\rm d}$ along with redshift (black line in Figure \ref{fig:Tdust_z_deltaMS}a) and an assumption of modified black body (Equation~\ref{eq:ModifiedBB}) to reproduce the $L_{\rm IR}$ detection limit for all of our data (solid curve). } 
\label{fig:LIR_vs_redshift}
\end{figure}

Figure \ref{fig:LIR_vs_redshift} shows the $L_{\rm IR}$ of our sample as a function of redshift. We note that the error measurements in sources with $z_{\rm FIR}$ estimates only represent statistical errors from the minimization procedure (\S\ref{subsubsec:MissingSources}) and do not include the systematic uncertainties associated with different sets of SED templates used in the fitting. For the 46 sources without radio or 24-$\micron$ identifications, their infrared luminosities would be log($L_{\rm IR}$)=12.3--13.7\,$\rm L_{\sun}$ if we place them at their plausible median redshift of $z=3$ (see \S\ref{subsec:UnidentifiedSources}) and scale their 450-$\micron$ flux densities to $L_{\rm IR}$ using the averaged ALESS SMG SED \citep{da-Cunha:2015aa}. They are shown with the orange box in Figure \ref{fig:LIR_vs_redshift}. In Figure \ref{fig:LIR_vs_redshift}, we also show the detection limit corresponding to a 4$\sigma$ limit of 2.6\,mJy (= 4 $\times$ 0.65\,mJy). To do this, we simply converted the 450-$\micron$ flux density limit to $L_{\rm IR}$ using the averaged ALESS SMG SED (\citealt{da-Cunha:2015aa}; dashed curve in Figure \ref{fig:LIR_vs_redshift}). 

%%%%%%%%%%%%%%%%%%%%%%%%%%%%%%%%%%%%%%%%%%%%%%%%%%%%%%%%%%%%%%%%%%%%%%

\subsection{Dust Properties} \label{subsec:DustProperties}

The shapes of SEDs at the FIR wavelengths ($\lambda$ $\geqslant$ 60\,$\micron$) are empirically found (e.g. \citealt{Magnelli:2012aa, Casey:2013aa, Roseboom:2013aa}) to be fairly well represented by a single modified blackbody function: 
\begin{equation}
S_\nu = S_{0} \left[ 1 - e^{-(\nu/\nu_{0})^{\beta}}\right] \frac{2 h \nu^{3}}{e^{\frac{h\nu}{kT_{\rm d}}}-1} .
\label{eq:ModifiedBB}
\end{equation}
Here $S_\nu$ is the flux density, $S_{0}$ is the normalization of the modified blackbody, $\nu_0$ (which we take to be c/100\,$\micron$; \citealt{Klaas:2001aa}) is the rest-frame frequency where the emission becomes optically thick, $\beta$ is the dust emissivity spectral index, and $T_{\rm d}$ is the dust temperature. We employed a $\chi^2$ SED fitting procedure to estimate the three unknown parameters, $S_{0}$, $T_{\rm d}$, and $\beta$. We set them as free parameters for the sources that have $\geqslant$ 3 photometric observations in FIR wavelength. There are 97 such sources in our sample. We inserted the median value of the emissivity index $\beta = 1.80\pm{0.03}$ from this sub-sample into the fitting for the 72 sources with two photometric observations in the FIR wavelength. Our typical $\beta$ value is consistent with previous observational studies, which suggest $\beta$ = 1.5--2 (e.g., \citealt{Magnelli:2012aa, Roseboom:2013aa}). In total, we determine the dust properties of the 169 sources that have optical redshifts, and their median dust temperature is $T_{\rm d} = 38.3^{+0.4}_{-0.9}$\,K with a 16th-to-84th percentile range of 30--50\,K. We do not attempt to constrain the FIR SEDs of sources without optical redshifts, since there is a degeneracy between redshift and dust temperature. We also can estimate the $L_{\rm IR}$ from the best-fit modified blackbody (black curves in Figure \ref{fig:SEDs}). The $L_{\rm IR}$ estimates from this method are slightly off by $0.8\pm0.2$\,dex, on average, compared to those from the template-based measurements (\S\ref{subsec:InfraredLuminosities}). This offset is expected, since the assumption of a single modified blackbody will lead to an underestimate in the mid-infrared (see fits in Figure \ref{fig:SEDs}). Therefore, we adopt the template-based measurements of $L_{\rm IR}$ in this work. 

Our median of $38.3^{+0.4}_{-0.9}$\,K is between the estimates from previous studies of SCUBA-2 450-$\micron$-selected samples ($\left\langle T_{\rm d}\right\rangle = 42 \pm 11\,$K, \citealt{Roseboom:2013aa} ; $\left\langle T_{\rm d}\right\rangle = 42 \pm 15\,$K, \citealt{Zavala:2018aa}) and ALMA-identified LABOCA 870-$\micron$-selected SMGs ($\left\langle T_{\rm d}\right\rangle = 33^{+3}_{-2}\,$K, \citealt{Simpson:2017ab}). This may not be consistent with the expectation that a longer selection waveband tends to select cooler sources (see also \citealt{Chapin:2009aa, MacKenzie:2016aa}). This may be explained by the correlation between $T_{\rm d}$ and $L_{\rm IR}$ (\S\ref{subsec:Td-LIR}) and the fact that our observations are more sensitive to low-luminosity systems. On the other hand, if we consider the error bars in all of these measurements, the differences among them are marginal.

%%%%%%%%%%%%%%%%%%%%%%%%%%%%%%%%%%%%%%%%%%%%%%%%%%%%%%%%%%%%%%%%%%%%%%

\subsection{Ultraviolet-continuum Slope and Ultraviolet Luminosity}  \label{subsection:UV}

The rest-frame ultraviolet-continuum slope ($\beta_{\rm UV}$) has been widely used to measure dust attenuation in galaxies \citep{Calzetti:1994aa, Meurer:1999aa}. The accurate broad-band photometry from the COSMOS2015 catalog provides us with reliable measurements of $\beta_{\rm UV}$. For each source with a redshift measurement, we selected the filters that are close to its rest-frame ultraviolet (1650 and 2300\,$\AA$; e.g. \citealt{Bouwens:2009aa, Bouwens:2014aa}). There is potential contamination from stellar or interstellar absorption features in this wavelength interval at rest-frame 2175\,$\AA$ \citep{Stecher:1965aa}. The 2175-$\AA$ absorption has been detected in star-forming galaxies up to $z\simeq2$ \citep{Noll:2005aa, Noll:2007aa, Noll:2009aa, Conroy:2010aa, Buat:2011aa, Buat:2012aa, Wild:2011aa, Kriek:2013aa} but is absent in local starburst galaxies \citep{Calzetti:1994aa} and the Small Magellanic Cloud (SMC; \citealt{Pei:1992aa, Gordon:2003aa}). It is unclear whether this 2175-$\AA$ feature is present in our sample due to the difficulty of observations on ultraviolet-absorption in the dusty population. Throughout this work, we do not apply any correction for this feature (similar to that in \citealt{Casey:2014aa} for dusty galaxies), and we leave this question to future ultraviolet studies of the dusty population. We calculated $\beta_{\rm UV}$ with
\begin{equation}
\beta_{\rm UV} = \frac{M_{\lambda_{1}}-M_{\lambda_{2}}} {2.5 \times \log(\lambda_{2}/\lambda_{1}) - 2}, 
\label{eq:BetaUV}
\end{equation}
where $M_{\lambda_{1}}$ and $M_{\lambda_{2}}$ are the magnitudes in certain passbands at wavelengths $\lambda_{1}$ and $ \lambda_{2}$ in the ultraviolet range. Table \ref{tbl:FiltersBetaUV} summarizes the wavebands that we adopted for Equation~\ref{eq:BetaUV}. At the same time, the rest-frame ultraviolet magnitudes at wavelengths $\lambda_{1}$ and $ \lambda_{2}$ from Table \ref{tbl:FiltersBetaUV} also allow us to estimate the ultraviolet luminosity ($L_{\rm UV}$) for our sample. In total, 163 sources have both $\beta_{\rm UV}$ and $L_{\rm UV}$ estimations. Considering that the uncertainties in the measured $\beta_{\rm UV}$ and $L_{\rm UV}$ can be propagated from the redshift uncertainty, we perturbed the redshift by using the redshift uncertainty in each source, assuming a Gaussian distribution. We repeated the procedure 100 times and re-estimated the values of $\beta_{\rm UV}$ and $L_{\rm UV}$. The standard deviation from these procedures was added in quadrature to the estimated $\beta_{\rm UV}$ and $L_{\rm UV}$ errors.

%%%%%%%%%%%%%%%%%%%%%%%%%%%%%%%%%%%%%%%%%%%%%%%%%%%%%%%%%%%%%%%%%%%%%%

\subsection{SFRs} \label{subsection:SFRs}

We follow the ultraviolet and FIR SFR estimation procedure in \cite{Kennicutt:2012aa}, which is well calibrated from a combination of SFR tracers at different wavelengths. We derive the unobscured SFR ($\rm SFR_{\rm UV}$) using:
\begin{equation}
\log \left(\frac {\rm SFR_{\rm UV}} {\rm M_{\sun}\,yr^{-1}}\right) = \log \left(\frac {L_{\rm UV}} {\mathrm{W}}\right) - 36.17.
\end{equation}
On the other hand, the obscured star formation activity ($\rm SFR_{\rm IR}$) is related to the integrated $L_{\rm IR}$ by
\begin{equation}
\log \left(\frac {\rm SFR_{\rm IR}} {\rm M_{\sun}\,yr^{-1}}\right) = \log \left(\frac {L_{\rm IR}} {\mathrm{W}}\right) - 36.41.
\label{eq:SFR_IR}
\end{equation}
We sum the obscured and unobscured SFRs to be the total SFR of our sample, as is usually done (i.e. assuming these are independent tracers of star formation). We note that differences in assumed IMF, SPS models, luminosity-to-SFR conversions, dust attenuations, and emission-line contributions can lead to differences in derived SFRs by as much as a factor of 3 \citep{Speagle:2014aa}.

We do not adopt the estimates of SFR in the COSMOS2015 catalog, since there is a notable discrepancy between the \texttt{LE~PHARE}-based SFR and our measurements. A similar finding was also mentioned in \cite{Casey:2013aa} and \cite{Elbaz:2018aa}, where the SED fitting underestimated the SFR of the dusty population, compared to the more direct estimate ($\rm SFR_{\rm UV}$~$+$~$\rm SFR_{\rm IR}$). In our case, the direct SFR measurements are, on average, $0.4^{+0.7}_{-0.5}$\,dex above the \texttt{LE~PHARE}-based determinations. The discrepancy between these two estimations is expected, since the \texttt{LE~PHARE}-based SFR is mainly derived from the dust-corrected ultraviolet flux, which highly depends on the assumption of dust extinction correction. Our sample is bright at the FIR wavelengths and known to be dusty. Hence, a method that works well on the bulk of the optical galaxy population does not necessarily work well on our 450-$\micron$-selected sources.

%%%%%%%%%%%%%%%%%%%%%%%%%%%%%%%%%%%%%%%%%%%%%%%%%%%%%%%%%%%%%%%%%%%%%%

\subsection{Radio Power at 1.4\,GHz} \label{subsec:RadioPower}
We compute the rest-frame 1.4\,GHz radio power using the following equation:
\begin{eqnarray}
\left(\frac{P_\mathrm{1.4\,GHz}}{\mathrm{W}\,\mathrm{Hz}^{-1}}\right) = 4\pi \left(\frac{d_{\rm L}}{\rm m}\right)^{2} \left(\frac{S_\mathrm{1.4\,GHz}}{\mathrm{W}\,\mathrm{m}^{-2}\,\mathrm{Hz}^{-1}}\right) \nonumber \\ \times (1+z)^{\alpha_{\rm r} -1}, 
\label{eq:RadioPower}
\end{eqnarray}
where $\alpha_{\rm r}$ is the radio spectral index and $d_{\rm L}$ is the luminosity distance. Out of the 134 sources identified with radio, 42 sources without AGN contamination have both 1.4-GHz and 3-GHz detections. For these sources, we can directly estimate their $\alpha_{\rm r}$ and $P_\mathrm{1.4\,GHz}$. We then adopt the median of their radio spectral index $\alpha_{\rm r} = -0.88^{+0.06}_{-0.02}$ from this sub-sample to extrapolate $S_\mathrm{1.4\,GHz}$ from $S_\mathrm{3\,GHz}$ for sources with only 3-GHz detections. Although our sample size is small, we find no evidence for redshift evolution in $\alpha_{\rm r}$, consistent with that typically found on star-forming galaxies \citep{Delhaize:2017aa}. Our adopted radio spectral index is consistent with previous studies of star-forming galaxies ($\alpha_{\rm r} = -0.8$, \citealt{Condon:1992aa}; $\alpha_{\rm r} = -0.7$, \citealt{Delhaize:2017aa}), \emph{Herschel}-250\,$\micron$ selected galaxies ($\alpha_{\rm r} = -0.75$, \citealt{Ivison:2010ab}), ALMA selected SMGs ($\alpha_{\rm r} = -0.79$, \citealt{Thomson:2014aa}; $\alpha_{\rm r} = -0.61$ to $-0.91$, \citealt{Thomson:2019aa}), and faint radio sources $S_\mathrm{1.4\,GHz} <$ 1\,mJy ($\alpha_{\rm r} = -0.67$, \citealt{Bondi:2007aa}; $\alpha_{\rm r} = -0.6$ to $-0.7$, \citealt{ Ibar:2009aa}) within the errors. 

\begin{figure}[!h]
\centering
%%% qIR v.s. z %%%
\includegraphics[width=\columnwidth]{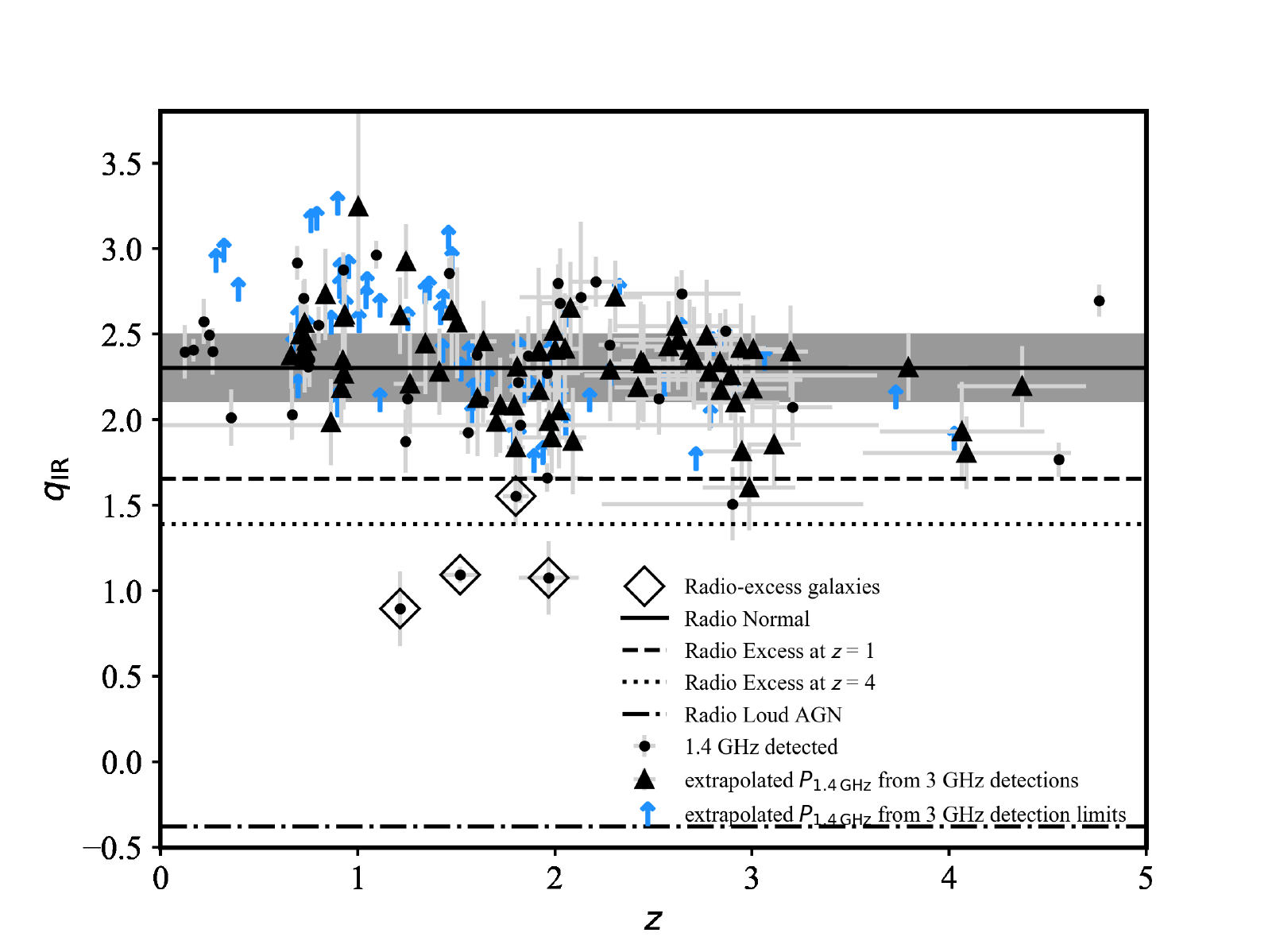}
\caption{Shown is $q_{\rm IR}$ vs. $z$ for our 450-$\micron$ sample. Our sample follows the tight local FIR--Radio correlation (solid line; \citealt{Condon:1992aa}), while the shaded area represents the $\pm1\sigma$ dispersion in the local correlation. The four sources enclosed by diamonds are classified as radio-excess galaxies, although none of them exceed the threshold for radio-loud AGNs (see \S \ref{subsec:AGNContamination}). The lower limits of $q_{\rm IR}$ for those sources without 3-GHz detections are shown as upward-pointing arrows. Their $S_\mathrm{1.4\,GHz}$ are extrapolated from the 5$\sigma$ detection limit of $S_\mathrm{3\,GHz} = 13\,\mu$Jy by assuming our typical $\alpha_{\rm r}$. }
\label{fig:qIR_vs_z}
\end{figure}

Figure \ref{fig:qIR_vs_z} shows $q_{\rm IR}$ versus $z$ for our sample, where $q_{\rm IR}$ is the ratio between the $L_{\rm IR}$ and $P_\mathrm{1.4\,GHz}$ \citep{Helou:1985aa}:

\begin{equation}
q_{IR} = \log \left(\frac{L_{\rm IR}}{3.75\,\times\,10^{12}\,\mathrm{W}}\right) \nonumber  - \log \left(\frac{P_\mathrm{1.4\,GHz}}{\mathrm{W}\,\mathrm{Hz}^{-1}}\right).
\label{eq:qIR}
\end{equation}
The $q_{\rm IR}$ values of our sample are nearly constant across redshift and agree well with the local FIR--radio correlation described in \citet{Condon:1992aa} (solid line in Figure \ref{fig:qIR_vs_z}). This also implies that the origin of both the radio and infrared emission of our sources is the same: star formation. In Figure \ref{fig:qIR_vs_z}, we also show the lower limits of $q_{\rm IR}$ for those sources without 3-GHz detections. We simply extrapolate their $S_\mathrm{1.4\,GHz}$ from the 5$\sigma$ detection limit of $S_\mathrm{3\,GHz} = 13\,\mu$Jy by assuming our typical $\alpha_{\rm r}$. Considering the dispersion in the local FIR--radio correlation (shaded area in Figure \ref{fig:qIR_vs_z}), we conclude that this sub-sample is in broad agreement with the normal galaxy population and that they should be detected in deeper radio surveys.

\section{The nature of SCUBA-2 450-$\micron$-selected sources} \label{sec:Results}

The main goals of this work are to characterize the 450-$\micron$-selected SMGs, to determine how they relate to ultraviolet/optical and other infrared/sub-millimeter selected populations, and to understand the role they play in the context of total star formation in the Universe. In this section, we investigate the correlations among the physical parameters measured in \S\ref{sec:PhysicalParameters} (summarized in Tables \ref{tbl:Catalog1} and \ref{tbl:Catalog2}) and compare them with results from the literature. Our sample comes from the deepest single-dish survey at 450\,$\micron$. This provides us a good opportunity to probe the physical properties of a fainter SMG population that was not previously possible, down to $L_{\rm IR} \simeq 10^{11}\,\rm L_{\sun}$. 

%%%%%%%%%%%%%%%%%%%%%%%%%%%%%%%%%%%%%%%%%%%%%%%%%%%%%%%%%%%%%%%%%%%%%%

\subsection{The star-formation main sequence}

\begin{figure*}
\centering
%%% SFR vs Mstar %%%
\includegraphics[width=0.7\paperwidth]{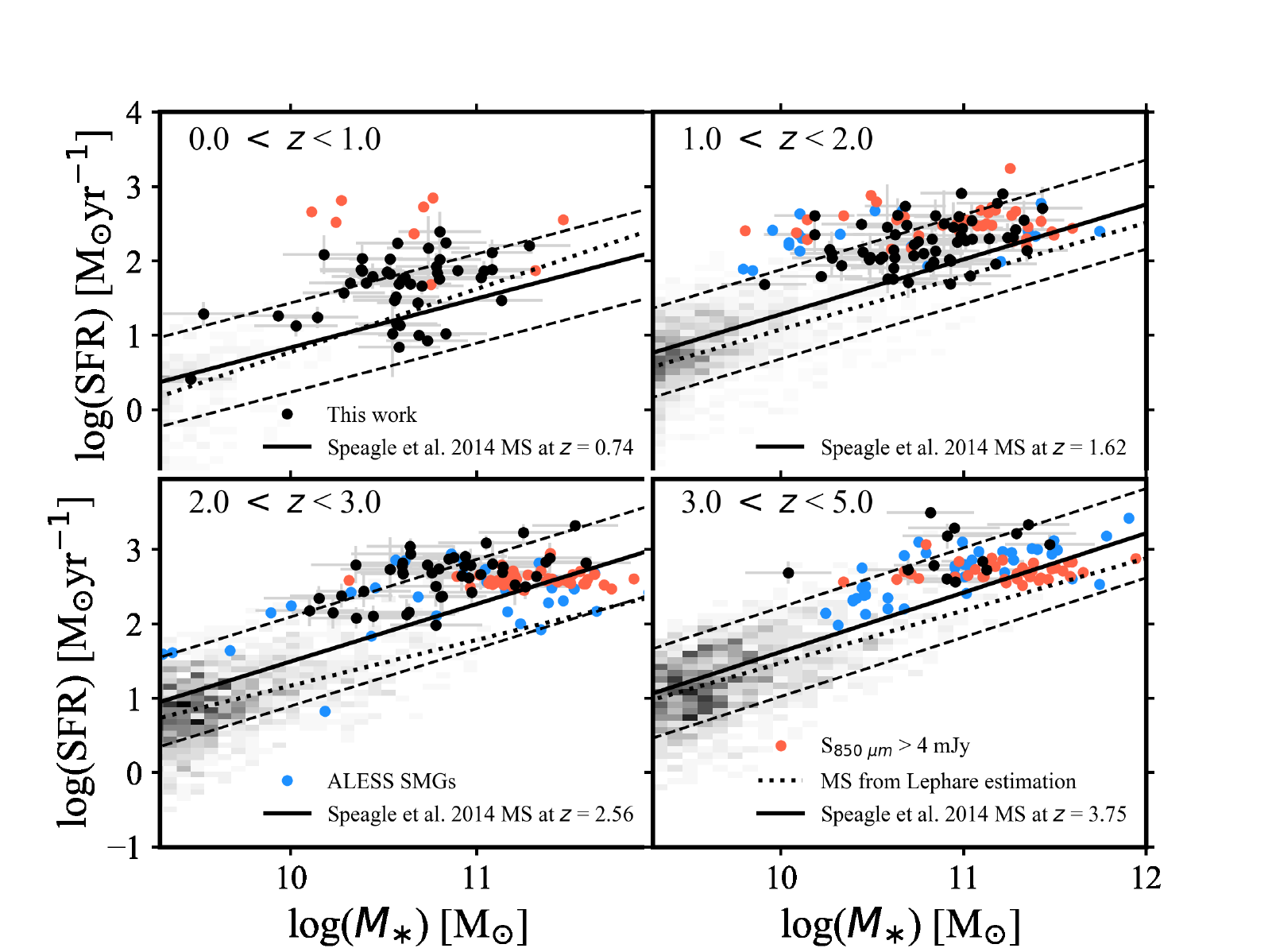}
\caption{The SFRs as a function of stellar mass for our sample. The gray scales and dotted lines represent the distributions and relations from normal star-forming galaxies in the COSMOS2015 catalog, respectively. The redshift-dependent relations from \citet{Speagle:2014aa} are shown with solid lines, while their $\pm0.6$\,dex ranges are shown with dashed lines. A sample of bright SCUBA-2 SMGs \citep[$S_\mathrm{850\,\micron} > 4$\,mJy; ][]{Michaowski:2017aa} is shown with red points, and ALESS SMGs \citep{da-Cunha:2015aa} are shown with blue points. }
\label{fig:SFR_vs_Mstar}
\end{figure*}

\begin{figure}
\centering
%%% Delta sSFR %%%
\includegraphics[width=\columnwidth]{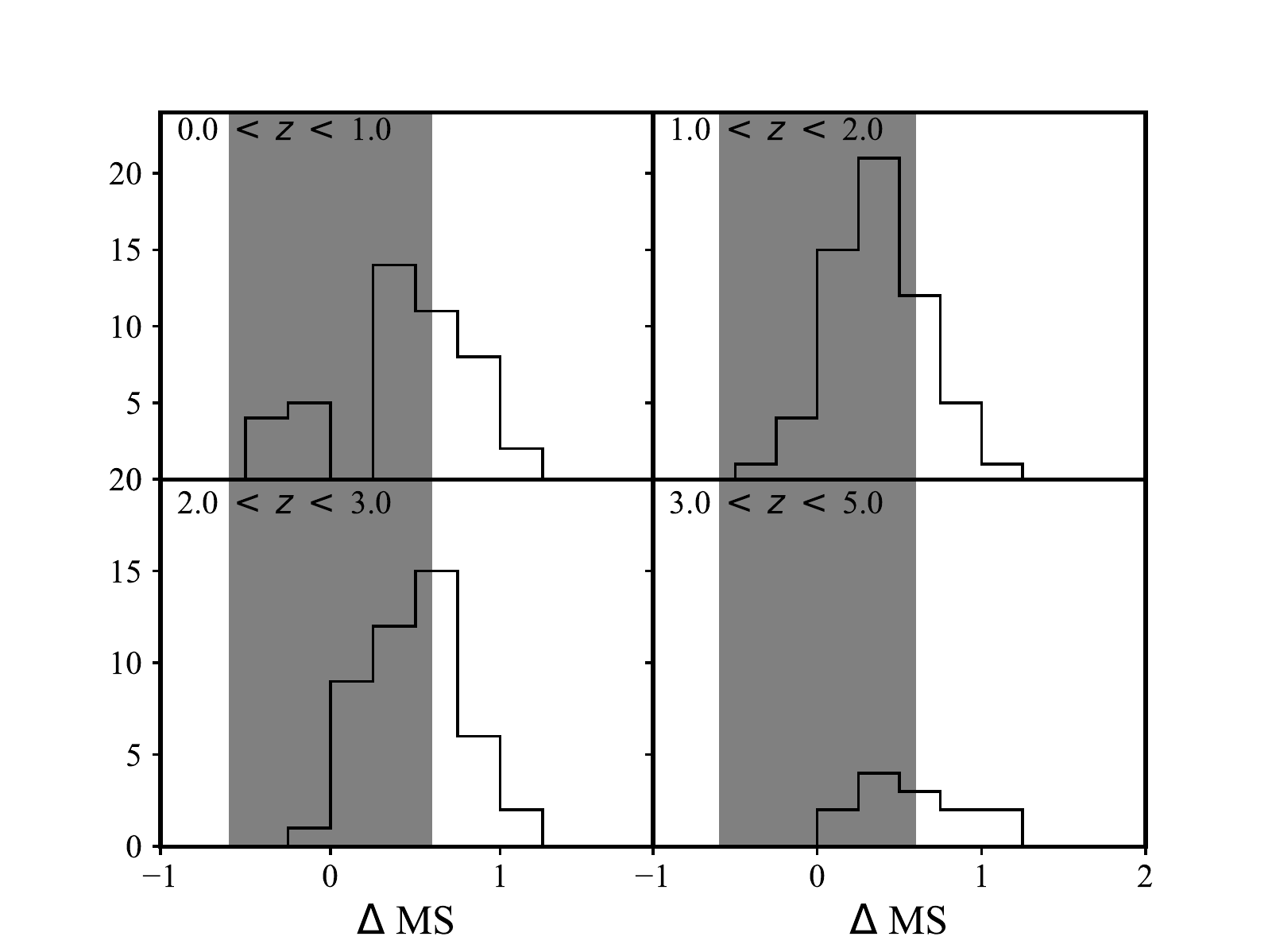}
\caption{Histogram of main-sequence deviation ${\Delta MS}$ of our sample in different redshift bins. The shaded areas show the range of 3$\sigma$ scatter ($\pm0.6$\,dex) of the \cite{Speagle:2014aa} relations. A significant fraction of our sample, 58/162 ($35^{+32}_{-25}$\%), can be classified as starburst galaxies.}
\label{fig:DeltasSFR}
\end{figure}

A tight relationship between SFR and stellar mass, the so-called star-formation ``main sequence,'' suggests that galaxies build up their stars on long timescales, likely a consequence of smooth gas accretion from the intergalactic medium. On the other hand, a ``starburst'' galaxy, which may be triggered by merging (or some other mechanism), may depart from the main sequence and exhibit enhanced sSFR. Based on observations of the nearby Universe, the star formation activity of luminous infrared galaxies (LIRGs; $10^{11}\,\rm L_{\sun} \lesssim$ $L_{\rm IR}$ $\lesssim 10^{12}\,\rm L_{\sun}$) and ultra-luminous infrared galaxies (ULIRGs; $10^{12}\,\rm L_{\sun} \lesssim$ $L_{\rm IR}$ $\lesssim 10^{13}\,\rm L_{\sun}$) are widely believed to be triggered by merger events \citep{Sanders:1988aa, Sanders:1996aa, Farrah:2001aa, Armus:2009aa, Swinbank:2010aa, Alaghband-Zadeh:2012aa, U:2012aa}. Various observations suggest that star formation in SMGs is merger-driven and that they may be scaled-up high-redshift analogs of local (U)LIRGs \citep{Smail:2004aa, Iono:2009aa, Ivison:2010aa}. Morphological analysis from \emph{Hubble Space Telescope} (\emph{HST}) near-infrared imaging also suggests that star formation correlates with galaxy interaction and merging activity \citep{Chen:2015aa, Chang:2018aa}. However, an alternative picture favoring disk star formation also exists. From studies using \emph{HST} optical imaging, a significant fraction of the 850-$\micron$-selected population is found to have clumpy disk-like stellar morphology \citep{Targett:2011aa, Targett:2013aa}. SMGs appear to have clumpy disk-like dust and gas based on 870-$\micron$ ALMA high-resolution dust imaging \citep{Hodge:2016aa}. To date, the triggering mechanism of SMGs remains inconclusive.

Figure~\ref{fig:SFR_vs_Mstar} presents the SFRs of our sample as a function of stellar masses (SFR--$M_{\ast}$ relation) in different redshift bins. We also show the redshift-dependent SFR--$M_{\ast}$ relations from \cite{Speagle:2014aa} computed using the median redshifts of our sources in the corresponding redshift bins (solid lines). \cite{Speagle:2014aa} included data from 25 different studies in the literature, which contained 64 star-formation main-sequence relations, to adjust each relation onto an empirically scaled correlation using a variety of conversion factors. The authors standardized the data to the Kroupa IMF \citep{Kroupa:2001aa, Kroupa:2003aa}, the SFR conversions in \cite{Kennicutt:2012aa}, and the BC03 SPS model. 

In Figure~\ref{fig:SFR_vs_Mstar}, we also show the SFR--$M_{\ast}$ relation for normal star-forming galaxies, which is defined through the condition specific SFR (sSFR $\equiv$ SFR/$M_{\ast}$) $\geqslant$ [3 $\times$ $t_{\rm H}(z)$]$^{-1}$ (similar to that in \citealt{Wuyts:2012aa} and \citealt{Chang:2013aa}), from the COSMOS2015 catalog (gray scales and dotted lines in Figure~\ref{fig:SFR_vs_Mstar}), where $t_{\rm H}(z)$ is the Hubble time at the redshift of a galaxy. For this purpose, we directly adopt the SFRs and stellar masses from the COSMOS2015 catalog. Readers need to keep in mind that, as mentioned in \S\ref{subsection:SFRs}, the \texttt{LE PHARE}-based SFR differs from our direct SFR measurements ($\rm SFR_{\rm UV}$ + $\rm SFR_{\rm IR}$) for the SMGs. For comparison, in Figure~\ref{fig:SFR_vs_Mstar}, we also show the bright SMGs ($S_\mathrm{850\,\micron} > 4$\,mJy) with identifications having at least one robust association with $p$-value $< 0.05$ at radio, 24\,$\micron$, or 8.0\,$\micron$ from \cite{Michaowski:2017aa} and ALESS SMGs from \cite{da-Cunha:2015aa}. In the redshift bins of $1 < z < 3$, where there are sufficient data points from \cite{Michaowski:2017aa}, \cite{da-Cunha:2015aa}, and us, our sample appears to lie between the bright SMGs and some of the faint SMGs from ALESS. This indicates that our data start to probe into the normal star-forming population out to $z \simeq 3$.

The scatter of the SFR--$M_{\ast}$ relation is often used for separating the main-sequence and starburst populations. The offset in the sSFR for each source from the main sequence is denoted as ${\Delta MS}$ (aka starburstiness, see \citealt{Elbaz:2011aa}) in this work. The main-sequence relation is relatively tight, with an intrinsic scatter that is approximately a constant of about 0.2\,dex over cosmic time \citep{Speagle:2014aa}. Using the definition of ${\Delta MS}$ $\geqslant$ $+0.6$\,dex (i.e., 3$\sigma$ away from the main sequence), we can identify starburst galaxies within our sample. This criterion is identical to that adopted by \citet{Rodighiero:2011aa}. In the following discussion, we only compare our results with the calibrated parameterization of the SFR--$M_{\ast}$ relations from \cite{Speagle:2014aa}, in which an extensive compilation of observations from ultraviolet to FIR were adopted.

Among the 165 sources in our sample that have both SFR and stellar mass estimates and do not host AGNs, a moderate fraction (58 sources, $35^{+32}_{-25}$\%) can be classified as starburst galaxies (Figure~\ref{fig:DeltasSFR}). The remaining sources are consistent with being on the main sequence. The fraction of starburst galaxies in our sample will decrease to $24^{+22}_{-17}$\% if we include the 18 sources without optical counterparts (\S \ref{subsec:CounterpartID}), the 46 sources without 24-$\micron$ or radio identifications (\S \ref{subsec:UnidentifiedSources}), and the nine sources without stellar-mass determinations (\S \ref{subsec:StellarMass}) by assuming that all of them are on the main sequence. A previous ALMA study of 870-$\micron$-selected galaxies \citep{da-Cunha:2015aa} showed that about half of the SMGs lie above the main sequence, while the other half are consistent with being on the massive end of the main sequence. On the other hand, some studies proposed that SMGs are just massive and highly star-forming main-sequence galaxies (SCUBA-2 SMGs; \citealt{Koprowski:2016aa, Michaowski:2017aa, Zavala:2018aa}; and ALMA 1.3-mm SMGs; \citealt{Dunlop:2017aa}). Our results show an intermediate main-sequence fraction that is between these two extremes and appear to be consistent with previous observations of SCUBA-2 450-$\micron$-selected galaxies from \citet[][26\%]{Roseboom:2013aa} and a recent study of ALMA-selected SMGs from \citet[][31\%]{Elbaz:2018aa}. We notice that the previous works are on relatively high-redshift ($z \simeq 2$) sources compared to ours, likely because most of them are 850-$\micron$-selected galaxies. Although \citealt{Zavala:2018aa} had both 450- and 850-$\micron$-selected galaxies, almost all of their sources have 850-$\micron$ detections. If we restrict ourselves to galaxies at $z > 1.5$, the starburst fraction is $36^{+22}_{-19}$\%. This value is still higher than that in \citet[][15\%]{Zavala:2018aa}. In any case, our work suggests that a notable fraction (50--85\%) of the 450-$\micron$ SMGs are consistent with lying on the main sequence. 

A recent study \citep{Sorba:2018aa} found that stellar masses from SED fitting can be underestimated and that the effect increases toward larger sSFR due to the outshining of stellar mass by regions of star formation, i.e., young stellar populations overpowering older stellar populations behind their bright flux (see also \citealt{Sorba:2015aa} and \citealt{Abdurrouf:2018aa}). This effect would shift all stellar masses to the right in Figure~\ref{fig:SFR_vs_Mstar}, but it would shift the high-sSFR outliers more than the galaxies on the main sequence: masses of galaxies on the main sequence with $\log$(sSFR) $\simeq$ 8.5 increase by $\simeq$ +0.05\,dex, but those with $\log$(sSFR) $\simeq$ 9.5 (at the edge of the main sequence) increase by $\simeq$ +0.5\,dex. However, it is unclear whether these corrections still hold in the dusty population. Therefore, in this work, we do not apply any correction from the literature. We expect this can be revealed by high-resolution observations with a kiloparsec scale.

Our sample shows that a notable fraction of the 450-$\micron$ SMGs are consistent with lying on the main sequence. A critical question exists regarding our observations. What is the main physical difference between the main-sequence SMGs and the optically selected normal star-forming galaxies? Some studies with morphological analyses of stellar structure suggest that the fraction of merger systems increases with the SFR or sSFR \citep{Hwang:2011aa, Hung:2013aa, Chang:2018aa}. However, the difference is statistically insignificant between the SMGs and normal star-forming galaxies (matched with SFR or sSFR) in their merger fractions \citep{Chang:2018aa}, indicating that merging events are probably not the only triggering mechanism for SMGs (see also \citealt{Hayward:2011aa}). Furthermore, we may also question the accuracy of the starburst fraction for SMGs. The exact locations of SMGs in the SFR--$M_{\ast}$ plane are highly dependent on the details of SFR and stellar mass estimations. A significant positional displacement between the optical stellar emission and corresponding ALMA 870-$\micron$ peaks has been found, suggesting that the majority of the dusty star-forming regions are not co-located with the unobscured stellar distribution \citep{Chen:2015aa}. Moreover, several spatially resolved studies of SMGs with ALMA reveal that the distribution of the gas emission is also spatially offset from unobscured stellar distribution \citep{Chen:2017aa, Calistro-Rivera:2018aa}. These findings caution against using global SED fitting routines or relying on stellar masses derived from them (e.g. \citealt{Laigle:2016aa}), particularly for dusty star-forming galaxies (DSFGs). High-resolution imaging is crucial for characterizing the properties of SMGs, including carrying out spatially resolved SED fitting and a better morphological description of the interstellar medium (gas or dust emission). 

%%%%%%%%%%%%%%%%%%%%%%%%%%%%%%%%%%%%%%%%%%%%%%%%%%%%%%%%%%%%%%%%%%%%%%

\subsection{$T_{\rm d}$--$L_{\rm IR}$ Correlation}\label{subsec:Td-LIR}

\begin{figure*}
\centering
%%% BBTemperature vs LIR %%%
\includegraphics[width=0.7\paperwidth]{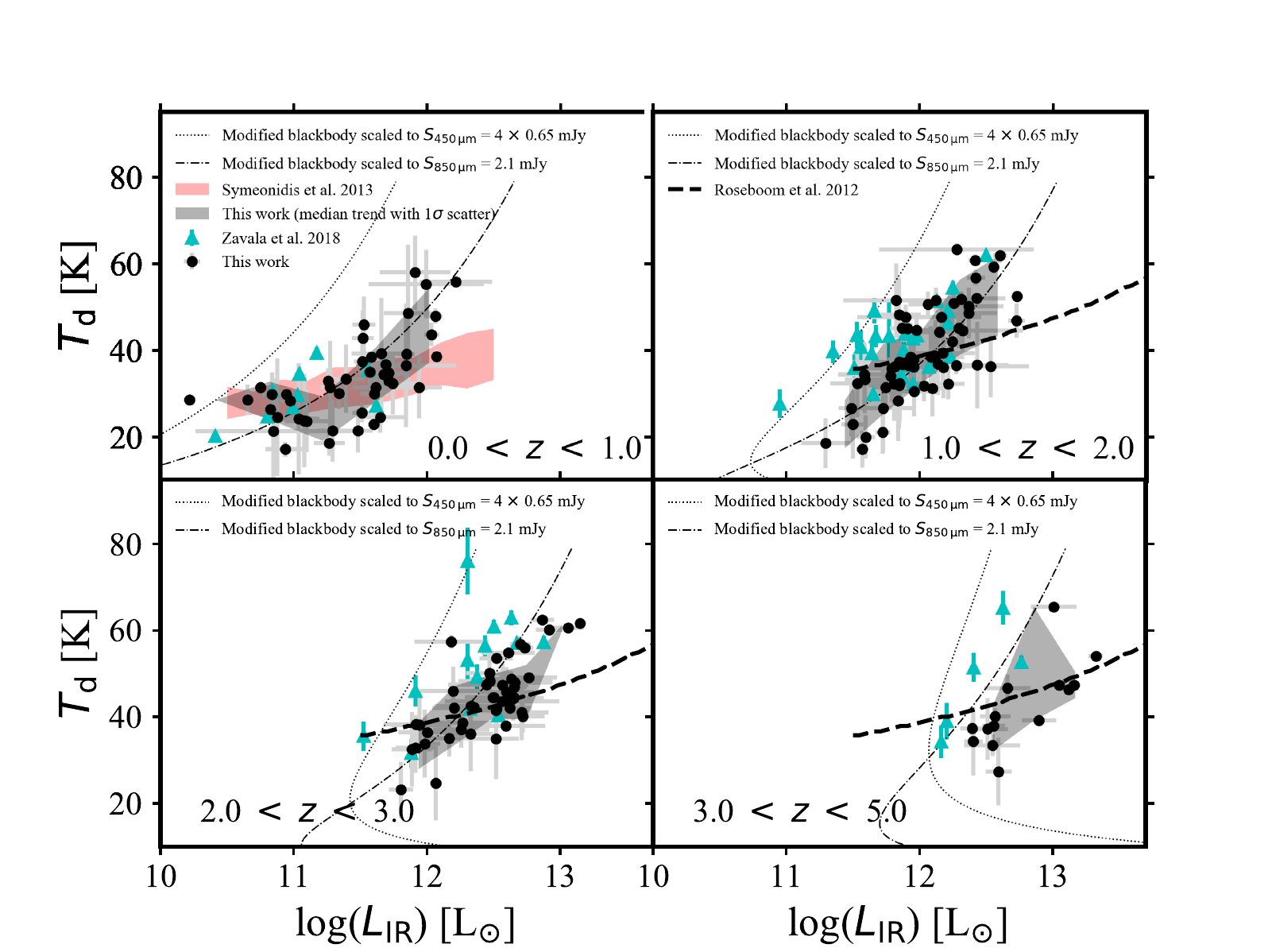}
\caption{$T_{\rm d}$ versus $L_{\rm IR}$ for our 450-$\micron$ sample (black circles) and other populations. The dark shaded areas show the running medians of our sample and the 1$\sigma$ scatters. The red shaded area shows the range of 1$\sigma$ scatter of the $z=0$--1 \emph{Herschel}/SPIRE-selected LIRGs/ULIRGs from \citet{Symeonidis:2013aa}. The dashed curves are the relation derived from $z > 1.5$ \emph{Herschel}/SPIRE-selected sources in \citet{Roseboom:2012aa}. The triangles are 450-$\micron$-selected SMGs from the shallower SCUBA-2 survey of \citet{Zavala:2018aa}. To demonstrate the effects of selection, we convert our flux detection limits to the $L_{\rm IR}$ sensitivity limits at a given $T_{\rm d}$ using a modified blackbody under detection limits corresponding to a 450-$\micron$ noise of 0.65\,mJy (dotted curves) and the 850-$\micron$ confusion limit of 2.1\,mJy (dash-dotted curves) in the middle of the redshift bins.}
\label{fig:BBTemperature_vs_LIR}
\end{figure*}

\begin{figure*}
\centering
%%% Tdust vs z vs deltaMS %%%
\includegraphics[width=0.7\paperwidth]{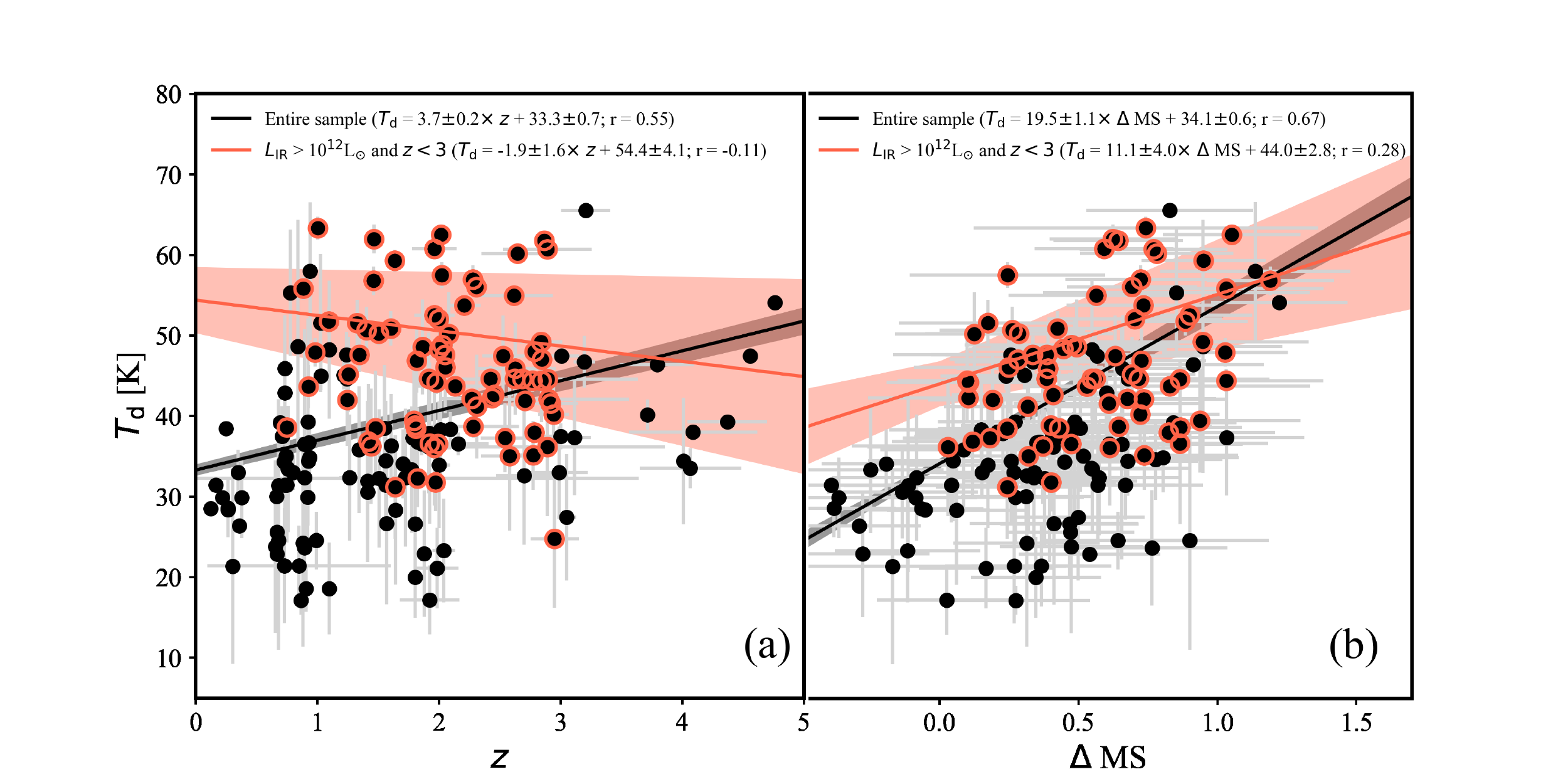}
\caption{(a) $T_{\rm d}$ vs. redshift. We find a moderate redshift dependence of dust temperature with a Pearson correlation coefficient of $r = 0.55$. We obtain almost no correlation ($r = -0.11$) on the $T_{\rm d}$--$z$ plane for the the sub-sample of $L_{\rm IR} > 10^{12}\,\rm L_{\sun}$ at $z<3$ (red symbols). This may reflect the selection effect that 450-$\micron$ observations are biased against low-$L_{\rm IR}$ (and thus low-$T_{\rm d}$) galaxies at high redshift. (b) $T_{\rm d}$ vs. ${\Delta MS}$. We also find moderate correlations between the ${\Delta MS}$ and $T_{\rm d}$ from our entire sample (black symbols) and the sub-sample of $L_{\rm IR} > 10^{12}\,\rm L_{\sun}$ at $z<3$ (red symbols) with Pearson correlation coefficients of $r = 0.67$ and 0.28, respectively. This finding supports the scenario that starburst galaxies have higher $T_{\rm d}$ that is driven by the enhanced star formation.}
\label{fig:Tdust_z_deltaMS}
\end{figure*}

\begin{figure}
\centering
%%% Tdust vs delta MS %%%
\includegraphics[width=\columnwidth]{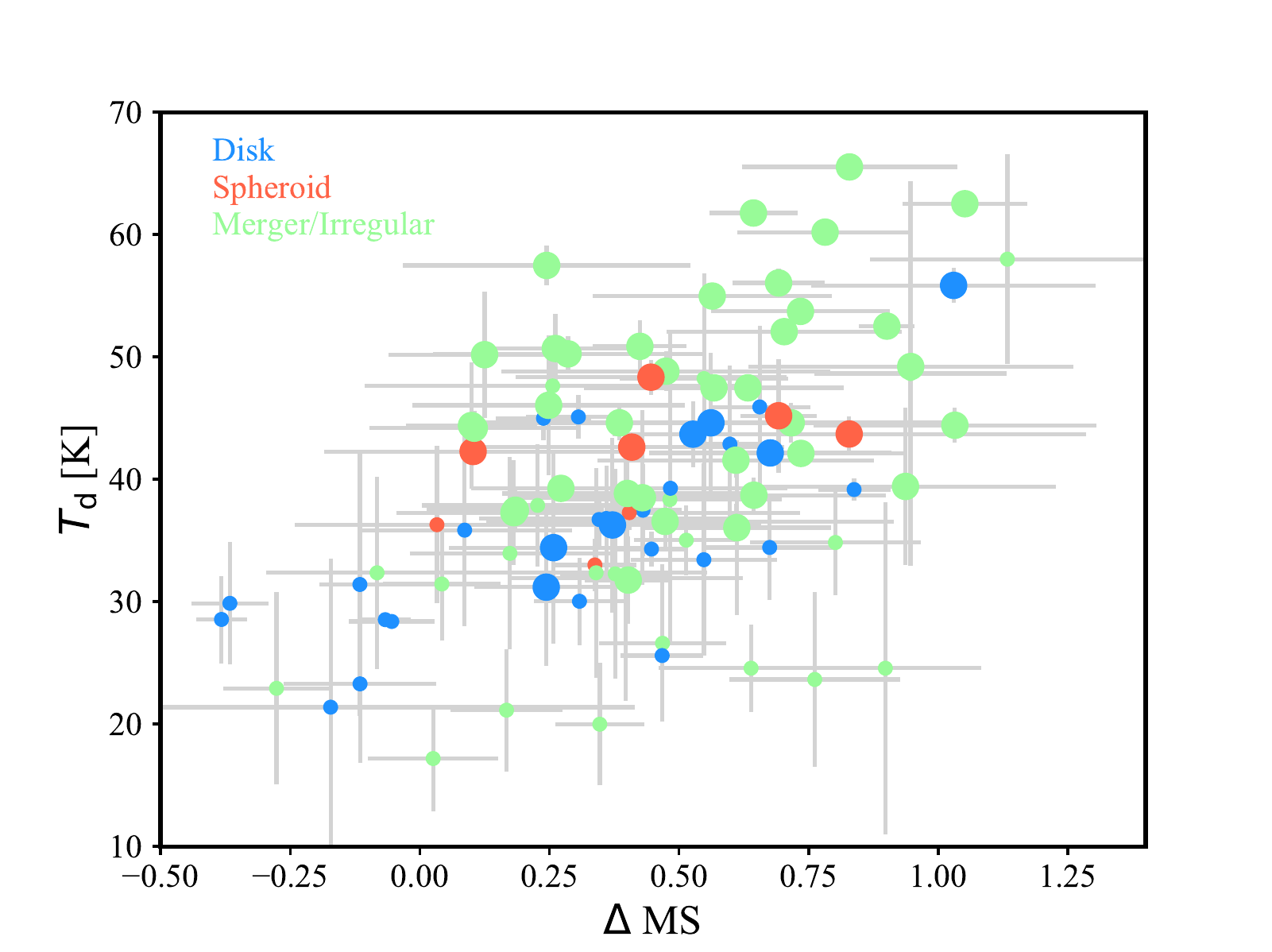}
\caption{Same as Figure \ref{fig:Tdust_z_deltaMS}b. Colors are the morphological classes from \cite{Chang:2018aa}. The sizes of the data points indicate $L_{\rm IR}$, where big circles are sources with $L_{\rm IR} > 10^{12}\,\rm L_{\sun}$. Sources that have merger or irregular features tend to have warmer $T_{\rm d}$ for a fixed $\Delta MS$. This finding further supports the scenario that mergers lead to an increase in $T_{\rm d}$. }
\label{fig:Tdust_vs_deltaMS(MorphClass)}
\end{figure}

\begin{figure*}
\centering
%%% IRX-beta %%%
\includegraphics[width=0.7\paperwidth]{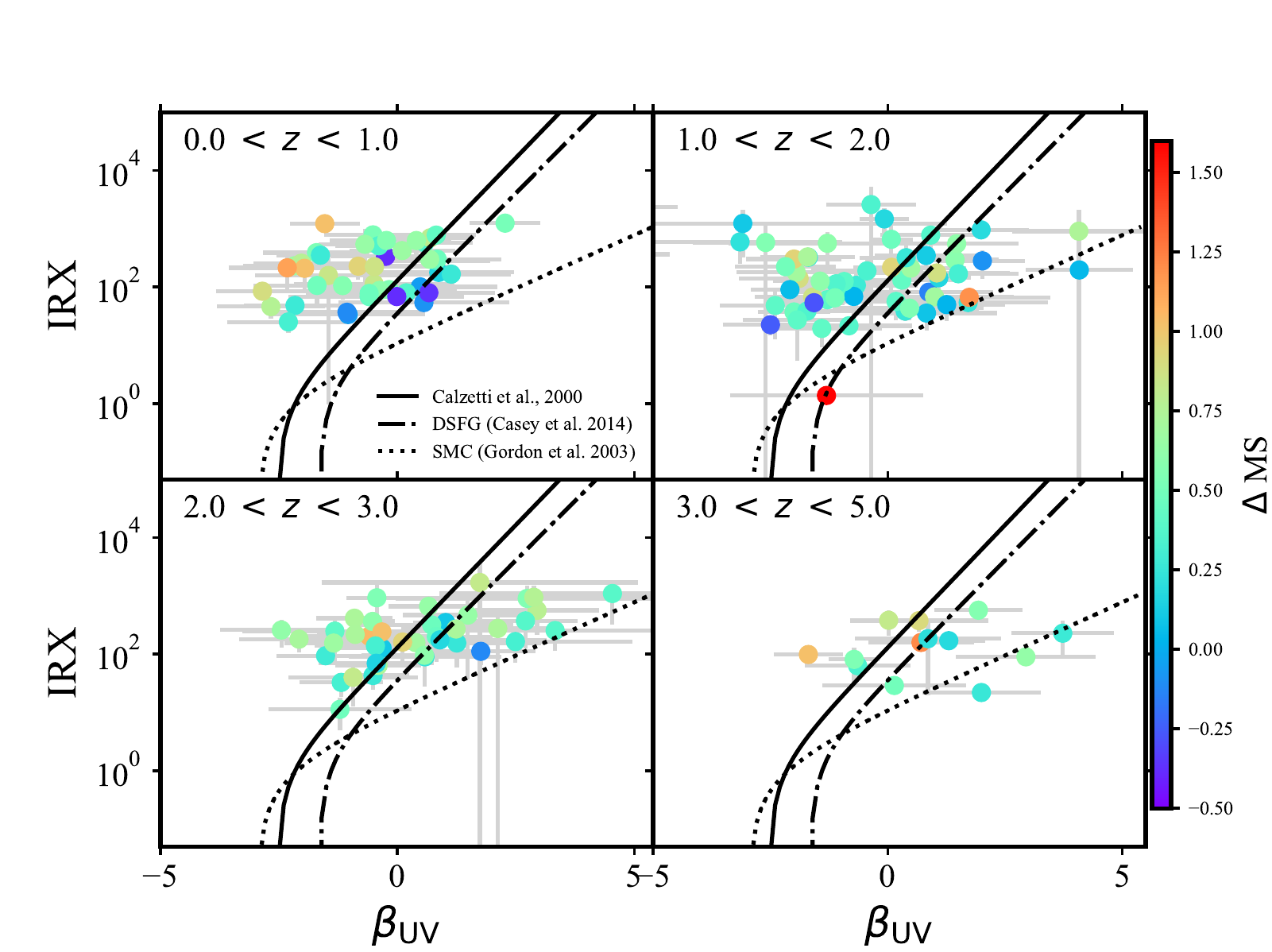}
\caption{The infrared excess (IRX) versus ultra-violet slope ($\beta_{\rm UV}$), color-coded by $\Delta MS$. Our sample shows no obvious correlation in the IRX--$\beta_{\rm UV}$ plane. For comparison, we plot attenuation curves from the SMC \citep[dotted lines; ][]{Gordon:2003aa}, local starburst galaxies \citep[solid lines; ][]{Calzetti:2000aa}, and local ($z < 0.085$) DSFGs \citep[dotted--dashed lines, ][]{Casey:2014aa}.}
\label{fig:IRX_vs_UVSlope}
\end{figure*}

Figure~\ref{fig:BBTemperature_vs_LIR} shows $T_{\rm d}$ versus $L_{\rm IR}$ for our sample in various redshift bins. The rising $T_{\rm d}$ with an increase in $L_{\rm IR}$ is expected, since the emission from the majority of dust is in equilibrium; therefore, the bulk of the infrared emission is well represented by the modified blackbody. However, the $T_{\rm d}$--$L_{\rm IR}$ correlation could be attributable partially to a real physical effect and partially to a selection effect. By comparison with our 450-$\micron$-selected sample, the typical SMG selection at 850-1200\,$\micron$ is known to be biased against very hot populations, since they are selected from the long-wavelength end of the Rayleigh-Jeans tail. To investigate this selection bias, we convert our flux detection limits to $L_{\rm IR}$ limits as functions of $T_{\rm d}$ using the modified blackbody in Equation~\ref{eq:ModifiedBB}. The 450-$\micron$ limits assuming a noise level of 0.65\,mJy are shown as the dotted curves, while the 850-$\micron$ limits assuming a confusion limit of 2.1\,mJy are shown as dotted--dashed curves. By comparing the detection limits at 450 and 850\,$\micron$, it is clear that the 850-$\micron$ selection effect biases the sample against hot sources and the 450-$\micron$ biases against cooler sources at high redshift. This bias becomes less apparent at high $L_{\rm IR}$, where sub-millimeter observations probe a larger range in dust temperature (see also \citealt{Casey:2012aa, Magnelli:2012aa, Swinbank:2014aa}). 

Our 450-$\micron$-selected sample, which probes the dust SED closer to its peak at $z \lesssim 3.5$, is less affected by the long-wavelength selection bias and potentially probes a large range in dust temperature. Indeed, our galaxies span a wide range of dust temperatures (20\,K $ \lesssim T_{\rm d} \lesssim$ 60\,K). For comparison, at $z<$ 1, we show the $T_{\rm d}$--$L_{\rm IR}$ relation from $z$ = 0--1 \emph{Herschel}/SPIRE-selected LIRGs and ULIRGs \citep{Symeonidis:2013aa} in Figure~\ref{fig:BBTemperature_vs_LIR}. A small offset may exist between our $T_{\rm d}$ estimates and theirs, since their $T_{\rm d}$ estimates were translated from the SED peaks using the Wien displacement law for the modified blackbody. In the high-redshift bins, we show the $T_{\rm d}$--$L_{\rm IR}$ relation from $z >$ 1.5 \emph{Herschel}/SPIRE-selected sources \citep{Roseboom:2012aa} and brighter SCUBA-2 450-$\micron$ sources \citep{Zavala:2018aa}. Overall, the medians of our sample (dark shaded areas in Figure \ref{fig:BBTemperature_vs_LIR}) are consistent with the previous studies within the measurement errors. At $z<1$, the distribution of our 450-$\micron$ sources appears to be different from the trend from \cite{Symeonidis:2013aa}, with higher $T_{\rm d}$ at high $L_{\rm IR}$ and a wider spread in $T_{\rm d}$ at lower $L_{\rm IR}$. Similar trends also appear to exist at $1<z<2$ and $2<z<3$ when comparing the distribution of our 450-$\micron$ sources with the relation of \cite{Roseboom:2012aa}. However, our measured values are consistent with the trends of \cite{Roseboom:2012aa} and \cite{Symeonidis:2013aa} under our current sample size and error bars. Therefore, we conclude that our galaxies overlap with all of these samples on the $T_{\rm d}$--$L_{\rm IR}$ plane. Despite a weak 450-$\micron$ selection effect in the $T_{\rm d}$--$L_{\rm IR}$ plane, we conclude that our sample is representative for SMGs of $L_{\rm IR} > 10^{12}\,\rm L_{\sun}$ over a wide redshift range, at least up to $z \simeq 3$.

We further examined the dependence of dust temperature on redshift and ${\Delta MS}$. To do this, we performed linear fits to the properties from our sample. In the fitting, we applied weights estimated by adding the uncertainties to the variables in quadrature. For estimating the weighted Pearson correlation coefficient, we took the weighted average and weighted sum when calculating the total sum of squares and the sum of squares of residuals. Our sample shows a moderate redshift dependence of dust temperature with a Pearson correlation coefficient of $r = 0.55$ (black line in Figure~\ref{fig:Tdust_z_deltaMS}a). This correlation is likely driven by the aforementioned selection effects at the low-luminosity end. If we restrict ourselves to sources with $L_{\rm IR} > 10^{12}\,\rm L_{\sun}$ at $z<3$ (luminosity and redshift ranges that are less affected by selection bias in our sample), we obtain almost no correlation on the $T_{\rm d}$--$z$ plane with a Pearson correlation coefficient of $r = -0.11$ (red line in Figure~\ref{fig:Tdust_z_deltaMS}a). This finding conflicts with previous studies where $T_{\rm d}$ was found to increase with redshift \citep{Magnelli:2013aa, Swinbank:2014aa, Bethermin:2015aa, Schreiber:2018aa, Zavala:2018aa}. On the other hand, our finding may be consistent with a recent study of high-resolution cosmological simulations where the mass-weighted dust temperature (based on radiative transfer modeling) does not strongly evolve with redshift over $z=2$--6 \citep{Liang:2019aa}. We suggest that the evolution derived in previous studies might be biased by the $T_{\rm d}$ selection effect (see also \citealt{Chapman:2004ab, Chapin:2009aa, MacKenzie:2016aa}) if they do not apply similar $L_{\rm IR}$ and $z$ cuts as we do.

Our entire sample and the sub-sample of $L_{\rm IR} > 10^{12}\,\rm L_{\sun}$ at $z<3$ show moderate correlations between ${\Delta MS}$ and $T_{\rm d}$ with Pearson correlation coefficients of $r = 0.67$ and 0.28, respectively (Figure \ref{fig:Tdust_z_deltaMS}b). This result is consistent with previous observations of \emph{Herschel}-selected dusty galaxies \citep{Magnelli:2012aa, Magnelli:2014aa}. This result is also in line with the semi-analytical model of hierarchical galaxy formation of \cite{Cowley:2017aa}, which suggests that starburst-dominated galaxies have generally hotter $T_{\rm d}$ driven by the enhanced star formation. However, these positive linear relationships between ${\Delta MS}$ and $T_{\rm d}$ are likely driven by the fact that ${\Delta MS}$ is proportional to SFR (i.e., $L_{\rm IR}$) and $L_{\rm IR}$ is correlated with $T_{\rm d}$ (Figure \ref{fig:BBTemperature_vs_LIR}). Indeed, we find moderate dependencies between the ${\Delta MS}$ and $L_{\rm IR}$ with Pearson correlation coefficients of $r = 0.53$ and 0.31 for our entire sample and the sub-sample of $L_{\rm IR} > 10^{12}\,\rm L_{\sun}$ at $z<3$, respectively. 

The large scatter of our sources in the ${\Delta MS}$--$T_{\rm d}$ plane could be caused by the uncertainties in the ${\Delta MS}$ and $T_{\rm d}$ measurements. To test this, we performed a simple Monte Carlo simulation, assuming perfect correlations (black and red lines in Figure \ref{fig:Tdust_z_deltaMS}b) with Pearson correlation coefficients of $r = 1.0$ and generating random realizations of ${\Delta MS}$ and $T_{\rm d}$ with the same sample size as our real data. We then perturbed the simulated ${\Delta MS}$ and $T_{\rm d}$ with the uncertainties in our real sample under the assumption of Gaussian distribution. We produced 100 realizations of these simulations and calculated their Pearson correlation coefficients. As expected, the mean values of the Pearson correlation coefficients from the iterations ($r = 0.68\pm0.04$ and $0.47\pm0.09$ for the black and red lines in Figure \ref{fig:Tdust_z_deltaMS}b, respectively) become lower and closer to the observed values. This implies that the intrinsic ${\Delta MS}$--$T_{\rm d}$ correlation appears to be stronger than the moderate observed correlations, which are strongly affected by measurement uncertainties. 

Figure~\ref{fig:Tdust_vs_deltaMS(MorphClass)} is a diagram similar to Figure~\ref{fig:Tdust_z_deltaMS}b with data points colored with \emph{HST} WFC3 morphological classes from \cite{Chang:2018aa} and sized with $L_{\rm IR}$. Ninety-seven sources have suitable CANDELS images to be classified. Galaxies with merger/irregular features have a median dust temperature of $T_{\rm d} = 40^{+4}_{-2}$\,K, which is warmer than galaxies with disk morphology (median $T_{\rm d} = 36\pm1$\,K), but the difference is marginal. We further performed the Kolmogorov--Smirnov test in the ${\Delta MS}$--$T_{\rm d}$ plane and the result shows $p$= 0.02, indicating that we can reject the null hypothesis of no difference between galaxies with merger/irregular features and galaxies with disk morphology. This is in line with three-dimensional dust radiative transfer calculations in hydrodynamic simulations of merging disk galaxies (e.g., \citealt{Hayward:2011aa}) and thus supports the scenario that the starbursts in SMGs are driven by mergers and that the more compact geometry in mergers leads to a sharp increase in $T_{\rm d}$ during the bursts.  

%%%%%%%%%%%%%%%%%%%%%%%%%%%%%%%%%%%%%%%%%%%%%%%%%%%%%%%%%%%%%%%%%%%%%%

\subsection{\rm IRX--$\beta_{\rm UV}$}

Detailed studies of dust attenuation, especially for the dusty population, will help in understanding the mechanism of the infrared reprocessed emission. Both $\beta_{\rm UV}$ and the ratio of $L_{\rm IR}/L_{\rm UV}$, often called ``IRX," are related to the amount of dust attenuation in galaxies. A correlation between $\beta_{\rm UV}$ and IRX is observed in local ultraviolet-bright starburst galaxies \citep{Meurer:1999aa, Calzetti:2001aa, Overzier:2011aa}. This correlation also seems to hold for high-redshift ultraviolet-selected star-forming systems at $z$ = 2--4 \citep{Reddy:2010aa, Heinis:2013aa, To:2014aa, Koprowski:2016ab, McLure:2018aa}. However, several studies have shown that some populations depart from the canonical IRX--$\beta_{\rm UV}$ relation. At low redshifts, LIRGs and ULIRGs are offset from the nominal IRX--$\beta_{\rm UV}$ relation, with larger IRX associated with bluer $\beta_{\rm UV}$ \citep{Goldader:2002aa, Howell:2010aa}. A similar trend is also observed in high-redshift DSFGs, which have bluer $\beta_{\rm UV}$ at a given IRX \citep{Oteo:2013aa, Casey:2014aa, Bourne:2017aa, Chen:2017aa}. On the other hand, at high redshift, rest optical-selected galaxies at $z \simeq 2$ \citep{Reddy:2018aa}, $z \gtrsim 5$ Lyman break galaxies \citep{Capak:2015aa}, and $z \simeq 7.5$ Lyman Break Galaxies \citep{Watson:2015aa} are observed to exhibit redder $\beta_{\rm UV}$ at given IRX values, which are more consistent with the SMC attenuation curve.

Many efforts have been made to explain this discrepancy, but the interpretation is still unclear. Geometrical effects have been proposed to explain the deviations between local ultraviolet-selected samples and infrared-luminous star-forming systems on the IRX--$\beta_{\rm UV}$ plane \citep{Goldader:2002aa, Chapman:2004aa, da-Cunha:2015aa, Narayanan:2018ab}. Furthermore, a prominent population of younger O and B stars with patchy dust geometry has been suggested to move galaxies above the canonical relation \citep{Casey:2014aa}. The intrinsic dust composition and interstellar medium properties will also impact the IRX--$\beta_{\rm UV}$ relation \citep{Safarzadeh:2017aa}. Differences in star formation history may also play some role. For instance, older or less massive stars contributing to the ultraviolet emission of galaxies tend to also drive galaxies below the nominal relation \citep{Kong:2004aa}. Some recent studies with galaxy formation simulations support all of these ideas \citep{Popping:2017aa, Narayanan:2018aa}. Also, several recent works suggest that a single dust attenuation law is incapable of explaining all galaxy populations on the IRX--$\beta_{\rm UV}$ plane \citep{Forrest:2016aa, Salmon:2016aa, Lo-Faro:2017aa, Corre:2018aa}.

It is clear that our sample does not follow a specific IRX--$\beta_{\rm UV}$ relationship (Figure \ref{fig:IRX_vs_UVSlope}). For comparison, in Figure~\ref{fig:IRX_vs_UVSlope}, we plot the attenuation curves for the SMC \citep{Gordon:2003aa}, local starburst galaxies \citep{Calzetti:2000aa}, and nearby ($z < 0.085$) DSFGs \citep{Casey:2014aa}. The majority of our sources are on or above the local DSFG relation and span a wide range of IRX values. This finding is consistent with earlier works on both local and $z \simeq 2$ dusty galaxies \citep{Howell:2010aa, Casey:2014aa} that the dust geometry, as well as dust mass and metallicity, could be contributing factors. A young, metal-poor galaxy like the SMC is thought to be less dusty and consequently fainter in infrared emission than starburst galaxies. Interestingly, most of our galaxies lie above the SMC relation, which is believed to be the limit for normal star-forming galaxies \citep{Boissier:2007aa, Buat:2010aa, Overzier:2011aa, Boquien:2012aa}. 

We find a weak trend that a galaxy with a higher $\Delta MS$ (color code in Figure~\ref{fig:IRX_vs_UVSlope}) tends to have a bluer $\beta_{\rm UV}$ compared to the \cite{Calzetti:2000aa} relation (with a Pearson correlation coefficient of $r = 0.19$). This finding appears to be consistent with the result in \citet{Kong:2004aa} that galaxies with more recent star formation (a higher proportion of young stars) will be intrinsically bluer for a fixed dust attenuation. However, the weak trend is only observed on sources at $z<1$ (top left panel of Figure~\ref{fig:IRX_vs_UVSlope}), in which the Pearson correlation coefficient is $r = 0.55$. In contrast, we obtain almost no correlation ($r = -0.06$) between $\Delta MS$ and the deviation from the nominal IRX--$\beta_{\rm UV}$ relation if we restrict ourselves to the galaxies at $1<z<5$. We attribute this to our small sample size, the large uncertainty in $\beta_{\rm UV}$, and/or the large uncertainty in stellar mass caused by the uncertain dust attenuation in the SED fitting. The large scatter of our sample on the IRX--$\beta_{\rm UV}$ plane may also partially explain the discrepancy between the \texttt{LE PHARE}-based SFR and the direct SFR measurements ($\rm SFR_{\rm UV}$ + $\rm SFR_{\rm IR}$), since the relationship between IRX and $\beta_{\rm UV}$ has been widely used as a calibration tool in SED fitting to infer dust obscuration, and thus SFR estimates. We expect this to be improved by future high-resolution imaging, which can enable spatially resolved SED fitting.

\section{Infrared LF} \label{sec:IR_LFs}

The infrared LF is an important measurement that can be directly related to the underlying obscured star formation. The evolution of the infrared LF can also provide strong constraints on the history of star formation in the Universe and on galaxy formation models. The LF, denoted by $\Phi (L)$ (in units of Mpc$^{-3}$ dex$^{-1}$), is defined to be the number of galaxies per unit luminosity per unit volume. Two estimations of LF are often adopted, and we adopt both in our studies.

%%%%%%%%%%%%%%%%%%%%%%%%%%%%%%%%%%%%%%%%%%%%%%%%%%%%%%%%%%%%%%%%%%%%%%
\begin{figure*}
\centering
%%% IR Luminosity Function %%%
\includegraphics[width=0.9\paperwidth]{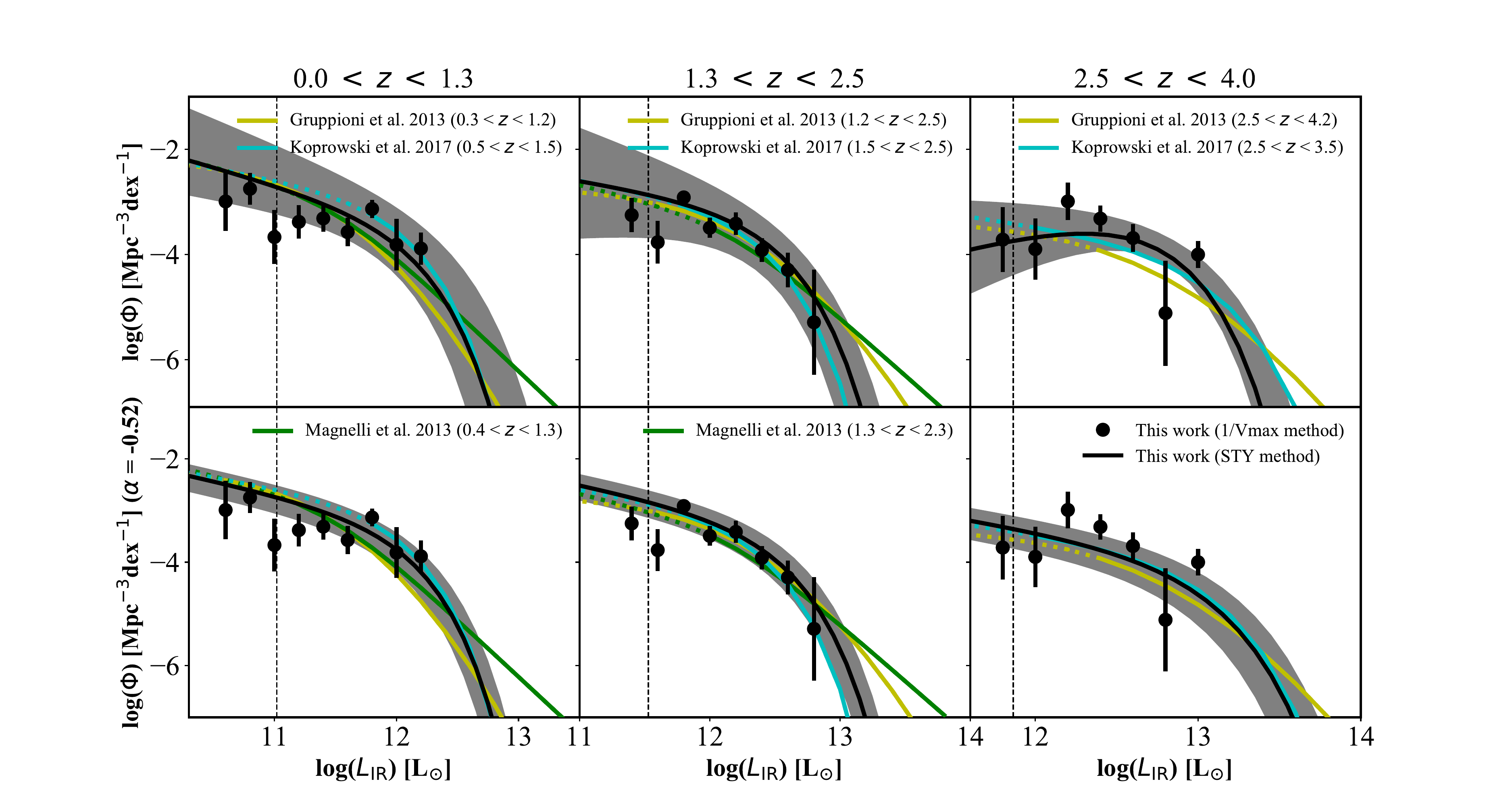}
\caption{ Upper panels: infrared LFs estimated using our $1/V_{\rm max}$ method (black points) and likelihood method (black curves) in three redshift bins. The uncertainties of the $1/V_{\rm max}$ LFs are calculated using the Poisson errors, while the errors of the likelihood LFs (dark shaded regions) are calculated as $|\Delta \chi^{2}| \leqslant 2.3 ( \pm 1\sigma )$. Lower panels: same as the upper panels, except that the likelihood LFs are estimated with a fixed $\alpha$ of $-0.5\pm0.7$ and their corresponding errors are calculated as $|\Delta \chi^{2}| \leqslant 1.0 ( \pm 1\sigma )$. We also present the LFs from \emph{Herschel}/PACS \citep{Magnelli:2013aa}, PEP-HerMES/\emph{Herschel} \citep{Gruppioni:2013aa} and the JCMT 850-$\micron$-selected sample \citep{Koprowski:2017aa}. The dashed curves represent published LFs that are extrapolated beyond their detection limits. The black dashed vertical lines show the median detection limit of our $L_{\rm IR}$ values for the corresponding redshift bins.} 
\label{fig:IRLuminosityFunction}
\end{figure*}

\begin{figure}
\centering
%%% IR Luminosity Function evolution %%%
\includegraphics[width=\columnwidth]{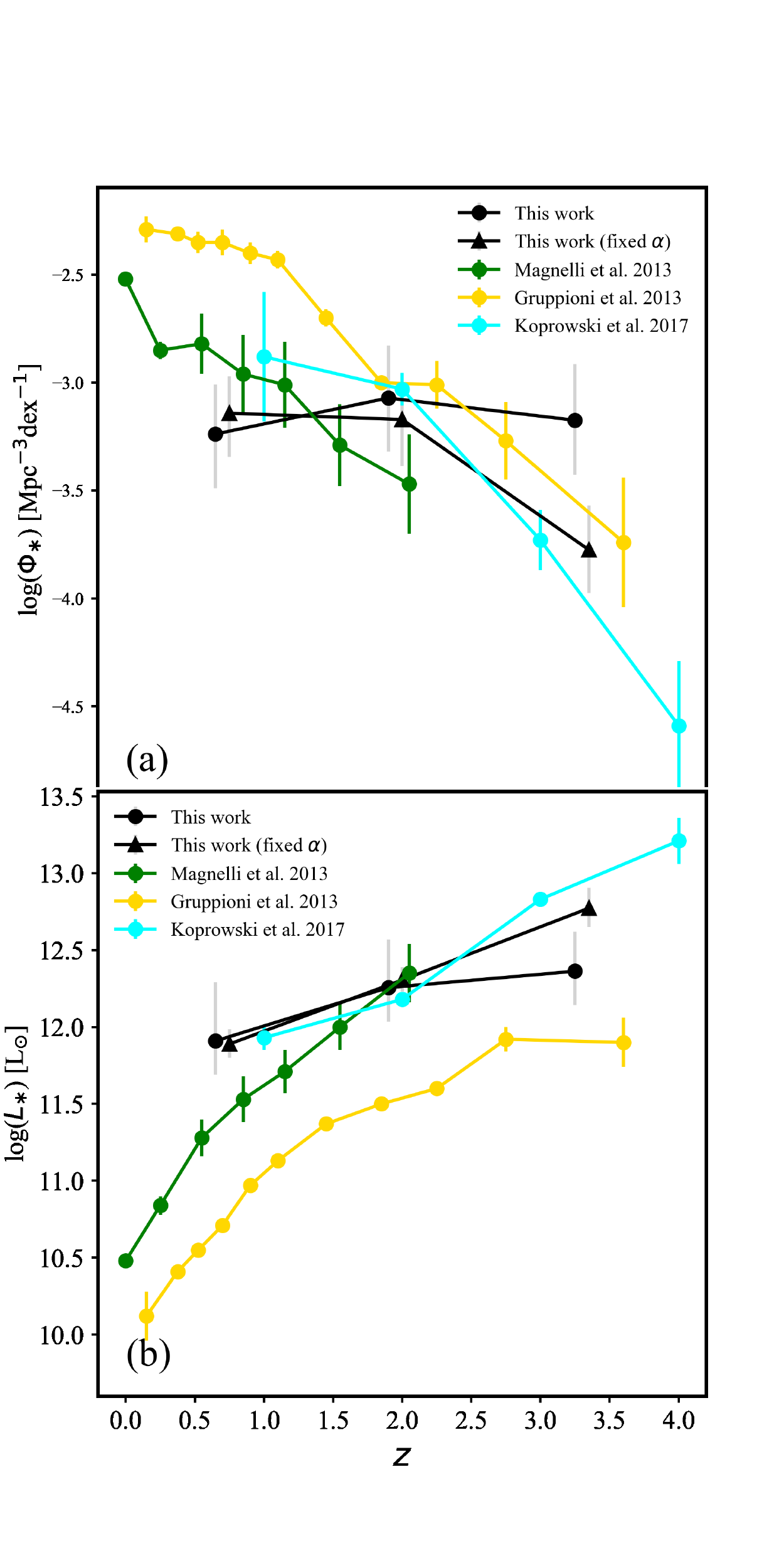}
\caption{Evolution of estimated (a) $\Phi_{\ast}$ and (b) $L_{\ast}$ values from the likelihood LFs. The evolution of $\Phi_{\ast}$ is more model-dependent and could be disrupted by the different assumptions on the shape of the LF or the adopted $\alpha$. Once the fits of the likelihood LFs are forced to have the same faint-end slope, $\Phi_{\ast}$ decreases with increasing redshift, and the trend is consistent with \emph{Herschel}-based \citep{Gruppioni:2013aa, Magnelli:2013aa} and SCUBA-2 850-$\micron$ \citep{Koprowski:2017aa} observations. The characteristic luminosity $L_{\ast}$ increases in cases of both free and fixed $\alpha$. The trend suggests that our observations are consistent with the ``cosmic downsizing'' scenario.}
\label{fig:IRLuminosityFunction_evolution}
\end{figure}

\subsection{1/V$_{\rm max}$ Method} \label{subsec:Vmax_Lfs}
The standard $1/V_{\rm max}$ method is a popular estimator for determining the LF and as a probe of evolution \citep{Schmidt:1968aa}. This method allows us to determine the LF directly from the data without any assumptions on the LF shape. The LF in a given luminosity and redshift bin is estimated as
\begin{equation}
\Phi (L, z) = \frac{1}{\Delta L} \sum\limits_{i}^{N} \frac{1 - s_{i}}{c_{i} \times V_{{\rm max},i}},
\end{equation}
where $\Delta L$ is the width of the luminosity bin, $s_{i}$ is the spurious fraction, and $c_{i}$ is the completeness at the flux level of the $i$-th galaxy. Also, $V_{{\rm max},i}$ is the maximum comoving volume over which the $i$-th galaxy can be detected,
\begin{equation}
V_{{\rm max},i} = \frac{\Omega_{i}}{4\pi} V_{z_{\rm max},i}, 
\end{equation}
where $\Omega_{i}$ is the effective solid angle of the survey and $V_{z_{\rm max},i}$ is the comoving volume at maximum redshift ($z_{\rm max}$) for which the $i$-th galaxy can be detected. By shifting the best-fit \texttt{LE PHARE} SED (\S\ref{subsec:InfraredLuminosities}) to a greater distance and comparing its observed 450-$\micron$ flux density with the survey detection limit, we can determine $V_{z_{\rm max},i}$ for each galaxy. Since the sensitivity of our map is inhomogeneous, each galaxy has its own corresponding survey solid angle $\Omega_{i}$. We therefore calculate the map area over which the $i$-th galaxy can be detected at 4$\sigma$. The results are presented as black circles in Figure \ref{fig:IRLuminosityFunction} and also summarized in Table \ref{tbl:LF(Vmax)}.  The errors on the LFs are calculated assuming Poissonian statistics without including the uncertainties from photometric redshift and SED template degeneracy. It is worth noting that there is a bias caused by the fact that we can only consider galaxies having optical redshift estimates. Nineteen sources in our sample have $z_{\rm FIR}$ estimates (\S\ref{subsubsec:MissingSources}), and they do not have reliable L$_{\rm IR}$ determinations. If we include these 19 $z_{\rm FIR}$ estimated sources, their effects are negligible in the LFs at $z<1.3$, while the LFs in the redshift bins of $1.3 < z < 2.5$ and $2.5 < z < 4.0$ will be enhanced, on average, by $0.12\pm0.06$ and $0.21\pm0.08$\,dex, respectively. 

%%%%%%%%%%%%%%%%%%%%%%%%%%%%%%%%%%%%%%%%%%%%%%%%%%%%%%%%%%%%%%%%%%%%%%

\subsection{Likelihood Method} \label{subsec:ML_Lfs}
We also adopt a parametric likelihood estimator in the form proposed by \citet*[hereafter STY]{Sandage:1979aa}, to model the infrared LF. This parametric technique assumes an analytical form for the LF and therefore does not require the binning of the data. The STY estimator can be constructed as follows. 

The probability density for a galaxy of luminosity $L_{i}$ to be detected at redshift $z_{i}$ in a luminosity-limited redshift survey is estimated as
\begin{equation}
p_{i} \propto \frac{\Phi (L_{i})}{\int_{L_{\min(z_{i})}}^{\infty} \Phi (L) dL}.
\label{eq:STYEstimator}
\end{equation}
The corresponding likelihood estimator is:
\begin{equation}
\mathcal{L} = \prod_{i=1}^{N} [p_{i}]^{\frac{1-s_{i}}{c_{i}}},
\end{equation}
where $c_{i}$ is the completeness, $s_{i}$ is the spurious fraction, and $L_{\min(z_{i})}$ is the minimum $L_{\rm IR}$ observable at redshift $z_{i}$ in a flux-limited sample. The product is made over $N$ galaxies of the sample in the redshift bin. The estimation of the minimum $L_{\rm IR}$ for our entire sample is difficult to determine, since no SED template can well represent all $L_{\rm IR}$ values over a large redshift range. An assumed SED template, which is represented by a single temperature, may lead to a potentially biased result because of the degeneracy between $T_{\rm d}$ and redshift. To remedy this, we adopt a similar procedure to that in \cite{Zavala:2018aa}. We adopt the evolution of $T_{\rm d}$ with redshift (black line in Figure \ref{fig:Tdust_z_deltaMS}a) and the assumption of a modified blackbody SED (Equation \ref{eq:ModifiedBB}) with a median emissivity index $\beta = 1.80$ to reproduce the luminosity detection limit as a function of redshift for all of our data (solid curve in Figure \ref{fig:LIR_vs_redshift}). We then take the interpolated value at a given $z_{i}$ from this function to calculate the $L_{\min(z_{i})}$ for the $i$-th galaxies. We note that the redshift evolution in $T_{\rm d}$ for our sample is consistent with being driven mainly by the selection effects (\S \ref{subsec:Td-LIR}). Therefore, our estimated $L_{\rm IR}$ limits take into account the selection bias and its effects on the averaged SED. We verify that a change in $T_{\rm d}$ of $^{+12}_{-8}$\,K (16th-to-84th percentile range for our sample) only leads to an uncertainty of $L_{\rm IR}$ limits by $^{+0.24}_{-0.18}$\,dex, on average, from $z=0$ to 5. We then minimize $-2ln(\mathcal{L})$, which can be taken as following the $\chi^{2}$ distribution for large-$N$ statistics (\citealt{Pearson:1900aa}; see also the review in \citealt{Cochran:1952aa}), by using the ``\texttt{minimize}'' algorithm in the \texttt{scipy} package.

We assume the classical LF form \citep{Schechter:1976aa},
\begin{equation}
\Phi (L) = \Phi_{\ast} \left(\frac{L}{L_{\ast}}\right)^{\alpha} \exp\left(\frac{- L}{L_{\ast}}\right),
\label{eq:SchechterFunction}
\end{equation}
where the parameters are the normalization $\Phi_{\ast}$, the characteristic luminosity $L_{\ast}$, and the faint-end slope $\alpha$. A further consideration is that the $\Phi_{\ast}$ value will be canceled in the STY estimation (Equation \ref{eq:STYEstimator}) and consequently has to be estimated independently. Here $\Phi_{\ast}$ can be recovered by integrating the obtained likelihood LF over the luminosity range of the survey and then equating it to the mean number density $\bar{n}$ of the observed galaxy sample,
\begin{equation}
\bar{n} = \Phi_{\ast} \int_{L_{\rm min}}^{\infty} \Phi (L) dL, 
\label{eq:MeanNumber1}
\end{equation}
where $L_{\rm min}$ is a minimum luminosity. In practice, we can ignore this $L_{\rm min}$, since the integration of this equation will cancel out in the following procedure.

The mean number density of galaxies at redshift $z$ also can be represented by
\begin{equation}
\bar{n} = \frac{n(z)}{\Omega(z) V(z) S(z)},
\label{eq:MeanNumber2}
\end{equation}
where $n(z)$ is the number of observed galaxies, $V(z)$ is the volume, and $\Omega (z)$ is the mean solid angle of our sample at redshift $z$. The quantity $S(z)$ is the selection function of the survey, given by
\begin{equation}
S(z) = \frac{\int_{L_{\min(z)}}^{\infty} \Phi (L) dL} {\int_{L_{min}}^{\infty} \Phi (L) dL},
\end{equation}
where $L_{\min(z)}$ is the minimum $L_{\rm IR}$ observable at redshift $z$ in a flux-limited sample. By combining Equations \ref{eq:MeanNumber1} and \ref{eq:MeanNumber2}, we can derive the normalization $\Phi_{\ast}$.

The results of the likelihood method are presented as black curves in Figure~\ref{fig:IRLuminosityFunction} and summarized in Table \ref{tbl:LF(likelihood)}. The constraint in the faint-end slope may be weak at our highest redshift bin ($2.5<z<4.0$), since our observations start to lose sensitivity for the $L_{\rm IR} < 10^{12}\,\rm L_{\sun}$ population at such a high redshift. Therefore, we fix $\alpha$ to $-0.5\pm0.7$, which is the $z<2.5$ average; refit the likelihood LFs to all three redshift bins; and show the results in the lower panels of Figure~\ref{fig:IRLuminosityFunction}. The errors on the likelihood LFs (dark shaded areas in Figure~\ref{fig:IRLuminosityFunction}) are calculated by taking $|\Delta \chi^{2}| \leqslant$ 2.3 (or $\leqslant$ 1.0), where the number 2.3 (or 1.0) corresponds to $\pm1\sigma$ \citep{Avni:1976aa} for two (or one, in the case of fixed $\alpha$) degrees of freedom.

%%%%%%%%%%%%%%%%%%%%%%%%%%%%%%%%%%%%%%%%%%%%%%%%%%%%%%%%%%%%%%%%%%%%%%
\subsection{Comparison with other observations}

It is interesting to compare our infrared LFs with previous studies. We plot the LFs from \emph{Herschel}/PACS \citep{Magnelli:2013aa}, \emph{Herschel} PEP-HerMES \citep{Gruppioni:2013aa}, and JCMT 850-$\micron$ \citep{Koprowski:2017aa} samples in Figure~\ref{fig:IRLuminosityFunction}. To adapt their results to our redshift bins, we simply take the mean value of $\Phi$ that is within or nearest to our redshift bins from the published estimations. The work of \cite{Magnelli:2013aa} is based on \emph{Herschel}/PACS. Their data do not extend to wavelengths longer than 160\,$\micron$ and their LF estimations are only for $z\lesssim 2.3$. For the work of \cite{Koprowski:2017aa}, we converted their rest-frame 250-$\micron$ luminosity into the total $L_{\rm IR}$ by using the averaged ALESS 870-$\micron$ SEDs \citep{da-Cunha:2015aa} for the same redshift bins as their LF estimates at $z < 1.5$, $1.5 < z < 2.5$, $2.5 < z < 3.5$, and $z > 3.5$. Our LFs are statistically consistent with all of the previous estimates within the uncertainties.

For the \emph{Herschel}-based LFs, \citet{Magnelli:2013aa} adopted a fixed power-law slope of $-0.60$, whereas \citet{Gruppioni:2013aa} used a flatter fixed slope of $-0.20$. In the work of \cite{Koprowski:2017aa}, they adopted a fixed faint-end slope of $-0.40$, based on the result from ALMA-1.3-mm selected SMGs at $1.5 < z < 2.5$ \citep{Dunlop:2017aa}. Our best-fit faint-end slope from the likelihood method at $1.3 < z < 2.5$ ($\alpha$ = $-0.4\pm0.5$) is more consistent with the ALMA result. In contrast, our measured faint-end slope at $2.5 < z < 4.0$ ($\alpha$ = $0.9^{+1.0}_{-1.2}$) is flatter than the assumed faint-end slope from the literature ($\alpha$ = $-0.20$, $-0.40$, and $-0.60$) or even our low-redshift measurements ($\alpha$ = $-0.6\pm0.4$ and $-0.4\pm0.5$). This may reflect the fact that our likelihood method, which takes the faint-end slope as a free parameter in determination, is less well-constrained at low luminosities where the numerous faint sources may lie beyond our current detection limit at high redshift. The future STUDIES survey with increased sensitivity will detect the fainter population and improve the faint-end slope estimations.

We compare the characteristic parameters of the various likelihood LFs at various redshifts in Figure \ref{fig:IRLuminosityFunction_evolution}. Our $\Phi_{\ast}$ estimated by the LF fit with $\alpha$ kept as a free parameter shows no evolution with redshift (Figure~\ref{fig:IRLuminosityFunction_evolution}a). The $\Phi_{\ast}$ estimation is more model-dependent and could be disrupted by the different assumptions on the shape of the LF or the adopted $\alpha$ (see also \citealt{Casey:2014ab}). Once the fits of likelihood LFs are forced to have the same faint-end slope, $\Phi_{\ast}$ decreases with increasing redshift, and the trend is consistent with previous studies based on \emph{Herschel} \citep{Gruppioni:2013aa, Magnelli:2013aa} and SCUBA-2 850-$\micron$ observations \citep{Koprowski:2017aa}. On the other hand, we find that as we increase the redshift, the characteristic luminosity $L_{\ast}$ increases in both cases of free or fixed $\alpha$ (Figure~\ref{fig:IRLuminosityFunction_evolution}b). The increase in $L_{\ast}$ with redshift suggests that our observations are consistent with the ``cosmic downsizing'' scenario (e.g., \citealt{Cowie:1996aa}), in which the contribution of luminous sources dominates in the early Universe, whereas the growth of the less luminous ones continues at lower redshifts. 

%%%%%%%%%%%%%%%%%%%%%%%%%%%%%%%%%%%%%%%%%%%%%%%%%%%%%%%%%%%%%%%%%%%%%%
\subsection{Comparison with models}

\begin{figure*}
\centering
%%% IR Luminosity Function compare with theoretical %%%
\includegraphics[width=0.9\paperwidth]{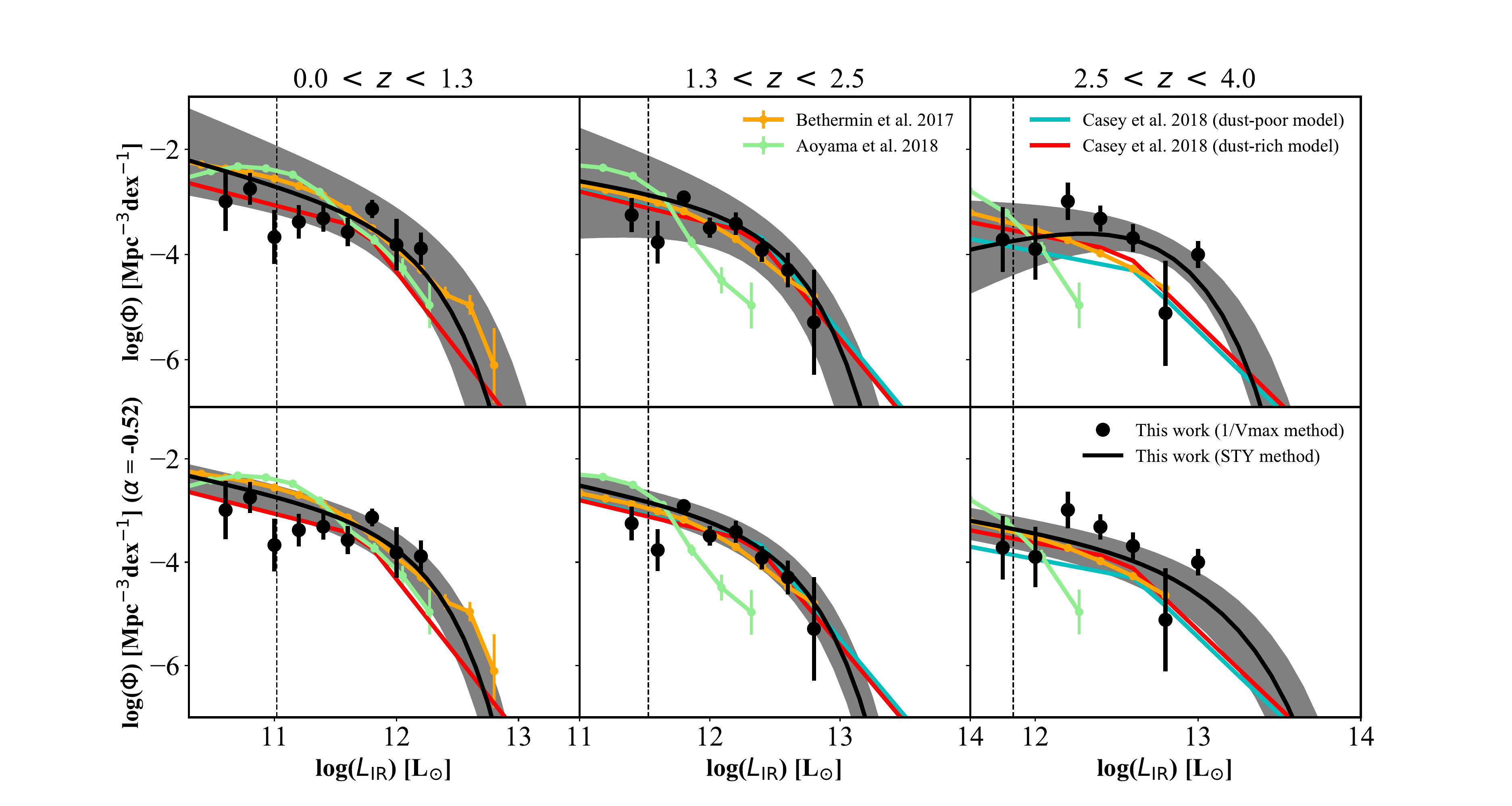}
\caption{ Same as Figure \ref{fig:IRLuminosityFunction}, except that we compare our LFs with results from theoretical studies. The LFs from \cite{Bethermin:2017aa} and \cite{Aoyama:2019aa} are shown in orange and light green, respectively. The results of the dust-poor model are shown in cyan and the dust-rich model in red \cite{Casey:2018aa}. }
\label{fig:IRLuminosityFunction_theoretical}
\end{figure*}

We compare our LFs with theoretical studies in Figure~\ref{fig:IRLuminosityFunction_theoretical}. In the work of \cite{Bethermin:2017aa}, they built a 2\,deg$^{2}$ dark matter simulation, called the Simulated Infrared Dusty Extragalactic Sky, for FIR-to-millimeter wavelengths. This is a phenomenological simulation based on the latest observational constraints on the stellar mass function, the main sequence of star-forming galaxies, and the evolution of SEDs. The authors claimed to reproduce the number counts from the FIR to the millimeter, the measured redshift distributions, and the evolution of the obscured SFR density (SFRD). Their work also described the disagreement between the number counts from single-dish instruments and interferometers. When the $\alpha$ is free to vary in the likelihood LF fitting, the only slight discrepancy between our results and the \cite{Bethermin:2017aa} model is at $z> 2.5$, where the model is, on average, $0.3^{+0.2}_{-0.1}$\,dex below our observations. Once the $\alpha$ is fixed in the LF fitting, our LFs are in broad agreement with their predicted LFs within the uncertainties. 

\citet{Aoyama:2019aa} performed a cosmological hydrodynamic simulation with dust evolution based on the GADGET-3 code (originally described in \citealt{Springel:2005aa}) to predict the cosmic dust abundances at various redshifts. In their simulation, they considered the distribution of dust grain size to be represented by two populations: large (0.1\,$\micron$ in radii) and small ($5\times10^{-3}$\,$\micron$ in radii) grains. Their simulation treats the enrichment of dust self-consistently with star formation and stellar feedback. Dust is generated by supernovae and AGB stars and can grow by accretion. Dust can also be destroyed by supernova shocks, coagulated in the dense ISM, or shattered in the diffuse ISM. Here we compare our results with their high spatial resolution simulation with $2\times512^{3}$ particles in a box size of 50\,$h^{-1}$\,Mpc. As shown in Figure~\ref{fig:IRLuminosityFunction_theoretical}, at $z<1.3$, their predicted LF is consistent with our measurements within the uncertainties. However, at $1.3 < z < 2.5$, their simulation significantly underpredicts the LFs by $1.1^{+0.4}_{-0.3}$\,dex in the free-$\alpha$ case and $1.1\pm0.2$\,dex in the fixed-$\alpha$ case at $L_{\rm IR} > 10^{12}\,\rm L_{\sun}$, on average. Their simulation does not even have sufficient data points in the bright-end at redshift bins of $2.5<z<4.0$ compared to our results. The authors attributed this to the lack of certain heating sources (e.g., AGN feedback or a top-heavy IMF) in their simulation, and it was partially due to the insufficient spatial resolution of their model. 

\cite{Casey:2018aa} explored two extreme evolution models: dust-poor and dust-rich. The DSFGs contribute negligibly ($<10\%$) in the early Universe ($z > 4$) in the dust-poor model, while DSFGs dominate ($>90\%$) the star formation in the early Universe in the dust-rich model. These models are based on the existing measurements of the infrared LFs and the existing empirical constraints on the dust SED characteristics of infrared-luminous galaxies (i.e., emissivity spectral index and mid-infrared power-law index) as a function of $L_{\rm IR}$ and $z$. Their simulation generated 1\,deg$^{2}$ synthesis maps with 0${\farcs}$5 pixel scale from 70\,$\micron$ to 2\,mm by injecting sources with densities determined from the projection of LFs and flux densities from inferred SEDs. They provide predictions of number counts from 70$\,\micron$ to 2\,mm, redshift distributions, and evolving galaxy LFs at both ultraviolet and infrared wavelengths. The predictions of LFs from their two models are consistent with our measurements at $z$=0--2.5. However, at $z > 2.5$, comparing to our LF fit with $\alpha$ as a free parameter, their dust-poor model underpredicts the infrared LF by $0.5\pm0.2$\,dex, while their dust-rich model underpredicts the LF by $(0.3\pm0.2)$\,dex, on average. The discrepancies still exist compared to our LF results with $\alpha$ fixed. Their dust-poor model underpredicts the LF by $0.6\pm0.2$\,dex, while their dust-rich model underpredicts the LF by $0.4\pm0.2$\,dex, on average. This may be simply due to the lack of data at higher redshifts in their models. The existing measurements of the infrared LFs in their work are mainly at $z<2$ and do not tightly constrain the shape of infrared LFs at high redshift.

In summary, the simulated LFs from models and observations appear to diverge for high-redshift bins ($z > 2.5$; $z > 1.3$ in the case of \citealt{Aoyama:2019aa}). These results may highlight the complexity for the models to interpret the high-redshift FIR observations and/or the difficulty for the observations to well constrain the LFs at high redshift. Nevertheless, our results seem to suggest that the models require some ingredient that produces more infrared-emitting galaxies at high redshift.

\section{The obscured star-formation history}\label{sec:SFRD}

\begin{figure*}
\centering
%%% SFRD %%%
\includegraphics[width=0.7\paperwidth]{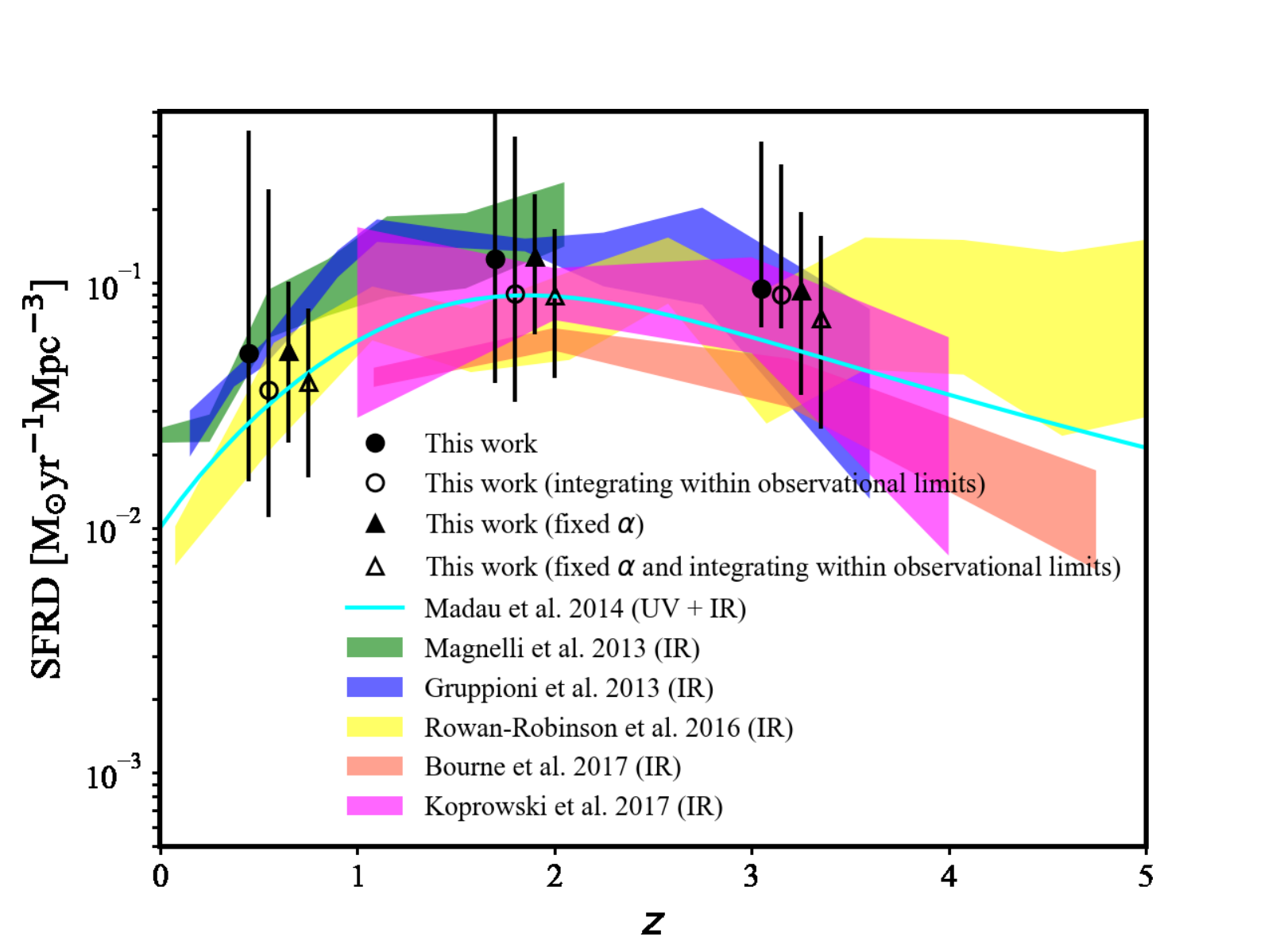}
\caption{ SFRD versus redshift. Our estimations are shown as black filled (integrated from $0.03\,L_{\ast}$ to $10^{13.5}\rm L_{\sun}$) and open (integrated from the minimum observational limits to $10^{13.5}\rm L_{\sun}$) symbols. The circles and triangles show the results from the cases of free and fixed $\alpha$, respectively. The horizontal displacements between the symbols are artificial, to avoid confusion. For comparison, we also show the \emph{Herschel}-based studies \citep{Gruppioni:2013aa, Magnelli:2013aa, Rowan-Robinson:2016aa}, as well as SCUBA-2-based studies \citep{Bourne:2017aa, Koprowski:2017aa}. The width of the shaded areas represents the range of $\pm1\sigma$ scatter of the corresponding data sets. The solid cyan curve shows the best-fit evolution function of SFRD from \citet{Madau:2014aa} using rest-frame far-ultraviolet or mid-to-FIR data from a variety of galaxy surveys. }
\label{fig:SFRD}
\end{figure*}

\begin{figure}
\centering
%%% SFRD %%%
\includegraphics[width=\columnwidth]{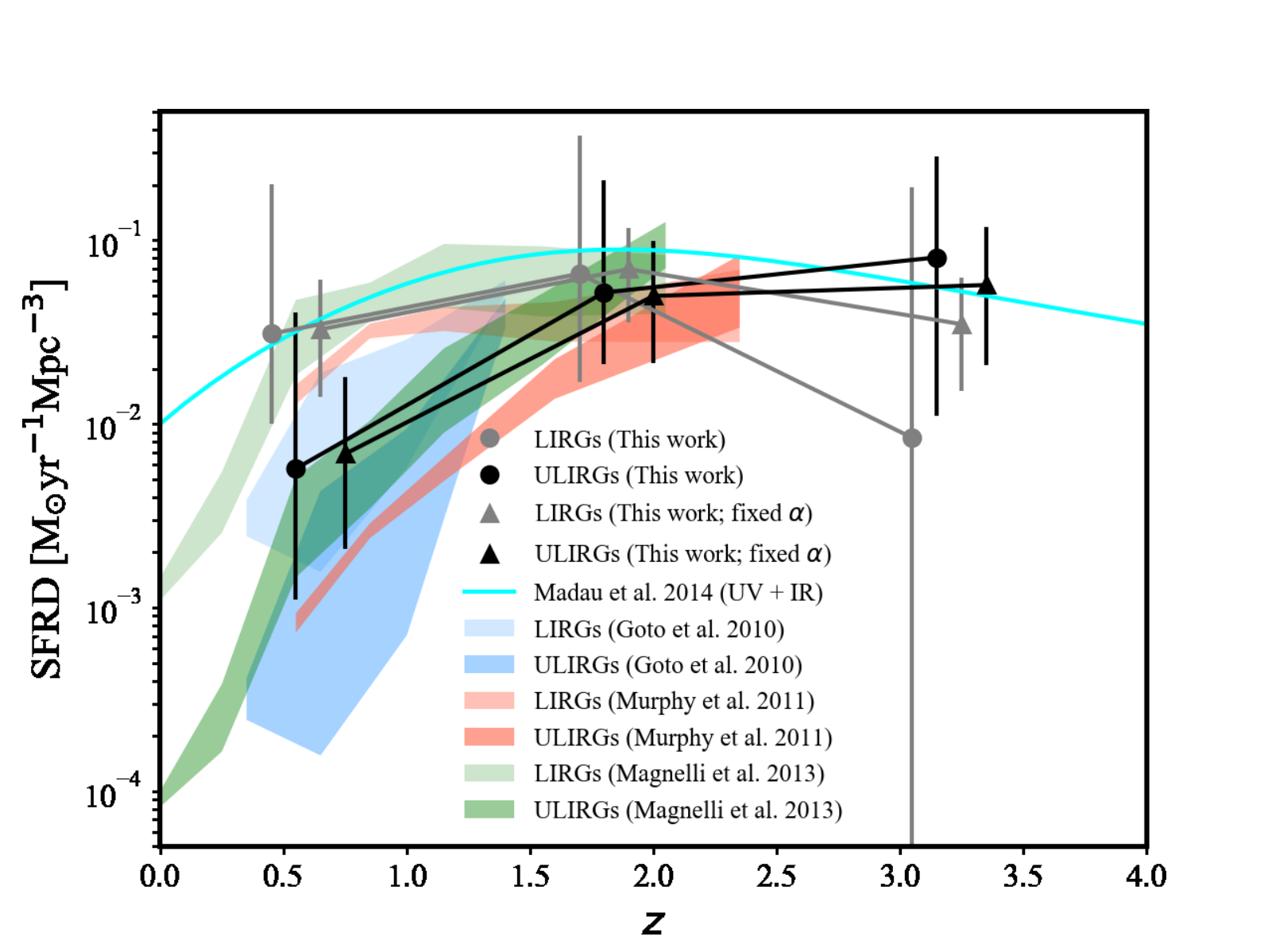}
\caption{SFRD versus redshift, breaking down by LIRG and ULIRG contributions. Our estimations are shown as gray points (LIRGs) and black points (ULIRGs). The horizontal displacement between the two measurements is artificial, to avoid confusion. The results from \citealt{Magnelli:2013aa}, \citealt{Goto:2010aa}, and \citealt{Murphy:2011ab} are shown as green, blue, and red shaded regions (1$\sigma$ scatter), respectively. The cyan curve shows the total SFRD from \citet{Madau:2014aa}.}
\label{fig:SFRD_bins}
\end{figure}

The constructed infrared LFs allow us to determine the redshift evolution of the obscured SFRD. By integrating our infrared LFs produced with the likelihood method, we obtained the infrared comoving luminosity densities. The integrated infrared comoving luminosity densities can then be converted to SFRDs using Equation~\ref{eq:SFR_IR}. The uncertainties of our SFRDs were derived by integrating the 1$\sigma$ upper- and lower-bounds on the likelihood LFs. The results are presented in Figure~\ref{fig:SFRD} and summarized in Table~\ref{tbl:SFRD}. In the figure, filled symbols show the results derived using integration limits of $L_{\rm min}= 0.03\,L_{\ast}$ to $L_{\rm max}=10^{13.5}\rm L_{\rm \sun}$, while open symbols show those derived by integrating from our minimum observational limits (vertical dashed lines in Figure~\ref{fig:IRLuminosityFunction}) to $L_{\rm max}=10^{13.5}\rm L_{\rm \sun}$. There are no significant differences between these two. We verify that the uncertainties of SFRDs will be lower, on average, by a factor of $\simeq$ 3.5 if we assume a fixed $\alpha$ in the LF estimations, which is the case in most other studies in the literature.

In Figure \ref{fig:SFRD}, we also show the SFRDs from the combined optical and infrared analyses in \citet[which is supposed to represent the total SFRD in the Universe]{Madau:2014aa} and various \emph{Herschel} \citep{Gruppioni:2013aa, Magnelli:2013aa} and SCUBA-2 studies \citep{Bourne:2017aa, Koprowski:2017aa}. \cite{Madau:2014aa} obtained measurements of the cosmic SFRDs based on rest-frame far-ultraviolet or mid-/FIR data from a variety of galaxy surveys (mostly post-2006). The surveys used in their work provided best-fit LF parameters, which allowed them to integrate the LF down to the same relative limiting luminosity. In the \emph{Herschel}-based studies, the SFRDs were derived by adopting fixed faint-end slopes of $\alpha=-0.20$ in \citet{Gruppioni:2013aa} and $\alpha=-0.60$ in \citet{Magnelli:2013aa} or fitting the infrared LF with the Saunders functional form \citep{Saunders:1990aa} in \cite{Rowan-Robinson:2016aa}. We reproduce the SFRDs of \cite{Koprowski:2017aa} by integrating the total infrared LFs that are calculated from their rest-frame 250-$\micron$ LFs and averaged ALESS SEDs \citep{da-Cunha:2015aa} for the same redshift bins of their LFs at $z < 1.5$, $1.5 < z < 2.5$, $2.5 < z < 3.5$, and $z > 3.5$. The SFRDs from \citet{Magnelli:2013aa}, \citet{Gruppioni:2013aa}, and us (filled symbols in Figure~\ref{fig:SFRD}) are higher than the others. This could be caused by the adopted limits of integration assumed in the different studies. Overall, all these measurements (including ours) are in broad agreement with each other between redshifts of 1 and 4, although this is partially due to the large uncertainties in all surveys. In summary, the majority of the SFRD is obscured over redshifts up to $z\sim$ 4. Some of these results show a potential SFRD peak at $z\sim1$--2. Our measurements at this moment do not yet have sufficient precision to confirm this peak, but again, we expect this to improve when STUDIES is complete. 

We also present the evolution of the SFRDs, breaking them down into LIRG and ULIRG contributions in Figure \ref{fig:SFRD_bins} and Table~\ref{tbl:SFRD}. Our measurements at $z=1$--2 are consistent with the estimates of the \emph{Akari} mid-infrared selected sample \citep{Goto:2010aa}, \emph{Spitzer} 24-$\micron$-selected galaxies \citep{Murphy:2011ab}, and \emph{Herschel}-selected sources \citep{Magnelli:2013aa}. By combining the measurements from the low-redshift sample, we find that the contribution of the ULIRG population to the SFRD rises dramatically from $z\simeq$ 0 to 2, $\propto (1+z)^{3.9\pm1.1}$ ($\propto (1+z)^{3.5\pm0.4}$ in the case of fixed $\alpha$ in the LF fit), and plays a dominant role at $z \gtrsim 2$. Our observations confirm the importance of luminous obscured star formation in the early Universe up to $z\sim$ 3.

\section{Summary}\label{sec:Summary}

By combining the SCUBA-2 data from the ongoing JCMT Large Program STUDIES and the archive in the COSMOS-CANDELS region, we have obtained the deepest to date 450-$\micron$ blank-field image, which has a 1$\sigma$ noise level of 0.65\,mJy in the deepest area. We detected 256 450-$\micron$ sources at S/N $> 4.0$ in an area of 300 arcmin$^{2}$, 192 of which have optical counterparts and abundant multi-wavelength photometric and spectroscopic data. Our main findings are the following.

\begin{itemize}

\item The median redshift of our sample with optical redshifts is $z = 1.79^{+0.03}_{-0.15}$ with a 16th-to-84th percentile range of 1.7--1.9. Their redshifts range from $z=0.12$ to 4.76, with the majority at $z \lesssim 3$. The median redshift will increase to $z = 1.9\pm0.1$ if we remove the suspected AGNs and assume that sources without reliable identifications in the optical are at $z=3$.

\item We investigated the relation between the total SFR and stellar mass. We conclude that our data start to probe into the normal star-forming population out to $z \simeq 3$. Around $35^{+32}_{-25}$\% of our sources with a lower limit of $24^{+22}_{-17}$\% are classified as starburst galaxies, while the rest are on the star-formation main sequence.

\item Our galaxies have a median dust temperature of ${T}_{\rm d} = 38.3^{+0.4}_{-0.9}$\,K with a 16th-to-84th percentile range of 30--50\,K and overlap with the ranges previously observed on SMGs out to $z \simeq 3$. After examining the $T_{\rm d}$--$L_{\rm IR}$ relation of our sources and our detection limits, we conclude that our sample is representative for SMGs of $L_{\rm IR} > 10^{12}\,\rm L_{\sun}$ over a wide redshift range, at least up to $z \simeq 3$.

\item We found a moderate correlation between $T_{\rm d}$ and $z$ for our entire sample. However, we obtained almost no correlation between $T_{\rm d}$ and $z$ if we restricted ourselves to sources with $L_{\rm IR} > 10^{12}\,\rm L_{\sun}$ at $z<3$. We suggest that the apparent $T_{\rm d}$--$z$ evolution of our sample and some previous studies may be caused by the selection effect that 450-$\micron$ biases against cooler sources at high redshift.

\item We found a moderate, positive correlation between ${\Delta MS}$ (deviation from the SFR--$M_{\ast}$ relation of the main sequence) and $T_{\rm d}$. Galaxies in our sample with mergers or irregular features also tend to have higher $T_{\rm d}$ at fixed ${\Delta MS}$. These findings are consistent with the simulations of merger-triggered SMGs, where the more compact geometries in star-forming galaxies lead to a sharp increase in $T_{\rm d}$ during the burst. 

\item Our sources span a wide range in IRX ($L_{\rm IR}/L_{\rm UV}$) and do not follow the tight IRX--$\beta_{\rm UV}$ relation that was observed in the local Universe. Almost all of our galaxies lie above the SMC relation that is believed to represent the limit of normal star-forming galaxies. 

\item We conducted direct ($1/V_{\rm max}$) and likelihood estimations of the infrared LFs. Our measurements are consistent with previous studies within the errors. Our sample size and depth at $z<2.5$ allow us to leave the faint-end slope as a free parameter, while at $z>2.5$ our measured faint-end slope is less well-constrained where more faint sources lie beyond our current detection limit. Our faint-end slope at $1.3 < z < 2.5$ ($\alpha = -0.4\pm0.5$) is consistent with recent ALMA-based estimations.

\item Our SFRD measurements are in broad agreement with previous studies. We find that the contribution of the ULIRG population to the SFRD rises rapidly from $z$ = 0 to $z\simeq2$ and remains dominant at $z \gtrsim 2$. Our observations confirm the importance of luminous obscured star formation in the early Universe up to $z\sim$ 3.

\end{itemize}

\acknowledgments

We thank the JCMT/EAO staff for observational support and the data/survey management and the anonymous referee for comments that significantly improved the manuscript. 
C.F.L., W.H.W., and Y.Y.C. acknowledge the grant support from the Ministry of Science and Technology of Taiwan (105-2112-M-001-029-MY3 and 108-2112-M-001-014-). 
The submillimeter observations used in this work include the STUDIES program (program code M16AL006), archival data from the S2CLS program (program code MJLSC01) and the PI program of \citet[][program codes M11BH11A, M12AH11A, and M12BH21A]{Casey:2013aa}. 
The James Clerk Maxwell Telescope is operated by the East Asian Observatory on behalf of the National Astronomical Observatory of Japan; the Academia Sinica Institute of Astronomy and Astrophysics; the Korea Astronomy and Space Science Institute; and the Operation, Maintenance and Upgrading Fund for Astronomical Telescopes and Facility Instruments, budgeted from the Ministry of Finance (MOF) of China and administrated by the Chinese Academy of Sciences (CAS), as well as the National Key R\&D Program of China (No. 2017YFA0402700). Additional funding support is provided by the Science and Technology Facilities Council of the United Kingdom and participating universities in the United Kingdom and Canada. 
I.R.S acknowledges support from STFC (ST/P000541/1). 
L.C.H. acknowledges support from the National Science Foundation of China (11721303 and 11991052) and National Key R\&D Program of China (2016YFA0400702). 
M.J.M acknowledges the support of the National Science Centre, Poland, through the SONATA BIS grant 2018/30/E/ST9/00208. 
Y.G. is partially supported by the National Key Basic Research and Development Program of China (grant No. 2017YFA0402704), NSFC grant Nos. 11861131007 and 11420101002, and the Chinese Academy of Sciences Key Research Program of Frontier Sciences (grant No. QYZDJ-SSW-SLH008). 
M.P.K. acknowledges support from the First TEAM grant of the Foundation for Polish Science No. POIR.04.04.00-00-5D21/18-00. 
X.S. acknowledges support from the NSFC (11573001 and 11822301). 
T.G. acknowledges the support by the Ministry of Science and Technology of Taiwan through grant108-2628-M-007-004-MY3.
H.D. acknowledges financial support from the Spanish Ministry of Science, Innovation and Universities (MICIU) under the 2014 Ramón y Cajal program RYC-2014-15686 and AYA2017-84061-P, the later one co-financed by FEDER (European Regional Development Funds).
U.D. acknowledges the support of STFC studentship (ST/R504725/1). 
J.B. acknowledges the support of STFC studentship (ST/S50536/1).

\software{XID+ \citep{Hurley:2017aa}, Scipy \citep{Jones:2001aa}, LE PHARE \citep{Arnouts:1999aa, Ilbert:2006aa}, SMURF \citep{Chapin:2013aa}, PICARD \citep{Jenness:2008aa}, SExtractor \citep{Bertin:1996aa}.}

\appendix
\setcounter{figure}{0}
\renewcommand\thefigure{\thesection.\arabic{figure}}

\section{Monte Carlo Simulation} \label{sec:MonteCarloSimulation}

\begin{figure*}
\gridline{ \fig{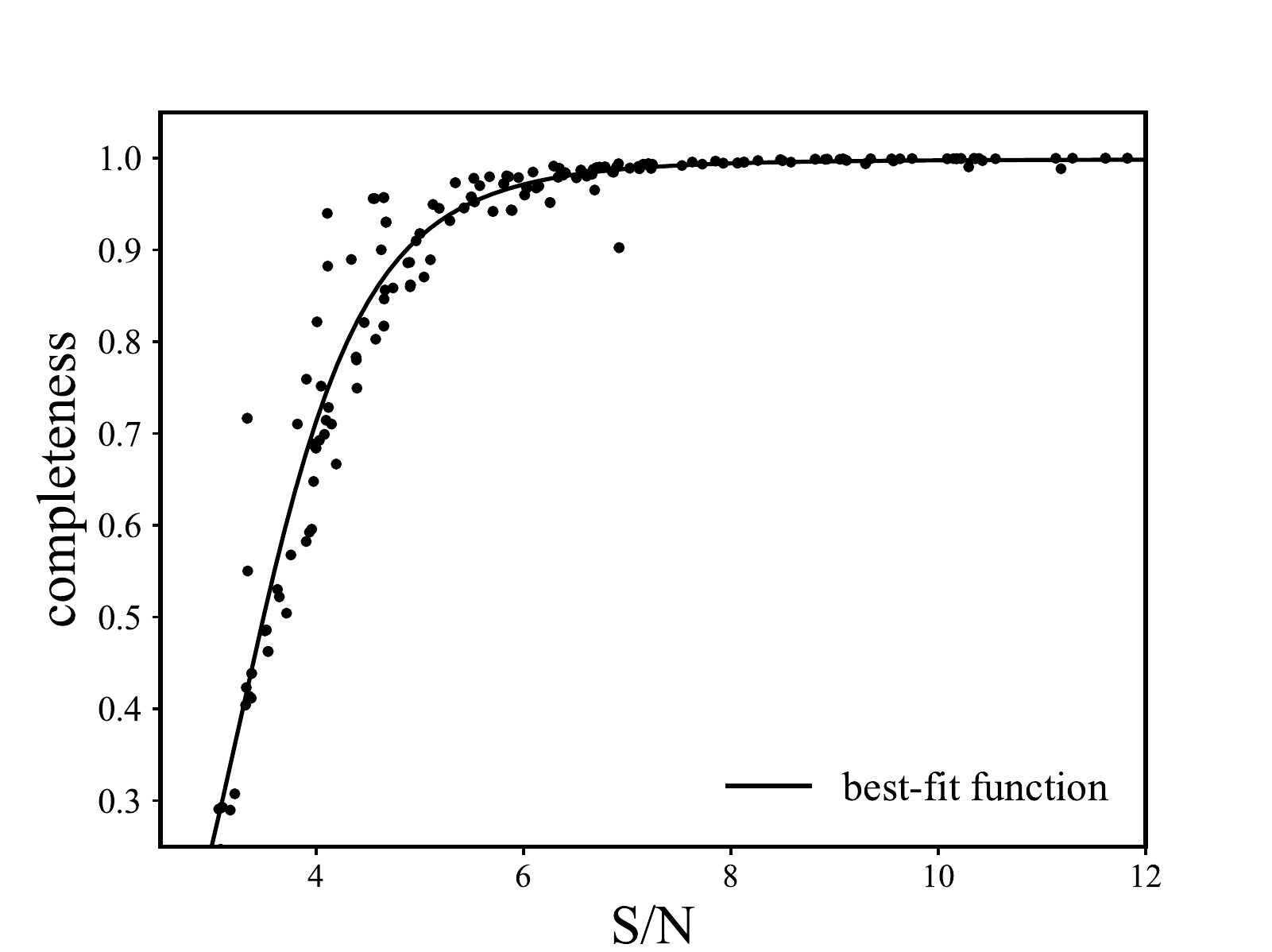}{0.37\textwidth}{(a)}\hspace{-2em}%
\fig{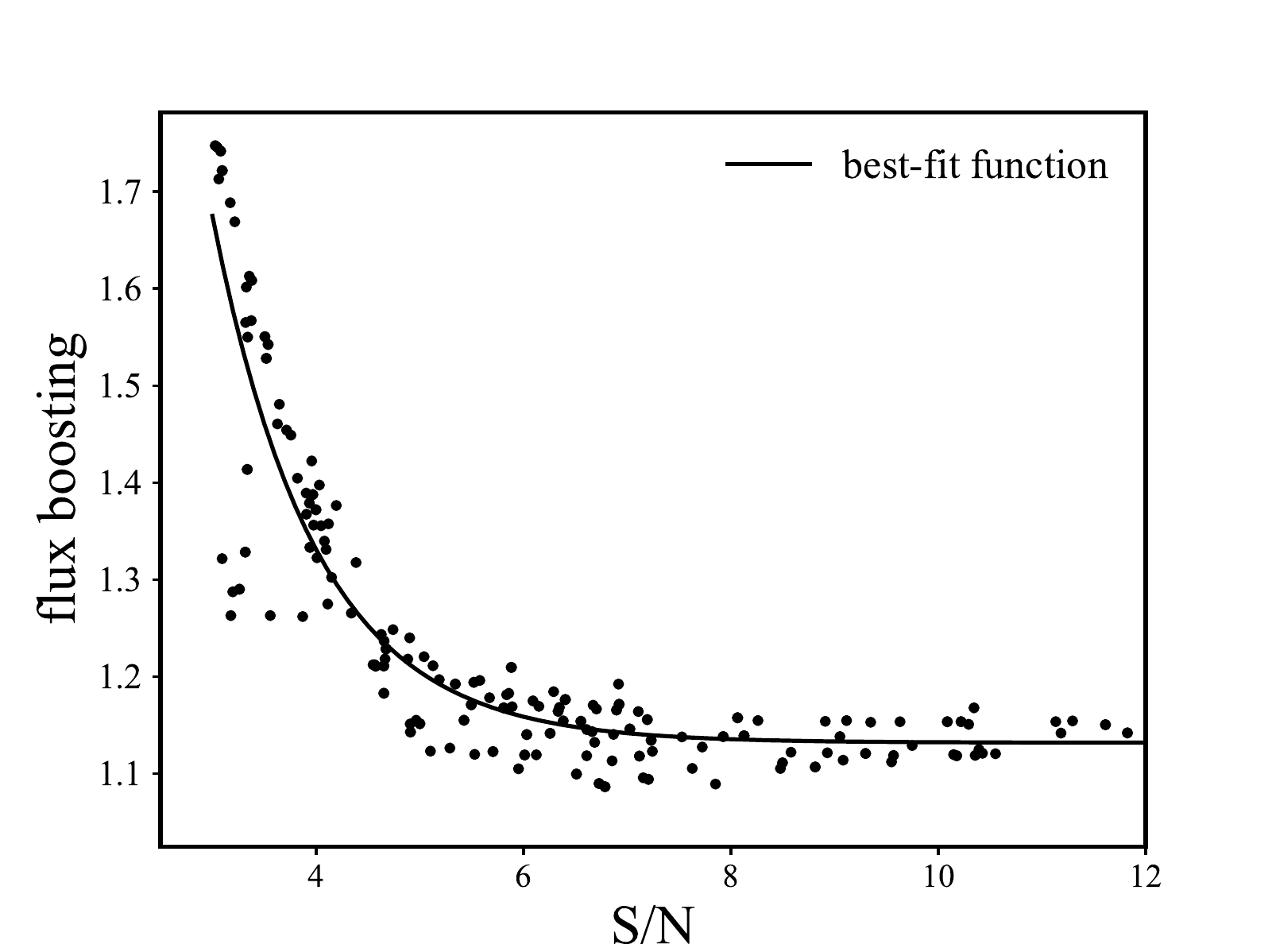}{0.37\textwidth}{(b)}\hspace{-2em}%
\fig{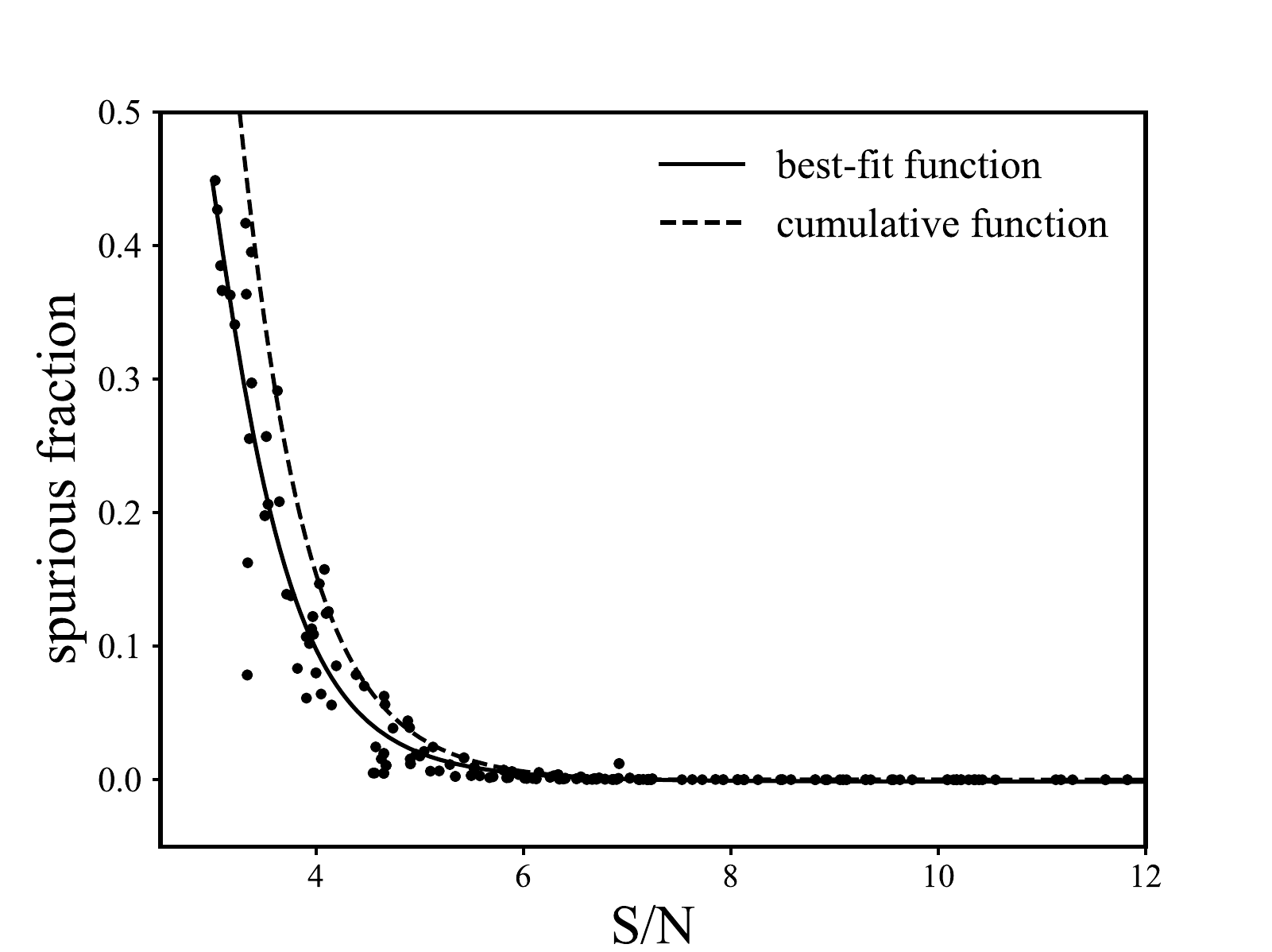}{0.37\textwidth}{(c)}
}
\caption{ (a) Completeness, (b) flux boosting, and (c) spurious source corrections from 200 source realizations as a function of S/N. }
\label{fig:CFScorrections_snr}
\end{figure*}

We performed Monte Carlo simulations to compute the detection completeness, flux boosting, and spurious fraction. We first generated a ``true noise'' map by using the jackknife technique (e.g., \citealt{Cowie:2002aa, Chapin:2013aa, Wang:2017aa}). We divided the individual scans into two interlacing halves, then co-added them separately. After that, they were subtracted from one another to make a clean removal of astronomical sources. The resultant map was scaled by $\sqrt{t_{1}t_{2}}/(t_{1} +t_{2})$, where $t_{1}$ and $t_{1}$ are the noise-weighted integration times of each pixel in each of the two half-maps. We verify that the r.m.s noise estimated from the true noise map is consistent with the instrumental noise calculated by SMURF. 

To recover the observational biases, we randomly inserted the scaled synthetic PSF into this jackknife map without any clustering and with intrinsic (corrected) counts (see below) in 1--50\,mJy. According to \citealt{Wang:2017aa}, there is no significant difference between their observed counts and the counts in the simulations of infrared-to-sub-millimeter extragalactic sky with clustering \citep{Bethermin:2017aa}. This indicates that the clustering of 450-$\micron$ sources on the scale of our beam size is likely to be weak. Because the effects of observational biases crucially depend on the intrinsic counts, we adopted an iterative procedure to determine the intrinsic counts from the observed raw counts. We fitted our observed raw counts with a Schechter function and took this to be the initial source counts. We then ran the source extraction on the simulated image and derived the output source counts. We estimated the ratio between the input and output source counts and used this to adjust the input counts for the next iteration. We then repeated the procedure 300 times. The first 100 simulations make the output counts converge to the observed raw counts. Utilizing the results from the remaining 200 simulations (including the position information and flux densities of the input and output sources), we can calculate the completeness, flux boosting, and spurious source corrections. We randomly choose 200 sources and show these bias effects as a function of S/N in Figure \ref{fig:CFScorrections_snr}. 

We estimated the expected positional errors from the Monte Carlo simulations. The mean positional offset between the input positions and the measured output positions is $\simeq$ 1.2$\arcsec$ for 4$\sigma$ sources, where the 90\% confidence interval is $\simeq$ 4$\arcsec$ (under a maximum search radius of 7.0$\arcsec$). Therefore, to estimate the completeness, we matched the sources between input and output catalogs using a search radius of 4$\arcsec$ from the input source positions. An input source without a match is considered to be undetected. The ratio between the numbers of matched output sources and the total number of input sources is the completeness factor (Figure \ref{fig:CFScorrections_snr}a). The completeness is about 73\%, 91\%, and 97\% at 4$\sigma$, 5$\sigma$, and 6$\sigma$, respectively.

To estimate the flux boosting caused by noise and the confusion of faint sources and the spurious fraction, we matched the sources in both the input and output catalogs by using a search radius of 4$\arcsec$ from the output source positions. For the flux-boosting estimation, we need to ensure that the input and output sources have similar flux densities. Therefore, we only consider matches when the flux densities of the input and output sources are within a factor of 2 of each other. When multiple input sources meet the above flux ratio criterion, the brightest one is considered the match. The mean output-to-input flux density ratio of matched sources is the flux-boosting factor (Figure \ref{fig:CFScorrections_snr}b). The flux-boosting corrections are about 1.3, 1.2, and 1.1 at 4$\sigma$, 5$\sigma$, and 6$\sigma$, respectively. On the other hand, an output source without a match, or where the flux densities of matched input and output sources are larger than a factor of 2 from each other, is considered as a spurious source (Figure \ref{fig:CFScorrections_snr}c). The spurious source fractions are 9\%, 2\%, and 0\% at 4$\sigma$, 5$\sigma$, and 6$\sigma$, respectively.

\section{Confusion Noises} \label{sec:ConfusionNoise}

\begin{figure}
\gridline{\fig{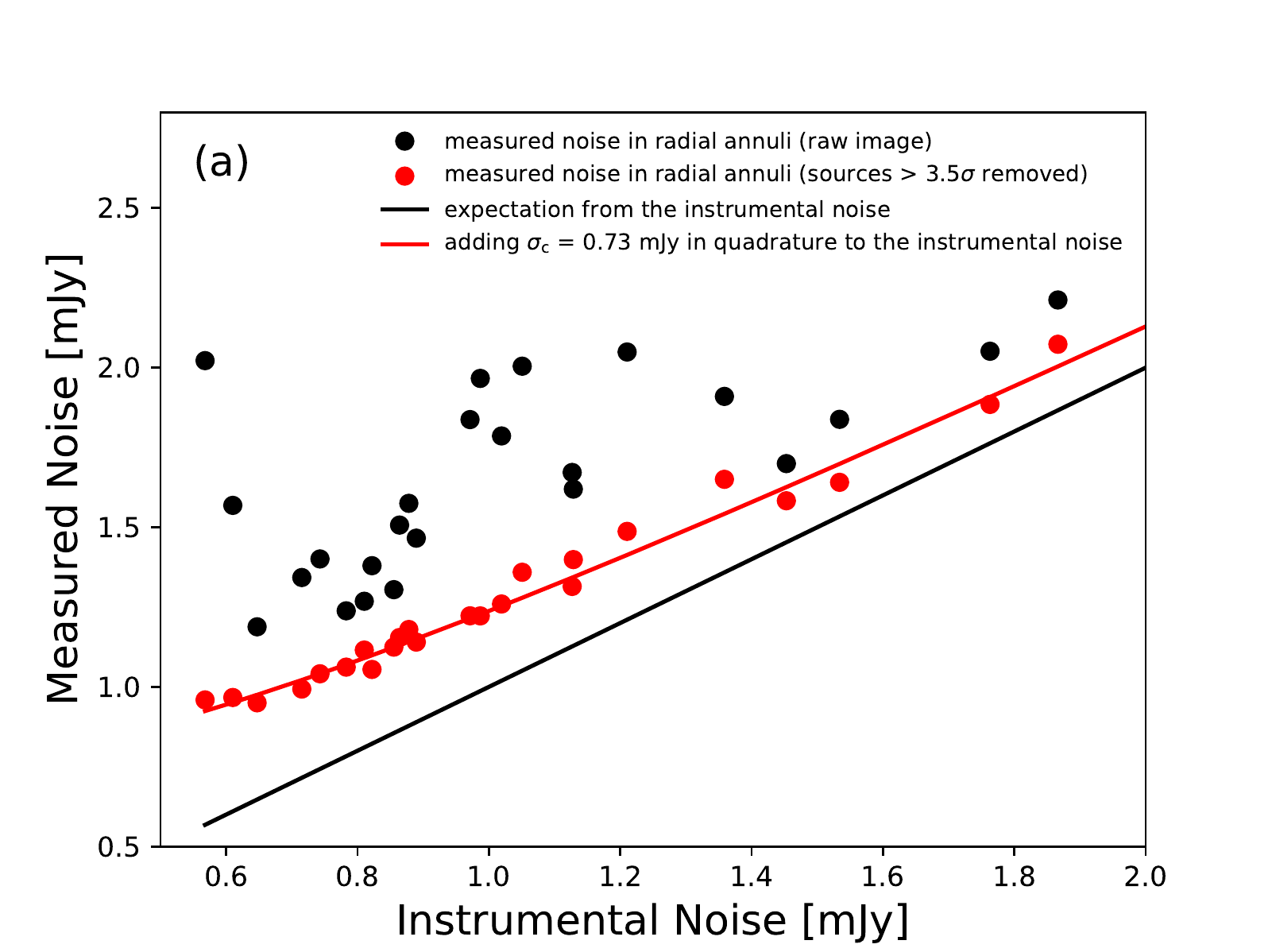}{0.5\textwidth}{}
\fig{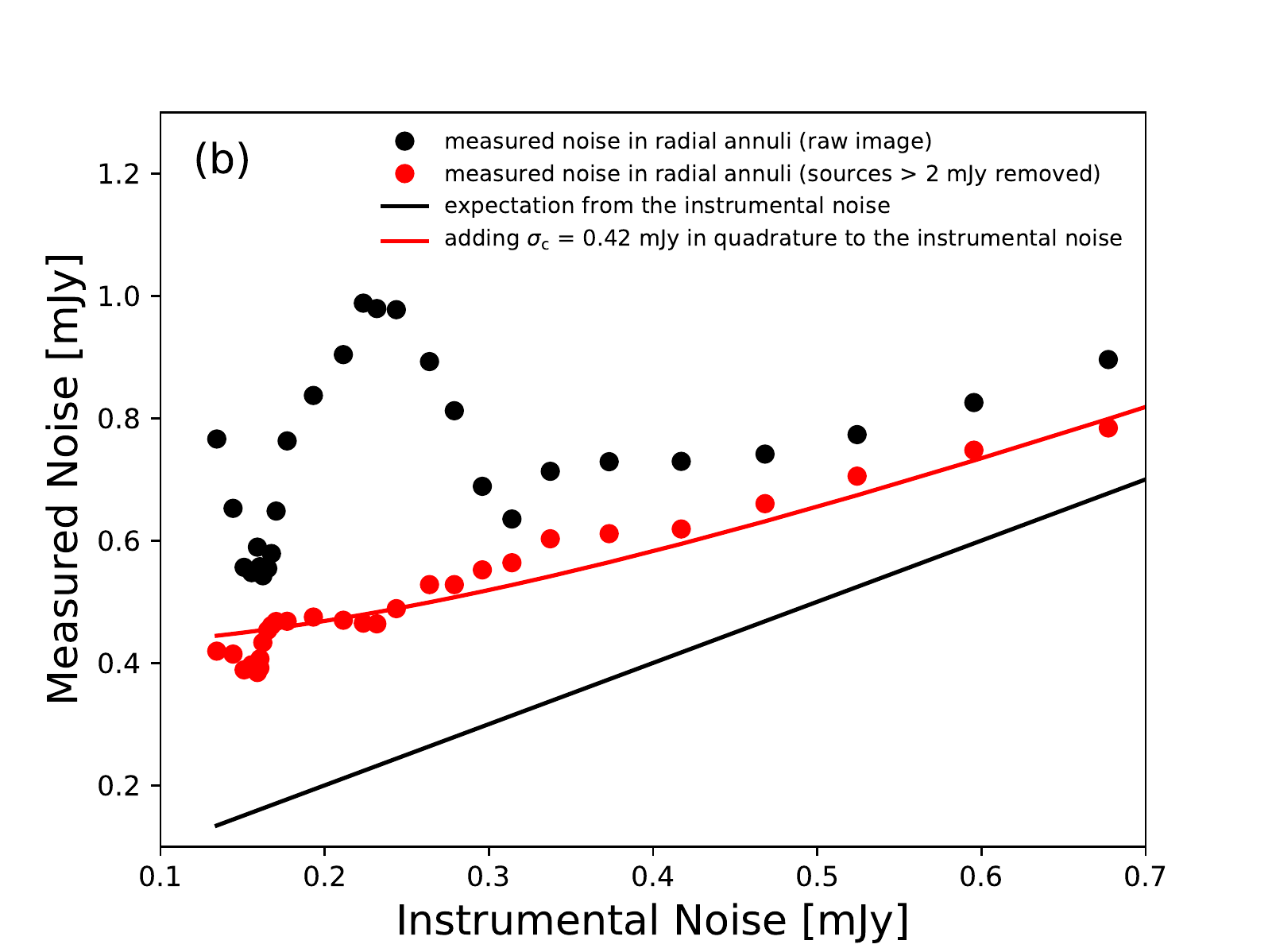}{0.5\textwidth}{}}
\caption{(a) Measured 450\,$\micron$ noise in radial annuli as a function of the mean instrumental noise in each annulus. The black points show the noise measured from the raw image, while the red points represent the flux dispersion measured in the image after the sources that are $\gt 3.5\sigma$ are removed. The black curve shows the mean instrumental noise, and the red curve shows the result when $\sigma_{\rm c}$ = 0.73\,mJy is added in quadrature to the instrumental noise. (b) Same as panel (a) but for 850\,$\micron$. The red points show the flux dispersion measured in the image after the sources brighter than 2\,mJy are removed. The red curve shows the result when $\sigma_{\rm c}$ = 0.42\,mJy is added in quadrature to the instrumental noise.}
\label{fig:ConfusionNoise}
\end{figure}

We estimate the confusion noises ($\sigma_{\rm c}$) at 450 and 850\,$\micron$ from the images by comparing the measured local flux density dispersions with their corresponding instrumental noises ($\sigma_{\rm i}$). In Figure \ref{fig:ConfusionNoise}, we show the measured noises in radial annuli around the map centers as functions of the mean instrumental noises in each annulus. The black points show the flux density dispersions measured from the raw images. The large variation in the measurement is mainly caused by the brighter sources. The red points represent the flux dispersion measured from the image with the bright sources removed (greater than $3.5\sigma$ for 450\,$\micron$ and brighter than 2\,mJy for 850\,$\micron$, respectively). As the instrumental noise ($\sigma_{\rm i}$, black curves in Figure \ref{fig:ConfusionNoise}) becomes smaller, the measured dispersions (with bright sources removed) should asymptotically approach $\sigma_{\rm c}$. We then minimize the $\chi^2$ in the function of $\sigma_{\rm c}^{2} = \sigma_{\rm total}^{2} - \sigma_{\rm i}^{2}$ and find the best-fit value of $\sigma_{\rm c}$. The red curves in Figure \ref{fig:ConfusionNoise} show the results of adding best-fit $\sigma_{\rm c}$ (0.73\,mJy for 450\,$\micron$ and 0.42\,mJy for 850\,$\micron$) in quadrature to $\sigma_{\rm i}$.

\begin{deluxetable*}{cccc}
\tablecaption{\label{tbl:OriginOfBands} Summary of Broad-band Photometry Used in This Work. }
\tablehead{ \colhead{ Instrument/Telescope } & \colhead{ Broad-band Filter } & \colhead{ $3\sigma$ Depth } & \colhead{ Origin Catalog } }
\startdata
GALEX & FUV & 25.5\,mag ($3\sigma$) & COSMOS2015  \\
  	    & NUV & 25.5\,mag ($3\sigma$) & COSMOS2015  \\
\hline
MegaCam/CFHT & $u^{\star}$ & 27.2\,mag ($3\sigma$) & COSMOS2015 \\
\hline
Suprime-cam/Subaru & $B$ & 27.6\,mag ($3\sigma$) & COSMOS2015  \\
 				 & $V$ & 26.9\,mag ($3\sigma$) & COSMOS2015  \\
 				 & $r$ & 27.0\,mag ($3\sigma$) & COSMOS2015  \\
 				 & $i^{+}$ & 26.9\,mag ($3\sigma$) & COSMOS2015  \\
 				 & $z^{++}$ & 26.4\,mag ($3\sigma$) & COSMOS2015  \\
\enddata
\tablenotetext{a} {Catalog is extracted by using \texttt{SExtractor} (see \S\ref{subsec:AncillaryData}).}
\tablenotetext{b} {Catalog is extracted by using \texttt{XID+} (see \S\ref{subsec:AncillaryData}).}
\tablenotetext{c} {See \S\ref{subsec:SourceExtraction}.}
\end{deluxetable*}

\startlongtable
\begin{longrotatetable}
\tabletypesize{\scriptsize}
\begin{deluxetable*}{llllllllllllll}
\tablecaption{\label{tbl:Catalog1} Derived Properties of VLA 3-GHz Identified 450-$\micron$ Sources.}
\tablehead{ 
\colhead{ R.A.$_\mathrm{450\,\micron}$\tablenotemark{a} } & \colhead{ Decl.$_\mathrm{450\,\micron}$\tablenotemark{a} } 
& \colhead{ R.A.$_\mathrm{3\,GHz}$ } & \colhead{ Decl.$_\mathrm{3\,GHz}$ }
& \colhead{ $S_\mathrm{450\,\micron}$\tablenotemark{b} } & \colhead{ $S_\mathrm{850\,\micron}$\tablenotemark{b} }
& \colhead{ $z$ } & \colhead{ flag$_{z}$\tablenotemark{c} } & \colhead{ log($M_{\ast}$)\tablenotemark{d} } & \colhead{ log($L_{\rm UV}$) }
& \colhead{ log($L_{\rm IR}$)\tablenotemark{d} } & \colhead{ $T_{\rm d}$ } & \colhead{ $\beta_{\rm UV}$ } & \colhead{ flag$_{\rm AGN}$\tablenotemark{e} } \\
\cline{1-14}
\colhead{ (J2000) } & \colhead{ (J2000) } &\colhead{ (J2000) } &\colhead{ (J2000) } & \colhead{ (mJy) } & \colhead{ (mJy) } & & & \colhead{ ($\rm M_{\sun}$) } & \colhead{ ($\rm L_{\sun}$) } & \colhead{ ($\rm L_{\sun}$) } & \colhead{ (K) } & & 
}
\startdata
09:59:57.23 & 2:21:27.47 & 09:59:57.11 & 2:21:27.46 & 13.94$\pm$4.30 & & 0.93$\pm$0.00 & 1 & 10.39$\pm$0.23 & 9.28$\pm$0.01 & 11.69$\pm$0.08 & 34.60$\pm$5.00 & -2.02$\pm$1.01 & \\
09:59:58.43 & 2:22:30.53 & 09:59:58.42 & 2:22:30.53 & 12.39$\pm$4.11 & & 0.90$\pm$0.03 & 2 & 10.71$\pm$0.24 & 9.59$\pm$0.04 & 11.83$\pm$0.03 & 44.98$\pm$1.95 & -2.18$\pm$1.26 & 1 \\
10:00:01.77 & 2:24:37.87 & 10:00:01.66 & 2:24:37.86 & 14.79$\pm$2.92 & 1.82$\pm$0.51 & 2.89$\pm$0.36 & 2 & 11.25$\pm$0.24 & 10.08$\pm$0.24 & 13.05$\pm$0.11 & 60.69$\pm$0.32 & 2.88$\pm$1.38 & \\
10:00:04.43 & 2:20:23.87 & 10:00:04.35 & 2:20:23.98 & 11.92$\pm$2.15 & 2.38$\pm$0.36 & 3.79$\pm$0.30 & 2 & 10.95$\pm$0.25 & 10.54$\pm$0.15 & 13.12$\pm$0.04 & 46.41$\pm$0.61 & 0.67$\pm$1.47 & \\
10:00:04.70 & 2:30:02.80 & 10:00:04.77 & 2:30:03.02 & 19.41$\pm$4.04 & 4.47$\pm$0.65 & 1.56$\pm$0.18 & 3 & & & 12.67$\pm$0.20 & & & \\
10:00:05.23 & 2:17:09.33 & 10:00:05.16 & 2:17:09.26 & 11.77$\pm$2.88 & 3.19$\pm$0.48 & 3.19$\pm$0.09 & 2 & 11.11$\pm$0.23 & 10.85$\pm$0.03 & 12.65$\pm$0.17 & 46.72$\pm$3.21 & -0.64$\pm$1.12 & \\
10:00:05.63 & 2:25:41.20 & 10:00:05.59 & 2:25:41.46 & 6.42$\pm$1.55 & & 1.00$\pm$0.00 & 1 & 10.58$\pm$0.22 & 9.32$\pm$0.02 & 12.28$\pm$0.58 & 63.35$\pm$1.43 & 4.07$\pm$1.43 & \\
10:00:08.17 & 2:26:44.80 & 10:00:08.04 & 2:26:44.82 & 18.95$\pm$1.27 & 14.29$\pm$0.20 & 4.25$\pm$0.04 & 3 & & & 13.04$\pm$0.08 & & & \\
\enddata
\tablenotetext{a} {The positional error at the 90\% confidence interval is $\simeq$ 4$\arcsec$ in radius (see more in \S\ref{sec:MonteCarloSimulation})}
\tablenotetext{b} {Here $S_\mathrm{450\,\micron}$ and $S_\mathrm{850\,\micron}$ are corrected for flux boosting.}
\tablenotetext{c} {1=spectroscopic redshift; 2=photometric redshift; 3=FIR redshift.}
\tablenotetext{d} {The uncertainties on $M_{\ast}$ and $L_{\rm IR}$ are derived  from \texttt{LE PHARE} by including the photometric errors in each waveband but without including the redshift uncertainties. }
\tablenotetext{e} {1=X-ray AGN; 2=mid-infrared AGN; 3=radio excess.}
\end{deluxetable*}
\end{longrotatetable}

%\startlongtable
\begin{longrotatetable}
\tabletypesize{\scriptsize}
\begin{deluxetable*}{llllllllllllll}
\tablecaption{\label{tbl:Catalog2} Derived Properties of MIPS 24-$\micron$ Identified 450-$\micron$ Sources.}
\tablehead{ 
\colhead{ R.A.$_\mathrm{450\,\micron}$\tablenotemark{a} } & \colhead{ Decl.$_\mathrm{450\,\micron}$\tablenotemark{a} } 
& \colhead{ R.A.$_\mathrm{24\,\micron}$ } & \colhead{ Decl.$_\mathrm{24\,\micron}$ }
& \colhead{ $S_\mathrm{450\,\micron}$\tablenotemark{b} } & \colhead{ $S_\mathrm{850\,\micron}$\tablenotemark{b} }
& \colhead{ $z$ } & \colhead{ flag$_{z}$\tablenotemark{c} } & \colhead{ log($M_{\ast}$)\tablenotemark{d} } & \colhead{ log($L_{\rm UV}$) }
& \colhead{ log($L_{\rm IR}$)\tablenotemark{d} } & \colhead{ $T_{\rm d}$ } & \colhead{ $\beta_{\rm UV}$ } & \colhead{ flag$_{\rm AGN}$\tablenotemark{e} } \\
\cline{1-14}
\colhead{ (J2000) } & \colhead{ (J2000) } &\colhead{ (J2000) } &\colhead{ (J2000) } & \colhead{ (mJy) } & \colhead{ (mJy) } & & & \colhead{ ($\rm M_{\sun}$) } & \colhead{ ($\rm L_{\sun}$) } & \colhead{ ($\rm L_{\sun}$) } & \colhead{ (K) } & & 
}
\startdata
09:59:52.70 & 2:27:48.67 & 09:59:52.64 & 2:27:48.86 & 14.67$\pm$4.86 & & 3.05$\pm$0.08 & 2 & 10.84$\pm$0.22 & 11.13$\pm$0.03 & 12.59$\pm$0.10 & 27.46$\pm$7.86 & 0.14$\pm$1.54 & \\
10:00:07.37 & 2:29:57.87 & 10:00:07.25 & 2:29:57.84 & 6.16$\pm$2.08 & & 0.75$\pm$0.00 & 1 & 11.08$\pm$0.22 & 9.14$\pm$0.01 & 11.94$\pm$0.17 & 31.45$\pm$7.14 & -0.23$\pm$1.20 & \\
10:00:08.97 & 2:19:16.13 & 10:00:08.77 & 2:19:16.16 & 10.62$\pm$1.72 & 1.52$\pm$0.26 & 1.46$\pm$0.00 & 1 & 10.18$\pm$0.22 & 10.60$\pm$0.00 & 12.42$\pm$0.07 & 56.81$\pm$1.79 & 1.74$\pm$1.73 & \\
10:00:12.30 & 2:23:34.27 & 10:00:12.35 & 2:23:34.34 & 2.98$\pm$0.98 & & 0.79$\pm$0.10 & 3 & & & 12.38$\pm$0.35 & & & \\
10:00:12.70 & 2:14:59.07 & 10:00:12.60 & 2:14:58.96 & 19.02$\pm$3.90 & 4.76$\pm$0.56 & 2.31$\pm$0.04 & 2 & 11.39$\pm$0.22 & 10.57$\pm$0.02 & 12.71$\pm$0.20 & 41.17$\pm$6.43 & -0.46$\pm$1.28 & \\
10:00:12.97 & 2:20:21.47 & 10:00:13.06 & 2:20:21.54 & 5.00$\pm$1.30 & & 0.38$\pm$0.00 & 1 & 10.83$\pm$0.22 & 8.93$\pm$0.01 & 10.84$\pm$0.07 & 29.87$\pm$5.00 & 0.66$\pm$1.39 & \\
10:00:13.77 & 2:17:10.40 & 10:00:13.67 & 2:17:10.30 & 8.27$\pm$1.69 & 3.10$\pm$0.26 & 2.89$\pm$0.97 & 2 & 11.26$\pm$0.24 & 9.46$\pm$0.29 & 12.32$\pm$0.17 & 36.18$\pm$8.57 & 11.51$\pm$0.54 & \\
10:00:14.17 & 2:18:12.53 & 10:00:14.17 & 2:18:12.53 & 5.14$\pm$1.58 & & 0.30$\pm$0.00 & 1 & 10.55$\pm$0.22 & 7.82$\pm$0.10 & 10.85$\pm$0.59 & 21.38$\pm$12.14 & & \\
\enddata
\tablenotetext{a} {The positional error at the 90\% confidence interval is $\simeq$ 4$\arcsec$ in radius (see more in \S\ref{sec:MonteCarloSimulation})}
\tablenotetext{b} {Here $S_\mathrm{450\,\micron}$ and $S_\mathrm{850\,\micron}$ are corrected for flux boosting.}
\tablenotetext{c} {1=spectroscopic redshift; 2=photometric redshift; 3=FIR redshift.}
\tablenotetext{d} {The uncertainties on $M_{\ast}$ and $L_{\rm IR}$ are derived  from \texttt{LE PHARE} by including the photometric errors in each waveband but without including the redshift uncertainties. }
\tablenotetext{e} {1=X-ray AGN; 2=mid-infrared AGN; 3=radio excess.}
\end{deluxetable*}
\end{longrotatetable}

\startlongtable
\begin{deluxetable*}{lllllllllll}
\centering
\tablecaption{\label{tbl:PhotometricCatalog1} Multi-wavelength Photometry for VLA 3-GHz Identified 450-$\micron$ Sources.}
\tablehead{ 
\colhead{ R.A.$_\mathrm{450\,\micron}$\tablenotemark{a} } & \colhead{ Decl.$_\mathrm{450\,\micron}$\tablenotemark{a} } & \colhead{ R.A.$_{opt}$ } & \colhead{ Decl.$_{opt}$ } 
& \colhead{ $r$} & \colhead{ $Ks$} & \colhead{ $S_\mathrm{3.6\,\micron}$} 
& \colhead{ $S_\mathrm{250\,\micron}$\tablenotemark{b} } & \colhead{ $S_\mathrm{450\,\micron}$\tablenotemark{c} } & \colhead{ $S_\mathrm{850\,\micron}$\tablenotemark{c} } 
& \colhead{ $S_\mathrm{3\,GHz}$} \\
\cline{1-11}
\colhead{ (J2000) } & \colhead{ (J2000) } &\colhead{ (J2000) } & \colhead{ (J2000) } & \colhead{ (mag) }  & \colhead{ (mag) } & \colhead{ (mag) } & \colhead{ (mJy) } & \colhead{ (mJy) } & \colhead{ (mJy) } & \colhead{ ($\mu$Jy) }
}
\startdata
09:59:57.23 & 2:21:27.47 & 09:59:57.10   & 2:21:27.45    & 24.20$\pm$0.04 & 20.98$\pm$0.02 & 20.02$\pm$0.01 & 20.75$\pm$2.64 & 13.94$\pm$4.30 &                & 33.8$\pm$2.9     \\
09:59:58.43 & 2:22:30.53 & 09:59:58.42   & 2:22:30.53    & 23.59$\pm$0.03 & 20.67$\pm$0.01 & 19.45$\pm$0.01 & 1.50$\pm$1.43  & 12.39$\pm$4.11 &                & 37.2$\pm$2.9     \\
10:00:01.77 & 2:24:37.87 & 10:00:01.67   & 2:24:37.87    &                & 22.80$\pm$0.08 & 21.37$\pm$0.01 & 30.76$\pm$2.62 & 14.79$\pm$2.92 & 1.82$\pm$0.51  & 53.1$\pm$3.5     \\
10:00:04.43 & 2:20:23.87 & 10:00:04.35   & 2:20:23.98    & 26.75$\pm$0.24 & 23.87$\pm$0.22 & 22.61$\pm$0.03 & 18.73$\pm$2.30 & 11.92$\pm$2.15 & 2.38$\pm$0.36  & 29.7$\pm$2.8     \\
10:00:04.70 & 2:30:02.80 & 10:00:04.80   & 2:30:03.02    &                & 22.46$\pm$0.07 & 20.49$\pm$0.01 & 20.51$\pm$2.85 & 19.41$\pm$4.04 & 4.47$\pm$0.65  & 48.1$\pm$3.4     \\
10:00:05.23 & 2:17:09.33 & 10:00:05.14   & 2:17:09.29    & 25.28$\pm$0.08 & 22.50$\pm$0.06 & 21.32$\pm$0.01 & 10.42$\pm$2.56 & 11.77$\pm$2.88 & 3.19$\pm$0.48  & 12.3$\pm$2.4     \\
10:00:05.63 & 2:25:41.20 & 10:00:05.61   & 2:25:41.46    & 24.37$\pm$0.04 & 20.74$\pm$0.01 & 20.01$\pm$0.01 & 42.31$\pm$2.25 & 6.42$\pm$1.55  &                & 11.5$\pm$2.3     \\
10:00:08.17 & 2:26:44.80 &               &               &                &                &                & 27.69$\pm$2.18 & 18.95$\pm$1.27 & 14.29$\pm$0.20 & 15.0$\pm$2.4     \\
\enddata
\tablenotetext{a} {The positional error at the 90\% confidence interval is $\simeq$ 4$\arcsec$ in radius (see more in \S\ref{sec:MonteCarloSimulation}).}
\tablenotetext{b} {Here $S_\mathrm{250\,\micron}$ are estimated from \texttt{XID+} with our 450-$\micron$ sources as positional priors (see \S\ref{subsec:AncillaryData}).}
\tablenotetext{c} {Here $S_\mathrm{450\,\micron}$ and $S_\mathrm{850\,\micron}$ are corrected for flux boosting.}
\end{deluxetable*}

\startlongtable
\begin{deluxetable*}{lllllllllll}
\tablecaption{\label{tbl:PhotometricCatalog2} Multi-wavelength Photometry for MIPS 24-$\micron$ Identified 450-$\micron$ Sources. }
\tablehead{ 
\colhead{ R.A.$_\mathrm{450\,\micron}$\tablenotemark{a} } & \colhead{ Decl.$_\mathrm{450\,\micron}$\tablenotemark{a} } & \colhead{ R.A.$_{opt}$ } & \colhead{ Decl.$_{opt}$ } 
& \colhead{ $r$} & \colhead{ $Ks$} & \colhead{ $S_\mathrm{3.6\,\micron}$}  & \colhead{ $S_\mathrm{24\,\micron}$}
& \colhead{ $S_\mathrm{250\,\micron}$\tablenotemark{b} } & \colhead{ $S_\mathrm{450\,\micron}$\tablenotemark{c} } & \colhead{ $S_\mathrm{850\,\micron}$\tablenotemark{c} }  \\
\cline{1-11}
\colhead{ (J2000) } & \colhead{ (J2000) } &\colhead{ (J2000) } & \colhead{ (J2000) } & \colhead{ (mag) }  & \colhead{ (mag) } & \colhead{ (mag) } & \colhead{ (mJy) } & \colhead{ (mJy) } & \colhead{ (mJy) } & \colhead{ (mJy) }
}
\startdata
09:59:52.70 & 2:27:48.67 & 09:59:52.67   & 2:27:48.92    & 24.67$\pm$0.05 & 22.32$\pm$0.05 & 20.83$\pm$0.01 & 0.26$\pm$0.01 & 3.41$\pm$2.34  & 14.67$\pm$4.86 &               \\
10:00:07.37 & 2:29:57.87 & 10:00:07.26   & 2:29:57.94    & 23.05$\pm$0.02 & 19.68$\pm$0.00 & 19.05$\pm$0.00 & 0.44$\pm$0.01 & 27.70$\pm$2.41 & 6.16$\pm$2.08  &               \\
10:00:08.97 & 2:19:16.13 & 10:00:08.76   & 2:19:16.13    & 22.73$\pm$0.02 & 20.38$\pm$0.00 & 19.17$\pm$0.01 & 0.41$\pm$0.01 & 23.34$\pm$2.58 & 10.62$\pm$1.72 & 1.52$\pm$0.26 \\
10:00:12.30 & 2:23:34.27 &               &               &                &                &                & 0.29$\pm$0.01 & 12.10$\pm$1.65 & 2.98$\pm$0.98  &               \\
10:00:12.70 & 2:14:59.07 & 10:00:12.60   & 2:14:58.94    & 24.83$\pm$0.06 & 21.18$\pm$0.01 & 20.02$\pm$0.01 & 0.33$\pm$0.02 & 24.16$\pm$3.20 & 19.02$\pm$3.90 & 4.76$\pm$0.56 \\
10:00:12.97 & 2:20:21.47 & 10:00:13.09   & 2:20:21.55    & 20.89$\pm$0.01 & 18.60$\pm$0.00 & 18.46$\pm$0.01 & 0.43$\pm$0.01 & 13.55$\pm$2.04 & 5.00$\pm$1.30  &               \\
10:00:13.77 & 2:17:10.40 & 10:00:13.69   & 2:17:10.30    &                & 23.24$\pm$0.06 & 21.60$\pm$0.01 & 0.16$\pm$0.02 & 5.73$\pm$2.49  & 8.27$\pm$1.69  & 3.10$\pm$0.26 \\
10:00:14.17 & 2:18:12.53 & 10:00:14.17   & 2:18:12.58    & 23.97$\pm$0.04 & 20.84$\pm$0.01 & 20.02$\pm$0.01 & 0.25$\pm$0.02 & 13.90$\pm$1.94 & 5.14$\pm$1.58  &               \\
\enddata
\tablenotetext{a} {The positional error at the 90\% confidence interval is $\simeq$ 4$\arcsec$ in radius (see more in \S\ref{sec:MonteCarloSimulation}).}
\tablenotetext{b} {Here $S_\mathrm{250\,\micron}$ are estimated from \texttt{XID+} with our 450-$\micron$ sources as positional priors (see \S\ref{subsec:AncillaryData}).}
\tablenotetext{c} {Here $S_\mathrm{450\,\micron}$ and $S_\mathrm{850\,\micron}$ are corrected for flux boosting.}
\end{deluxetable*}

\begin{deluxetable*}{llllll}
\tablecaption{\label{tbl:PhotometricCatalog3} List of 450-$\micron$ Sources without VLA 3\,GHz and MIPS 24\,$\micron$ Counterparts.}
\tablehead{ 
\colhead{ R.A.$_\mathrm{450\,\micron}$ } & \colhead{ Decl.$_\mathrm{450\,\micron}$ } 
& \colhead{ $S_\mathrm{250\,\micron}$\tablenotemark{a} } & \colhead{ $S_\mathrm{450\,\micron}$\tablenotemark{b} } & \colhead{ $S_\mathrm{850\,\micron}$\tablenotemark{b} } \\
\cline{1-2}
\colhead{ (J2000) } & \colhead{ (J2000) } & \colhead{ (mJy) } & \colhead{ (mJy) } & \colhead{ (mJy) } 
}
\startdata
10:00:09.90 & 2:26:47.07 & 0.49$\pm$0.61  & 3.28$\pm$1.16   &               \\
10:00:12.30 & 2:23:32.53 & 3.94$\pm$1.62  & 5.00$\pm$1.01   & 1.67$\pm$0.15 \\
10:00:14.30 & 2:23:33.73 & 6.35$\pm$1.91  & 3.06$\pm$0.90   &               \\
10:00:14.70 & 2:28:52.13 & 3.98$\pm$1.69  & 3.77$\pm$0.96   &               \\
10:00:16.03 & 2:24:38.93 & 11.43$\pm$1.59 & 3.75$\pm$0.88   & 2.86$\pm$0.14 \\
10:00:17.10 & 2:17:09.47 & 5.53$\pm$1.99  & 4.90$\pm$1.58   &               \\
10:00:17.23 & 2:25:41.33 & 5.79$\pm$1.45  & 5.93$\pm$0.90   & 2.23$\pm$0.14 \\
10:00:18.17 & 2:25:43.73 & 2.61$\pm$1.42  & 2.95$\pm$0.89   &               \\
\enddata
\tablenotetext{a} {Here $S_\mathrm{250\,\micron}$ are estimated from \texttt{XID+} with our 450-$\micron$ sources as positional priors (see \S\ref{subsec:AncillaryData}).}
\tablenotetext{b} {Here $S_\mathrm{450\,\micron}$ and $S_\mathrm{850\,\micron}$ are corrected for flux boosting.}
\end{deluxetable*}

\begin{deluxetable*}{ccc}
\tablecaption{\label{tbl:FiltersBetaUV} Broad-band Filters Used to Derive $\beta_{\rm UV}$ and Ultraviolet Luminosity.}
\tablehead{ \colhead{ Redshift Range } & \colhead{ $M_{\lambda_{1}}$ } & \colhead{ $M_{\lambda_{2}}$ } }
\startdata
0.0--0.5 & $\rm FUV_{1542}$ & $\rm NUV_{2314}$ \\
0.5--1.0 & $\rm NUV_{2314}$ & $u^{\star}_{3823}$ \\
1.0--1.5 & $B_{1542}$ & $V_{5478}$ \\
1.5--2.0 & $V_{5478}$ & $r_{6289}$ \\
2.0--2.5 & $r_{6289}$ & $i^{+}_{7684}$ \\
2.5--3.5 & $i^{+}_{7684}$ & $z^{++}_{9106}$ \\
3.5--4.5 & $z^{++}_{9106}$ & $J_{12535}$ \\
4.5--5.0 & $z^{++}_{9106}$ & $(J_{12535}+H_{16453})/2$ \\
5.0--6.0 & $J_{12535}$ & $H_{16453}$ \\
\enddata
\end{deluxetable*}

\begin{deluxetable*}{cccc}
\centering
\tablecaption{\label{tbl:LF(Vmax)} Infrared LF from 1/V$_{\rm max}$ Method.\tablenotemark{a}}
\tablehead{ \colhead{ log$(L_{\rm IR}/\rm L_{\sun})$ } & \multicolumn{3}{c}{log$(\Phi/\rm Mpc^{-3}\,dex^{-1})$} \\
\cline{2-4}
\colhead{} & \colhead{$0.0 < z < 1.3$} & \colhead{$1.3 < z < 2.5$} & \colhead{$2.5 < z < 4.0$} }
\startdata
10.6 & -2.99 $\pm$ 0.56 (2)& & \\
10.8 & -2.75 $\pm$ 0.30 (7)& & \\
11.0 & -3.67 $\pm$ 0.51 (3)& & \\
11.2 & -3.47 $\pm$ 0.33 (6)& & \\
11.4 & -3.31 $\pm$ 0.25 (8)& -3.25 $\pm$ 0.33 (6) & \\
11.6 & -3.60 $\pm$ 0.28 (11)& -3.77 $\pm$ 0.41 (5)& \\
11.8 & -3.13 $\pm$ 0.17 (12)& -2.91 $\pm$ 0.12 (17)& -3.72 $\pm$ 0.62 (2) \\
12.0 & -4.16 $\pm$ 0.56 (2)& -3.49 $\pm$ 0.19 (15)& -3.90 $\pm$ 0.58 (2) \\
12.2 & -3.89 $\pm$ 0.31 (5)& -3.41 $\pm$ 0.21 (9)& -2.99 $\pm$ 0.36 (3) \\
12.4 & & -3.92 $\pm$ 0.23 (12)& -3.32 $\pm$ 0.25 (10)\\
12.6 & & -4.29 $\pm$ 0.33 (7)& -3.69 $\pm$ 0.27 (11) \\
12.8 & & -5.29 $\pm$ 1.00 (1)& -5.12 $\pm$ 1.00 (1)\\
13.0 & & & -4.00 $\pm$ 0.26 (5)\\
\enddata
\tablenotetext{a}{The values in the parentheses in each column describe the number of sources in each luminosity and redshift bin.}
\end{deluxetable*}

\begin{deluxetable*}{cccccc}
\centering
\tablecaption{\label{tbl:LF(likelihood)} Parameter for the Best-fit Infrared LF from The Likelihood Method.}
\tablehead{ \colhead{redshift range} & \colhead{log$(L_{\ast}/\rm L_{\sun})$} & \colhead{log$(\Phi_{\ast}/\rm Mpc^{-3}\,dex^{-1})$} & \colhead{$\alpha$} & \colhead{log$(L_{\ast}/\rm L_{\sun})$\tablenotemark{a}} & \colhead{log$(\Phi_{\ast}/\rm Mpc^{-3}\,dex^{-1})$\tablenotemark{a}} }
\startdata
$0.0 < z < 1.3$ & 11.91$^{+0.32}_{-0.22}$ & $-3.24^{+0.19}_{-0.36}$ & $-0.64^{+0.44}_{-0.40}$ & $-3.14^{+0.18}_{-0.29}$ & 11.89$^{+0.10}_{-0.08}$ \\
$1.3 < z < 2.5$ & 12.26$^{+0.27}_{-0.17}$ & $-3.07^{+0.17}_{-0.36}$ & $-0.39^{+0.54}_{-0.52}$ & $-3.17^{+0.17}_{-0.30}$ & 12.31$^{+0.08}_{-0.08}$\\
$2.5 < z < 4.0$ & 12.36$^{+0.31}_{-0.19}$ & $-3.17^{+0.20}_{-0.38}$ & $0.86^{+1.04}_{-1.16}$ & $-3.77^{+0.18}_{-0.29}$ & 12.77$^{+0.14}_{-0.12}$\\
\enddata
\tablenotetext{a} {Derived with a fixed faint-end slope.}
\end{deluxetable*}

\begin{deluxetable*}{ccccccc}
\tablecaption{\label{tbl:SFRD} SFRDs Derived from the Likelihood LFs}
\tablehead{ \colhead{Redshift} & \colhead{SFRD\tablenotemark{a}} & \colhead{SFRD (LIRGs)} & \colhead{SFRD (ULIRGs)} & \colhead{SFRD\tablenotemark{a}$^{,}$\tablenotemark{b}} & \colhead{SFRD (LIRGs)\tablenotemark{b}} & \colhead{SFRD (ULIRGs)\tablenotemark{b}} \\
\cline{2-7}
& \multicolumn{6}{c}{($\rm M_{\sun}\,yr^{-1}\,Mpc^{-3}$)}}
\startdata
$0.0 < z < 1.3$ & 0.052$^{+0.464}_{-0.004}$ & 0.031$^{+0.214}_{-0.025}$ & 0.006$^{+0.050}_{-0.005}$ & 0.053$^{+0.047}_{-0.031}$ & 0.033$^{+0.027}_{-0.019}$ & 0.007$^{+0.011}_{-0.005}$\\
$1.3 < z < 2.5$ & 0.126$^{+0.905}_{-0.099}$ & 0.066$^{+0.518}_{-0.053}$ & 0.052$^{+0.227}_{-0.039}$ & 0.128$^{+0.107}_{-0.074}$ & 0.070$^{+0.048}_{-0.038}$ & 0.050$^{+0.052}_{-0.031}$\\
$2.5 < z < 4.0$ & 0.095$^{+0.272}_{-0.041}$ & 0.008$^{+0.181}_{-0.015}$ & 0.081$^{+0.191}_{-0.069}$ & 0.094$^{+0.103}_{-0.059}$ & 0.035$^{+0.027}_{-0.020}$ & 0.057$^{+0.062}_{-0.036}$\\
\enddata
\tablenotetext{a} {The results are derived from our best-fit likelihood LFs with integration limits of $L_{\rm min}= 0.03\,L_{\ast}$ to $L_{\rm max}=10^{13.5}\rm L_{\sun}$.}
\tablenotetext{b} { Derived with fixed faint-end slope.}
\end{deluxetable*}

\clearpage

\bibliography{references.bib}

\end{document}